\documentclass[runningheads]{llncs}
\usepackage[utf8]{inputenc}
\usepackage{amssymb}
\usepackage{amsmath}
\usepackage{latexsym}
\usepackage{textcomp}
\usepackage{graphicx}
\usepackage{xspace}
\usepackage{paralist} 
\usepackage{wrapfig}
\usepackage{lineno}
\usepackage{hyperref}
\usepackage{cleveref}
\usepackage{marvosym}
\usepackage{subfigure}

\usepackage{lscape}
\usepackage{pdflscape}
\usepackage{hhline}

\usepackage{rotating}

\usepackage{multirow}
\usepackage{multicol}
\usepackage{color, colortbl}

\usepackage{calc}
\usepackage{tabularx}

\usepackage{caption} 
\definecolor{lightcyan}{rgb}{0.88,1,1}
\definecolor{antiquewhite}{rgb}{0.98, 0.92, 0.84}

\newcommand{\cl}{\textsc{ChordLink}\xspace}
\newcommand{\nt}{\textsc{NodeTrix}\xspace}
\newcommand{\nl}{\textsc{NodeLink}\xspace}

\newcommand{\rci}{{\textsc{RCI-NodeTrix}}\xspace}

\newcommand{\myparagraph}[1]{\smallskip\noindent\textbf{\boldmath #1}}

\Crefname{figure}{Fig.}{Figs.}

\graphicspath{{figures/}}

\begin{document}
\title{A User Study on Hybrid Graph Visualizations\thanks{This work is partially supported by: $(i)$ MIUR, grant 20174LF3T8 ``AHeAD: efficient Algorithms for HArnessing networked Data'', $(ii)$ Dipartimento di Ingegneria - Universit\`a degli Studi di Perugia, grants RICBA19FM: ``Modelli, algoritmi e sistemi per la visualizzazione di grafi e reti'' and RICBA20EDG: ``Algoritmi e modelli per la rappresentazione visuale di reti''.}}

\author{
	Emilio Di Giacomo\texorpdfstring{ \href{https://orcid.org/0000-0002-9794-1928}{\protect\includegraphics[scale=0.45]{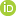}}}{} \and Walter Didimo\texorpdfstring{ \href{https://orcid.org/0000-0002-4379-6059}{\protect\includegraphics[scale=0.45]{orcid}}}{}
	\and \\Fabrizio Montecchiani\texorpdfstring{ \href{https://orcid.org/0000-0002-0543-8912}{\protect\includegraphics[scale=0.45]{orcid}}}{} \and Alessandra Tappini\texorpdfstring{ \href{https://orcid.org/0000-0001-9192-2067}{\protect\includegraphics[scale=0.45]{orcid}}}{}
	\textsuperscript{(\Letter)}
}

\date{}

\institute{
Department of Engineering, University of Perugia, Italy\\
	\email{
		\{emilio.digiacomo,walter.didimo,fabrizio.montecchiani,\\alessandra.tappini\}@unipg.it
	}
}

\maketitle

\begin{abstract}

Hybrid visualizations mix different metaphors in a single layout of a network.
In particular, the popular \nt model, introduced by Henry, Fekete, and McGuffin in 2007, combines node-link diagrams and matrix-based representations to support the analysis of real-world networks that are globally sparse but locally dense. That idea inspired a series of works, proposing variants or alternatives to \nt. We present a user study that compares the classical node-link model and three hybrid visualization models designed to work on the same types of networks. The results of our study provide interesting indications about advantages/drawbacks of the considered models on performing classical tasks of analysis. At the same time, our experiment has some limitations and opens up to further research on the subject.    

\end{abstract}

\section{Introduction}\label{se:introduction}
Many real-world networks, in a variety of application domains, exhibit a heterogeneous structure with a double nature: they are globally sparse but locally dense, i.e., they contain  \emph{clusters} of highly connected nodes (also called \emph{communities} in social network analysis) that are loosely connected to each other (see, e.g.,~\cite{fortunato-10,gn-02,pom-09}).  
Examples include social and financial networks~\cite{DBLP:journals/widm/BediS16,DBLP:journals/vlc/DidimoLM14,okk-04,DBLP:conf/icic/WuHPL10}, as well as biological and information networks~\cite{DBLP:journals/computer/FlakeLGC02,DBLP:conf/cibb/MahmoudMRR13}. 

\begin{figure}[tb]
	\centering
	\subfigure[\nl]{%
		\centering
		\includegraphics[width=0.47\columnwidth]{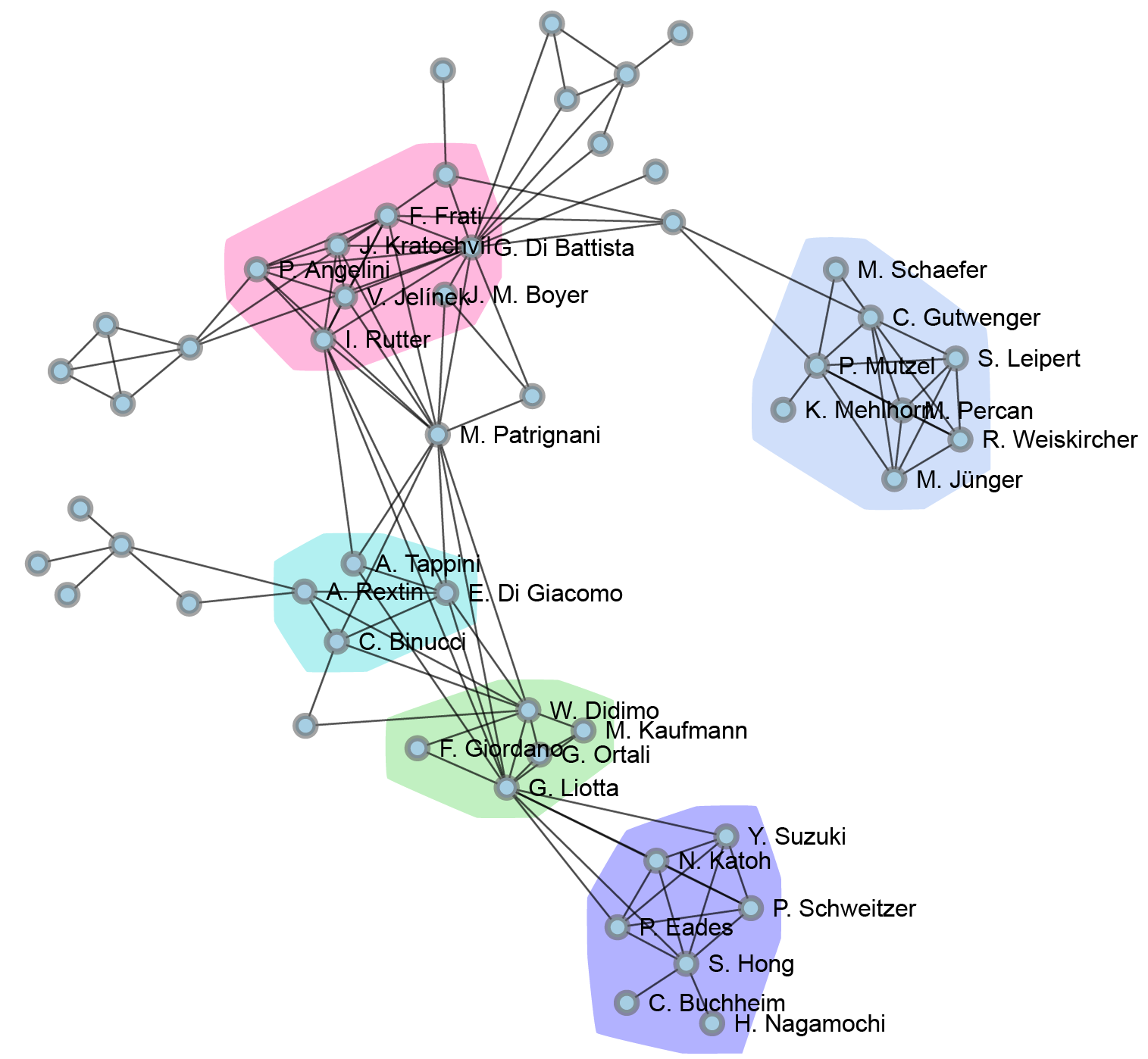}
		\label{fi:tutorial-nl}
	}\hfill
	\subfigure[\cl]{%
		\centering
		\includegraphics[width=0.47\columnwidth]{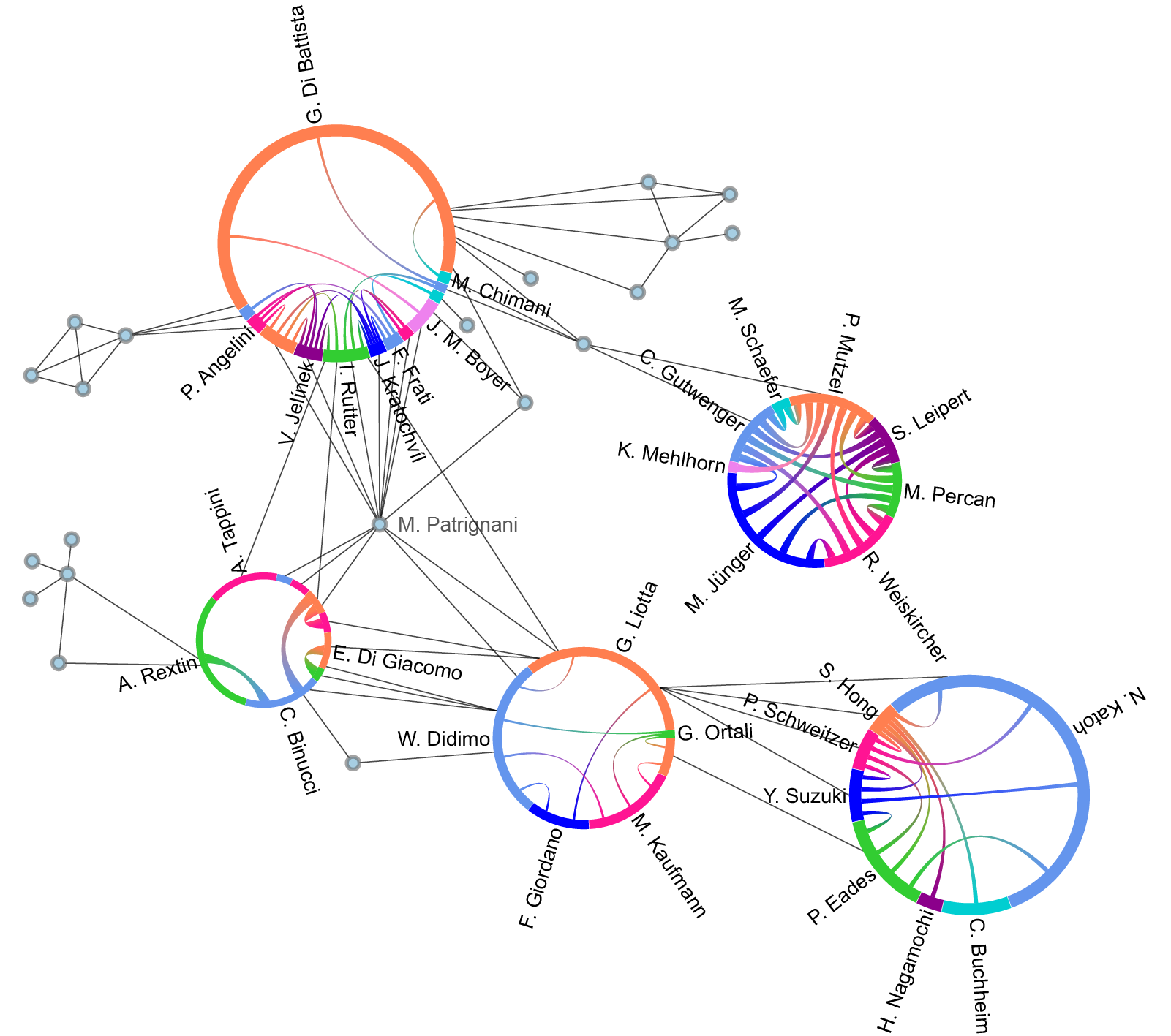}
		\label{fi:tutorial-cl}
	}\hfill
	\subfigure[\nt]{%
		\centering
		\includegraphics[width=0.47\columnwidth]{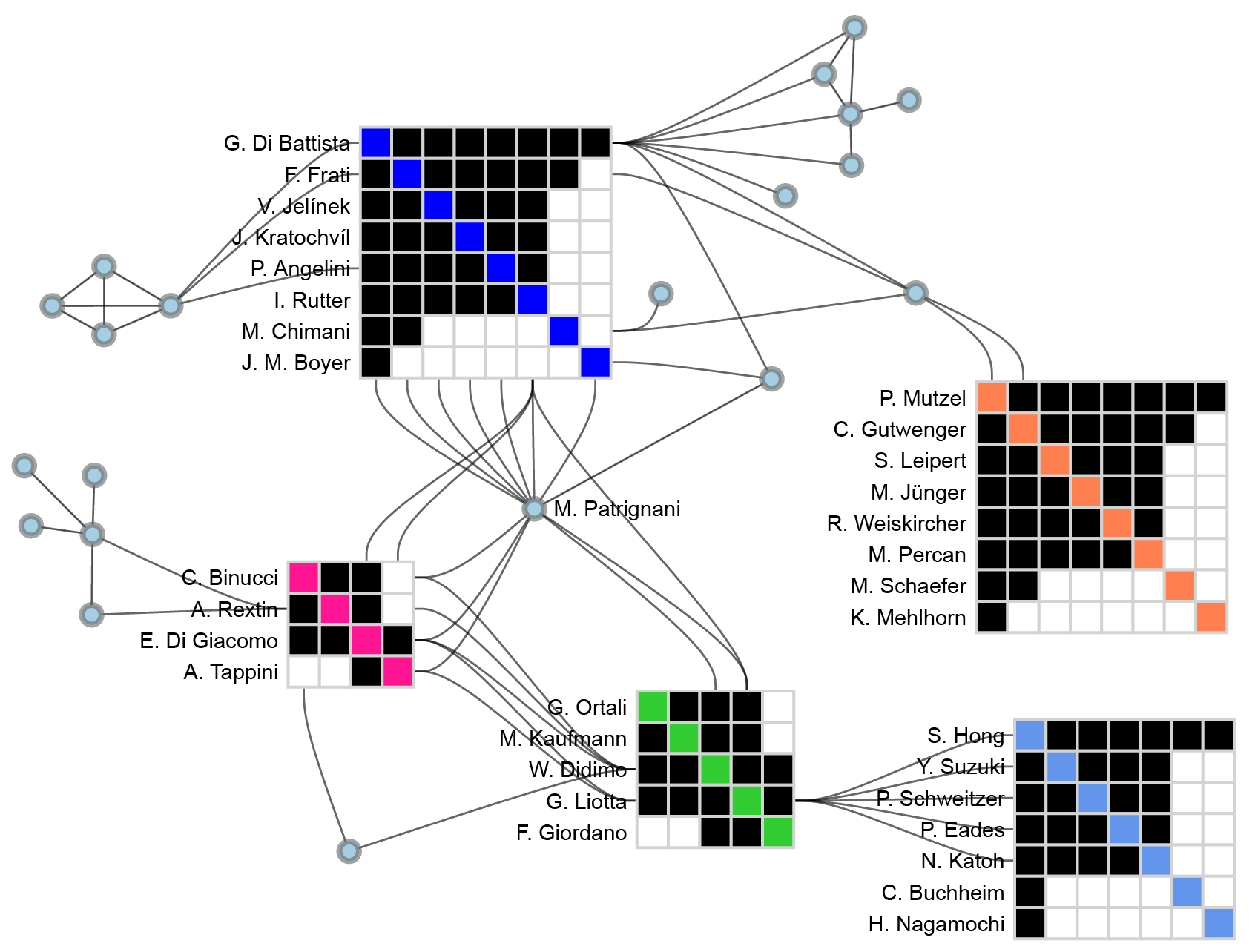}
		\label{fi:tutorial-nt}
	}\hfill
	\subfigure[\rci]{%
		\centering
		\includegraphics[width=0.47\columnwidth]{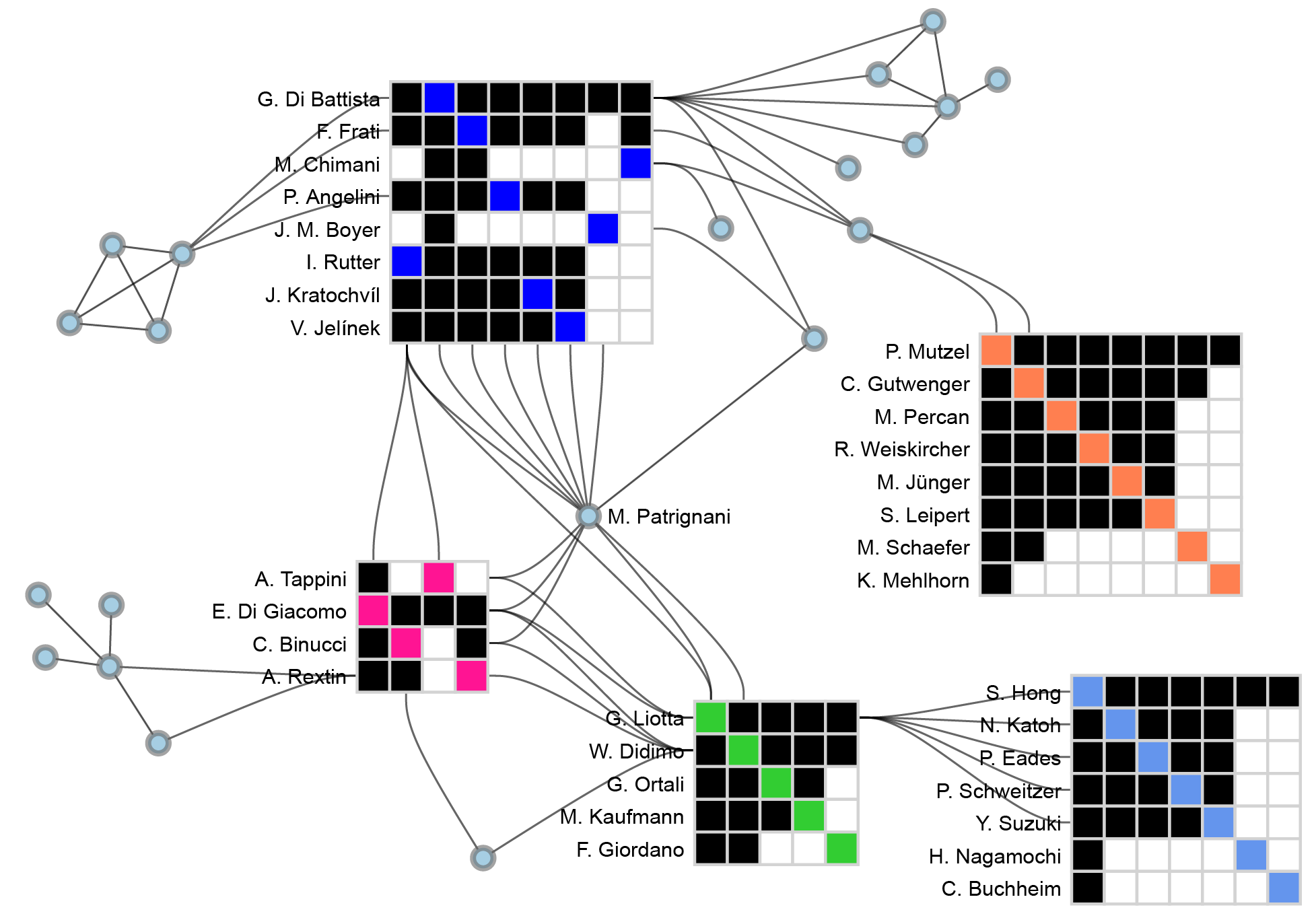}
		\label{fi:tutorial-rci}
	}
	
	\caption{The same clustered network with our four visualization models.}\label{fi:vis-models}
\end{figure}

The visualization of such networks through classical node-link diagrams is often unsatisfactory, due to the visual clutter caused by the high number of edges in the dense portions of the network (Fig.~\ref{fi:tutorial-nl}). To overcome this limit, \emph{hybrid visualizations} have been proposed. A hybrid visualization combines different graph visualization models in a unique drawing, with the aim of conveying the high-level cluster structure of the network and, at the same time, facilitating in the analysis of its communities. One of the seminal ideas in this regard is the \nt model, introduced by Henry, Fekete, and McGuffin~\cite{hfm-dhvsn-07}, which adopts a node-link diagram to represent the (sparse) global structure of the network, and a matrix representation for denser subgraphs identified and selected by the user (Fig.~\ref{fi:tutorial-nt}). After the introduction of \nt, hybrid visualizations have become an emerging topic in graph drawing and network visualization, and inspired an array of both theoretical and application results (see, e.g.,~\cite{addfpr-ilrg-17,aehkklnt-tcipap-alg2020,admpt-hgvcl-tvcg20,bpl-vpoot-swis2013,bbdlpp-valg-11,ddfp-cnrcg-jgaa-17,dllrt-kpprhp-walcom19,dlprt-ntptsc-19,hbf-ircsnund-tvcg2008,lrt-gpthec-19,lrt-sfohpt-tcs2021,ysdtlt-bhbnvcunr-tvcg2017}). 

\myparagraph{Contribution.} Motivated by the growing interest in hybrid visualizations, this paper focuses on network layouts with a given set of clusters, and addresses two broad research questions: \textsf{RQ1 --} ``Are hybrid visualizations more effective than node-link diagrams for the visual analysis of clustered networks?''; \textsf{RQ2 --} ``When considering specific tasks of analysis, are there differences in terms of response time or accuracy among different hybrid visualization models?'' 

To investigate these questions, we designed a user study that compares three hybrid visualization models and the classical node-link model. Namely, we considered two hybrid models that are designed to work on similar types of networks: the aforementioned \nt model~\cite{hfm-dhvsn-07} and the \cl model~\cite{admpt-hgvcl-tvcg20}, which represents clusters as chord diagrams instead of adjacency matrices (Fig.~\ref{fi:tutorial-cl}). Additionally, we considered the \rci model~\cite{lrt-sfohpt-tcs2021}, a variant of \nt that adopts independent orderings for the matrix rows and columns so to reduce crossings between inter-cluster edges (Fig.~\ref{fi:tutorial-rci}). 

To the best of our knowledge, our study is the first that addresses research question \textsf{RQ1}, and that considers \textsf{RQ2} for hybrid visualizations that adopt different styles to represent clusters. Our work is also motivated by open questions from~\cite{admpt-hgvcl-tvcg20,lrt-sfohpt-tcs2021}, namely:~\cite{admpt-hgvcl-tvcg20} suggests to perform a user study to compare \cl and other hybrid visualizations;~\cite{lrt-sfohpt-tcs2021} asks what is the impact of reducing crossings between inter-cluster edges at the expenses of independent row/column orderings in \nt.
The results of our study provide some hints about the usefulness of hybrid visualizations in the execution of topology-based tasks with respect to node-link diagrams. At the same time, our experiment has some limitations and opens up to new research to further investigate the subject.

The paper is structured as follows. \Cref{se:related-work} briefly surveys the scientific literature related to our work. \Cref{se:design} explains in detail the design of our user study and describes the rationale behind each of our choices. \Cref{se:results} discusses both the quantitative and qualitative results of our experiment, as well as its limitations. \Cref{se:conclusions} lists some future research directions. All the experimental data are available at {\small \url{http://mozart.diei.unipg.it/tappini/hybridUserStudy/}}.

\section{Related Work}\label{se:related-work}

The focus of our work is on hybrid graph representations that mix different visual metaphors to visually convey both the global structure of a sparse network and its locally dense subgraphs. In this direction,
Henry et al.~\cite{hfm-dhvsn-07} introduce the \nt model for social network analysis in one of the most cited papers of the InfoVis conference~\cite{citevis}; the model is implemented in a system where the user can select (dense) portions of a node-link diagram to be represented as adjaceny matrices. 
\nt visualizations have also been exploited to analyze other real-world graphs, such as ontology graphs~\cite{bpl-vpoot-swis2013} and brain networks~\cite{ysdtlt-bhbnvcunr-tvcg2017}. 

Angori et al.~\cite{admpt-hgvcl-tvcg20} introduce the \cl model. Similarly to \nt, this model is designed to work in a system where the user can visually identify and select clusters on a node-link diagram; differently from \nt, the selected cluster regions are represented as chord diagrams. \cl aims to represent all edges as geometric links and to preserve the drawing outside clusters by possibly duplicating nodes within a cluster (but each node can only appear in at most one cluster, as for \nt).

Our study focuses on comparing \nt and \cl, as they are conceived to work on networks with similar structure and within systems with similar characteristics. Our study also considers the \rci model~\cite{lrt-sfohpt-tcs2021}, a variant of \nt that allows independent orderings of the rows and columns in a matrix, to possibly reduce crossings between inter-cluster edges.

For social network analysis, the \nt model has also been proposed with a variant that considers ``overlapping clusters'', i.e., where a node can occur in multiple clusters at the same time~\cite{hbf-ircsnund-tvcg2008}. This kind of node duplication may help in the execution of community-related tasks, but sometimes interferes with other graph readability tasks.

Batagelj et al.~\cite{bbdlpp-valg-11} propose a system where the user can choose to represent each cluster according to a desired drawing style. Differently from \nt and \cl, this system is designed to automatically compute a set of clusters that guarantees desired properties (e.g., planarity) for the graph of clusters and adopts an orthogonal drawing convention (instead of a straight-line node-link diagram) to represent the outside of the clusters.
Hybrid visualizations have also been exploited in the context of dynamic network analysis (see, e.g.,~\cite{DBLP:journals/cgf/BeckBDW17,DBLP:journals/tvcg/HadlakSS11,DBLP:journals/tvcg/RufiangeM13}). We finally mention several theoretical results on hybrid visualizations that concentrate on the complexity of minimizing the number of inter-cluster edge crossings (see, e.g.,~\cite{ad-bcpg-2020,addfpr-ilrg-17,aehkklnt-tcipap-alg2020,bdg-ckmepd-esa2019,ddfp-cnrcg-jgaa-17,dllrt-kpprhp-walcom19,dlprt-ntptsc-19,lrt-gpthec-19,lrt-sfohpt-tcs2021}).

\smallskip Our work falls into the research line devoted to the design of user experiments in graph drawing and network visualization. 
We recall here the contributions that are mainly related to our study; refer to~\cite{bhwpwh-saeegv-ieeeaccess2021} for a comprehensive survey on the subject. 
There is a series of works that compare node-link diagrams with matrix-based representations (see, e.g.,~\cite{DBLP:conf/vissoft/AbuthawabehBZD13,DBLP:conf/chi/AlperBRIF13,DBLP:conf/sg/ChristensenBWR14,DBLP:journals/cgf/DidimoKMT18,gfc-rgunmrcesa-infvis2005,kec-mnldwvrbvcm-infvis2006,DBLP:journals/cgf/OkoeJ15,ojk-nlamoqni-tvcg2019}). An insight that seems to emerge from these studies is that node-link diagrams have usually better performance on topology and connectivity tasks when graphs are not too large and dense, while matrices perform better on group tasks. Our study does not aim to further compare node-link and matrix representations, but rather to investigate hybrid visualizations that mix these two, or others, drawing conventions.

In the context of hybrid graph visualizations, Henry and Fekete~\cite{DBLP:conf/interact/HenryF07} conduct a user study on MatLink, a model that combines adjacency matrices overlaid with node-link diagrams using curvature for the links. They find that MatLink outperforms the two individual metaphors (node-link diagrams and adjacency matrices) for most of the considered tasks, including path-related tasks, where matrices are usually worse than node-link. However, differently from our study, ~\cite{DBLP:conf/interact/HenryF07} does not focus on the visualization of clustered networks. Henry et al.~\cite{hbf-ircsnund-tvcg2008} present a user study aimed to understand whether node duplication for non-disjoint clusters improves the performance of \nt for some types of tasks.
Since the majority of hybrid visualizations are designed to deal with disjoint clusters, our study focuses on this setting; moreover, we consider tasks that are mostly different from those addressed in~\cite{hbf-ircsnund-tvcg2008}.

 \section{Study Design}\label{se:design}
  



This section describes in detail the design of our user study. Our target population are researchers and analysts (including practitioners, academics, and students) that make use of network visualization to accomplish tasks of analysis on real-world networks.
In the following we discuss the visualization models, the tasks and the hypothesis, the stimuli, and the experimental procedure.

\subsection{Visualization Models}\label{sse:models}
The conditions compared in our study are four different models for the visualization of undirected clustered networks, where subsets of nodes are grouped into clusters (see \Cref{fi:vis-models} for an illustration). We consider networks with neither self-loops nor multiple edges. An edge connecting two nodes in the same cluster is an \emph{intra-cluster edge}; every other edge is an \emph{inter-cluster edge}. The models~are:

\medskip\noindent{-- \nl (NL).} This is the classical node-link model, where nodes are represented as small disks and edges are straight-line segments connecting their end-nodes. In this model, we visually highlight each cluster through a colored convex region that includes all the nodes in the cluster. 
%

\smallskip\noindent{-- \cl (CL).} This model has been introduced in~\cite{admpt-hgvcl-tvcg20,admpt-clnhvm-19}. Nodes outside clusters and inter-cluster edges are drawn as in the \nl model. Clusters are represented as chord diagrams. A node in a cluster may have multiple copies, each represented as a colored circular arc along the circumference of the chord diagram; all copies of the same node have the same color. An intra-cluster edge is drawn as a ``ribbon'' connecting two of the copies representing its end-nodes.

\smallskip\noindent{-- \nt (NT).} This is the model introduced in~\cite{hfm-dhvsn-07}; each cluster~$C$ of size~$n$ is represented by a (symmetric) $n \times n$ adjacency matrix. 
Nodes outside clusters, and edges between them, are drawn as in \nl. An inter-cluster edge having an end-node $v$ in a cluster $C$ is drawn as a curve incident to the row or to the column associated with $v$, on one of the sides of the matrix representing~$C$.

\smallskip\noindent{-- \rci (RC).} This is a variant of the \nt model, introduced in~\cite{lrt-sfohpt-sofsem2020,lrt-sfohpt-tcs2021}. The difference with the \nt model is that in each adjacency matrix, the row and the column associated with the same node may have different indices, in order to save some crossings between inter-cluster edges. As a consequence the matrices may not be symmetric.  

\myparagraph{Rationale.} Among the various types of hybrid visualizations described in the literature, we selected NT and CL as they are designed to work similarly within visualization systems for the analysis of real-world networks. In particular, we exploited the system in~\cite{admpt-hgvcl-tvcg20}, which implements both these models in a unique interface, where the implementation of NT reflects that given in~\cite{nt-system} by the authors of~\cite{hfm-dhvsn-07}. The system in~\cite{admpt-hgvcl-tvcg20} allows direct support for clustered drawings in the NL model and makes it possible to create drawings in all the supported models by defining the same set of clusters on the same node-link diagram. For the purposes of our experiment, we enriched the system with the RC model.

\newcommand{\lee}{\textsf{LeeTax}\xspace}
\newcommand{\amar}{\textsf{AmarTax}\xspace}

\subsection{Tasks}\label{sse:tasks}
We defined six different tasks, listed in \Cref{ta:tasks}. We classify each task according to the taxonomy by Lee et al.~\cite{lpsfh-ttgv-beliv2006}, which we refer to as \lee. Moreover, following the taxonomy by Amar et al.~\cite{aes-llcaaiv-infovis2005}, which we refer to as \amar, we indicate the low-level visual analytics operations needed to execute each~task.

%

\newcommand{\PreserveBackslash}[1]{\let\temp=\\#1\let\\=\temp} 
\newlength{\tablength}
\setlength{\tablength}{\textwidth}
\newlength{\celllength}
\setlength{\celllength}{2.2cm}
\newlength{\celllengthmore}
\setlength{\celllengthmore}{3.3cm}
\newlength{\firstcell}
\setlength{\tabcolsep}{3pt}
\setlength{\firstcell}{\textwidth - \celllength - \celllengthmore - \arrayrulewidth*4 - \tabcolsep*6}
\renewcommand{\arraystretch}{1.5}

\begin{table}[tb]
	\centering
	\begin{tabular*}{\tablength}{|m{\firstcell}|m{\celllength}|m{\celllengthmore}|}
		\hline
		\rowcolor{antiquewhite}{\bf Task} & {\bf \lee} & {\bf \amar} \\
		\hline
		\textsf{T1}. Is there an edge that links the two highlighted nodes? & topology-based (adjacency) & retrieve value \\
		\hline
		\textsf{T2}. Which of the two highlighted nodes has higher degree? & topology-based (adjacency) & retrieve value;\par sort \\
		\hline
		\textsf{T3}. Is there a path of length at most $k$ that connects the two highlighted nodes? & topology-based (connectivity) & retrieve value;\par compute derived value;\par filter \\
		\hline
		\textsf{T4}. Which of the following three node labels appear in the highlighted portion of the network? & attribute-based (on the nodes) & retrieve value;\par filter \\
		\hline
		\textsf{T5}. What is the denser$^*$ cluster between the two highlighted? & overview & filter;\par compute derived value;\par sort \\
		\hline
		\textsf{T6}. How many edges directly connect the two highlighted parts of the drawing? & overview & filter;\par compute derived value \\
		\hline
		\multicolumn{3}{|l|}{\scriptsize{$^*$The cluster density is the ratio between the number of edges and the number nodes in a cluster}}\\
		\hline
	\end{tabular*}
\caption{Tasks used in our study.}\label{ta:tasks}
\end{table}

\myparagraph{Rationale.} 
We designed the user study with a set of tasks that requires to explore the drawing locally and globally. Moreover, each task is easy to explain, it can be executed in a reasonably short time, and it can be easily measured. Concentrating on representative tasks is a common approach for this kind of experiments (e.g.,~\cite{DBLP:books/daglib/0030807}), which supports generalizability to more complex tasks that include these representatives as subroutines. Most of our tasks have already been used in previous graph visualization user studies (e.g.,~\cite{ojk-nlamoqni-tvcg2019,DBLP:journals/vlc/Purchase98,phnk-ulgd-gd2012,DBLP:journals/tvcg/SaketSKB14}) and they cover all task categories in the taxonomy of Lee et al.~\cite{lpsfh-ttgv-beliv2006}, with the exception of the browsing category. We excluded the latter because it requires the users to interact with the visualization and we decided to avoid interaction to keep the test execution as simple as possible and avoid possible confounding factors. According to the top-level task classification by Burch et al.~\cite{bhwpwh-saeegv-ieeeaccess2021}, all our tasks are \emph{interpretation tasks}, as our goal is to evaluate the differences of the considered visualization models in terms of readability, understandability, and effectiveness. About task~\textsf{T5}, we point out that there are two commonly used definitions for the density of a graph with $n$ nodes and $m$ edges: $d_1 = \frac{m}{n}$ and $d_2 = \frac{2m}{n(n-1)}$. We adopted definition $d_1$ for two reasons: (a) it is simpler to explain to a user; (b) according to previous research work~\cite{10.1145/1168149.1168167}, $d_1$ is a better descriptor of the complexity of real-world networks. Indeed, the visual perception of the density of a cluster region is affected by the number of nodes in the cluster; if a drawing contains two clusters with different sizes, the largest one may be perceived as a denser portion of the drawing, even if it has lower density according~to~$d_2$. 

\subsection{Hypotheses}\label{sse:hypotheses}
Similarly to previous works (e.g.,~\cite{hbf-ircsnund-tvcg2008,ojk-nlamoqni-tvcg2019}), we define our hypotheses based on tasks, structuring them according to the task categories of \lee.

\smallskip\noindent \textbf{H1:} On topology-based tasks (\textsf{T1}, \textsf{T2}, \textsf{T3}), we expect that \nl outperforms hybrid visualizations in terms of response time. On the other hand, we expect hybrid visualizations to have a lower error rate than \nl, and \cl to behave better than \nt and \rci.  

\smallskip\noindent \textbf{H2:} On attribute-based tasks (\textsf{T4}), we expect \nt and \rci to outperform the other two models in terms of response time and error~rate. 

\smallskip\noindent \textbf{H3:} On overview tasks (\textsf{T5}, \textsf{T6}), we expect hybrid visualizations to perform better than \nl in terms of both response time and error rate. Among the hybrid visualizations, we expect \nt and \rci to be better than \cl, especially when one needs to estimate cluster~density.

\myparagraph{Rationale.} 
%
%
About \textbf{H1}, our expectations in terms of response time are motivated by the fact that \nl is quite intuitive and widely used. Moreover, hybrid visualizations intrinsically require to switch from a visualization metaphor to another during the visual exploration, which may represent a cognitive effort. Concerning the error rate, we think that, by reducing the visual clutter, hybrid visualizations may be able to avoid visual ambiguities (such as edges that are almost collinear) and therefore may better support topology-based tasks.  Also, since topology-based tasks are known to be harder when dealing with matrices, we expect \cl to have better performance than \nt and \rci in terms of error rate.  
About \textbf{H2}, we believe that placing labels on a matrix side is more effective than placing them around chord diagrams or near nodes in a node-link diagram. In chord diagrams labels may be harder to read due to their rotation, while in node-link diagrams they may be hidden by edges. 
About \textbf{H3}, we expect hybrid visualizations to behave better than \nl due to their capability to represent clusters more clearly. For tasks that require to estimate cluster density, both \nt and \rci have the advantage that the proportion between black (edges) and white (non-edges) cells immediately conveys the density of a cluster; the same estimation in \cl is more difficult due to node duplication, which may give the impression that a cluster is sparser than~it~actually~is. 

\subsection{Stimuli}\label{sse:stimuli}

Our experimental objects are three real-word networks of small/medium size. 
The first one, \textsf{weavers}, is an animal social network with 64 nodes and 177 edges, describing the interactions of a colony of weavers in the usage of nests~\cite{https://doi.org/10.1111/ele.12320,nr}. The second one, \textsf{e.coli}, is a biological network with 97 nodes and 212 edges that describes transcriptional interactions in the Escherichia coli bacterium~\cite{Mangan11980}. The third one, \textsf{dblp}, is a co-authorship network obtained from the DBLP repository~\cite{dblp} by searching for the keyword ``network visualization'' and considering only the largest connected component, which has 118 nodes and 322 edges.

For each of the four visualization models described in~\Cref{sse:models}, we produced a diagram of the three networks above. The diagrams for the \nl model are computed through the force-directed algorithm available in the D3 library~\cite{d3}. Starting from these drawings, we defined some geometric clusters with the technique based on the $K$-means algorithm~\cite{k-means} described in~\cite{admpt-hgvcl-tvcg20}. As explained in \Cref{sse:models}, the system presented in~\cite{admpt-hgvcl-tvcg20} is then used to compute the~diagrams in the \cl, \nt, and \rci models with the same sets of clusters. Further details about the stimuli creation are in~\Cref{app:stimuli}. 

Each of the $12$ stimuli obtained by applying each of the $4$ conditions (models) to the $3$ experimental objects (networks) is used in all of the $6$ tasks described in \Cref{sse:tasks}, for a total of $4 \times 3 \times 6 = 72$ trials. For \textsf{T1}, \textsf{T2}, and \textsf{T3}, we highlighted the node labels with a yellow background; to help the user to locate the nodes, we also put a red cross close to the clusters containing them. For \textsf{T4} and \textsf{T6}, we highlighted the regions of interest by enclosing them inside a colored polygonal area. Finally, for \textsf{T5} we indicated the two clusters of interest with large red labels. The trials for the network \textsf{weavers} are shown in~\Cref{app:trials}.

%

%
%

\myparagraph{Rationale.} The visualization models that we compare are suitable for networks with up to few thousand nodes and edges, while for significantly larger networks ad-hoc techniques are required that typically reduce the amount of displayed information. The choice of using networks with few hundred elements avoids an excessive burden for the participants. Namely, we wanted that each trial could be executed in a reasonable amount of time without an excessive fatigue and that the whole test could be completed in about 30 minutes.
Further, since we decided to show static images (without zoom), the whole picture of the network should be displayed with a level of zoom that keeps the labels readable. 
Since hybrid visualizations are mainly used to visualize networks that are globally sparse but locally dense, we selected three networks that exhibit this structure. Moreover, we designed the specific trials so that the user was required to explore both the sparse parts of the network, represented by the node-link metaphor, and the dense parts, represented in different ways depending on the model.



\subsection{Experimental Setting and Procedure}\label{sse:procedure}

We designed a between-subject experiment where each participant was exposed to one of the four conditions and hence to $18$ trials. The users executed the test fully on-line. The questionnaire was prepared using the LimeSurvey tool ({\small \url{https://www.limesurvey.org/}}) and is structured as follows. First, some information about the user are collected, namely: gender, age, educational level, expertise in graph visualization, screen size, and possible color vision deficiency. Then, the visualization model to be assigned to the user is decided in a round robin fashion. Based on this assignment, a video tutorial is presented, followed by a training phase in which the user has to answer a trial for each task with an explanatory feedback in case of wrong answer. Next, the $18$ trials are presented in random order. Finally, the user is asked for some qualitative feedback: two Likert scale questions about the aesthetic quality of the drawings and about the easiness of the questions, plus an optional free comment. While no time limit was given to complete the test, the participants were asked to answer each question as fast as they could but, at the same time, trying to be accurate. 
For each user, we collected the answers and the time spent on each question. 
We recruited the participants with announcements to the \textsf{gdnet}, \textsf{ieee\_vis}, \textsf{infovis} mailing lists and to the computer engineering students of the universities of Perugia and Roma~Tre. The actual experiment was preceded by a pilot study (see \Cref{app:pilot}).


\myparagraph{Rationale.} As previously explained, exposing the users to all four conditions would imply each user solving $72$ trials. We believe that keeping the same level of attention in such a long experiment is difficult, and may cause many participants prematurely quitting the test. Besides such undesired fatigue effect, a within-subject design would also imply that each user sees the same experimental object $24$ times, which makes it difficult to avoid the learning effect. Hence,  we adopted a between-subject design, where each participant is exposed to only one condition. This choice limited the number of trials per user to $18$, thus mitigating both the fatigue and the learning effect, which is further counteracted by presenting the trials in a random order. Finally, since the test also includes a video tutorial and a training phase to make the user familiar with the given visualization model, an additional advantage of the between-subject design is that these phases can be focused on one model only.
%
About the execution of the experiment, we opted for a fully on-line test for two reasons: (i) the difficulties to perform a controlled in-person experiment due to the COVID-19 pandemic; (ii) the possibility of recruiting a larger number of participants that better represent our target population, through announcements on the aforementioned~mailing~lists. 

\section{Study Results}\label{se:results}

\myparagraph{Participants.}
%
%
We collected questionnaires from 89 participants. We discarded seven tests for various reasons, reported in \Cref{app:discarded}.
Of the remaining 82 tests, 19 were for \cl and 21 for each of the other models.
Regarding the participants, 66 (80.49\%) were males, 15 (18.29\%) were females,~and~1~(1.22\%) preferred not to answer. The majority of them (82.72\%) were aged below 40. 85.37\% of the participants has at least a Bachelor's degree, with 34.15\% of them having a doctoral degree. 62.2\% of the participants declared a medium or high familiarity with graph visualization and 68.29\% used a screen of size at least~15''. \Cref{fi:demo} in \Cref{app:charts} summarizes the data about the participants.

\smallskip
\myparagraph{Quantitative Results.}
We compared the performance of the four models in terms of error rate and response time. For \textsf{T1}--\textsf{T5}, the error rate of a user is the ratio between the number of wrong answers and the total number of questions. Recall that there are three questions per task and that in \textsf{T4} the user has to find three labels for each question. About \textsf{T6}, the error on a question is computed as $1-\frac{1}{1+|u-r|}$, where $u$ is the value given by the user and $r$ is the correct value; the error rate for \textsf{T6} is the average of the errors on the three questions of the task.

\setlength{\tablength}{\textheight}
\newlength{\cellone}
\newlength{\celltwo}
\newlength{\cellthree}
\newlength{\celllarge}
\setlength{\cellone}{0.8cm}
\setlength{\celltwo}{1.25cm}
\setlength{\cellthree}{1.25cm}
\setlength{\tabcolsep}{2pt}
\setlength{\celllarge}{\tablength - \cellone - \celltwo*2 -\cellthree*6 - \arrayrulewidth*11 - \tabcolsep*20}
\renewcommand{\arraystretch}{1.5}

\begin{sidewaystable}	
	\begin{tabular*}{\tablength}{|>{\PreserveBackslash\centering}m{\cellone}
			|>{\PreserveBackslash\centering}m{\celllarge}
			|>{\PreserveBackslash\centering}m{\celltwo}
			|>{\PreserveBackslash\centering}m{\celltwo}
			|>{\PreserveBackslash\centering}m{\cellthree}
			|>{\PreserveBackslash\centering}m{\cellthree}
			|>{\PreserveBackslash\centering}m{\cellthree}
			|>{\PreserveBackslash\centering}m{\cellthree}
			|>{\PreserveBackslash\centering}m{\cellthree}
			|>{\PreserveBackslash\centering}m{\cellthree}|}
		\hline
		\rowcolor{antiquewhite} &  & \multicolumn{2}{c|}{\cellcolor{antiquewhite}\textbf{Kruskal-Wallis}} &  \multicolumn{6}{c|}{\textbf{Pairwise comparisons ($p$-value)}}  \\
		\hhline{|>{\arrayrulecolor{antiquewhite}}-|-|>{\arrayrulecolor{black}}-|-|-|-|-|-|-|-|}
		\rowcolor{antiquewhite}\multirow{-2}{*}{\textbf{Task}} & \multirow{-2}{*}{\textbf{Models ranked by average error rate}} &  $H(3)$ &  $p$-value &  NL-CL &  NL-NT &  NL-RC & CL-NT & CL-RC & NT-RC \\
		\hline
		\textsf{T1} & CL (0.175), NL (0.191), RC (0.254), NT (0.350)  & \textbf{8.471} & \textbf{0.037} & 1.000 & 0.179 & 1.000 & \textbf{0.036} & 1.000 & 0.549  \\
		\hline
		\textsf{T2} & CL (0.140), NL (0.143), RC (0.191), NT (0.238) & 2.419  & 0.490  & - & - & - & - & - & - \\
		\hline
		\textsf{T3} & CL (0.140), RC (0.143), NT (0.270), NL (0.333) & \textbf{10.882} & \textbf{0.012} & \textbf{0.035} & 0.788 & \textbf{0.024} & 1.000 & 1.000 & 1.000 \\
		\hline
		\textsf{T4} & CL (0.105), NT (0.143),  RC (0.206), NL (0.302) & 6.703 & 0.082 & - & - & - & - & - & - \\
		\hline
		\textsf{T5} & NT (0.270), NL (0.302), CL (0.386), RC (0.400) & 2.407 & 0.492 & - & -  & - & - & - & - \\
		\hline
		\textsf{T6} & RC (0.350), CL (0.386), NT (0.418), NL (0.429) & 0.478 & 0.924 & - & -  & - & - & - & - \\
		\hline
		\hline
		\rowcolor{antiquewhite} &  & \multicolumn{2}{c|}{\cellcolor{antiquewhite}\textbf{Kruskal-Wallis}} &  \multicolumn{6}{c|}{\textbf{Pairwise comparisons ($p$-value)}}  \\
		\hhline{|>{\arrayrulecolor{antiquewhite}}-|-|>{\arrayrulecolor{black}}-|-|-|-|-|-|-|-|}
		\rowcolor{antiquewhite}\multirow{-2}{*}{\textbf{Task}} & \multirow{-2}{*}{\textbf{Models ranked by average response time}} &  $H(3)$ &  $p$-value &  NL-CL &  NL-NT &  NL-RC & CL-NT & CL-RC & NT-RC \\
		\hline
		\textsf{T1} &  NL (16.64), NT (19.91), RC (23.47), CL (26.12) & \textbf{20.084} &  \textbf{$<$0.001} &  \textbf{0.000} & 1.000 & \textbf{0.031} & \textbf{0.016} & 0.832 & 0.720 \\
		\hline
		\textsf{T2} & NL (25.23), CL (36.35), NT (37.63), RC (39.95) & \textbf{12.632} & \textbf{0.006} & 0.058 & 0.061 & \textbf{0.006} & 1.000 & 1.000 & 1.000 \\
		\hline
		\textsf{T3} &  NL (24.58), RC (39.56), NT (42.68), CL (47.69) & \textbf{19.533} & \textbf{$<$0.001} & \textbf{0.000} & \textbf{0.005} & \textbf{0.011} & 1.000 & 1.000 & 1.000 \\
		\hline
		\textsf{T4} &  NT (35.88), RC (37.04), NL (42.36), CL (50.60) & \textbf{9.793} & \textbf{0.020} & 1.000 & 0.579 & 1.000 & \textbf{0.018} & 0.143 & 1.000 \\
		\hline
		\textsf{T5} &  NL (29.71), NT (37.77), CL (42.30), RC (44.24) & 3.657 & 0.301 & - & - & - & - & - & - \\
		\hline
		\textsf{T6} & RC (31.56), NL (34.07), CL (35.83), NT (38.27) & 1.265 & 0.737 & - & - & - & - & - & - \\
		\hline
	\end{tabular*}
	\caption{Results for error rate (top) and response time (bottom) for each task. \label{ta:errorate-time}}
\end{sidewaystable}

By performing the Shapiro-Wilk test with significance level $\alpha{=}0.05$, we found that data were not normally distributed. Hence, we performed the non-parametric Kruskal–Wallis test with significance level $\alpha{=}0.05$,  which is suitable for comparing multiple independent samples. We finally performed post-hoc pairwise comparisons by using Bonferroni corrections. (See also~\cite{opac-b1097694,thode-02}.) 

\begin{figure}
	\fbox{\includegraphics[width=0.48\textwidth,trim=20mm 25mm 25mm 20mm, clip]{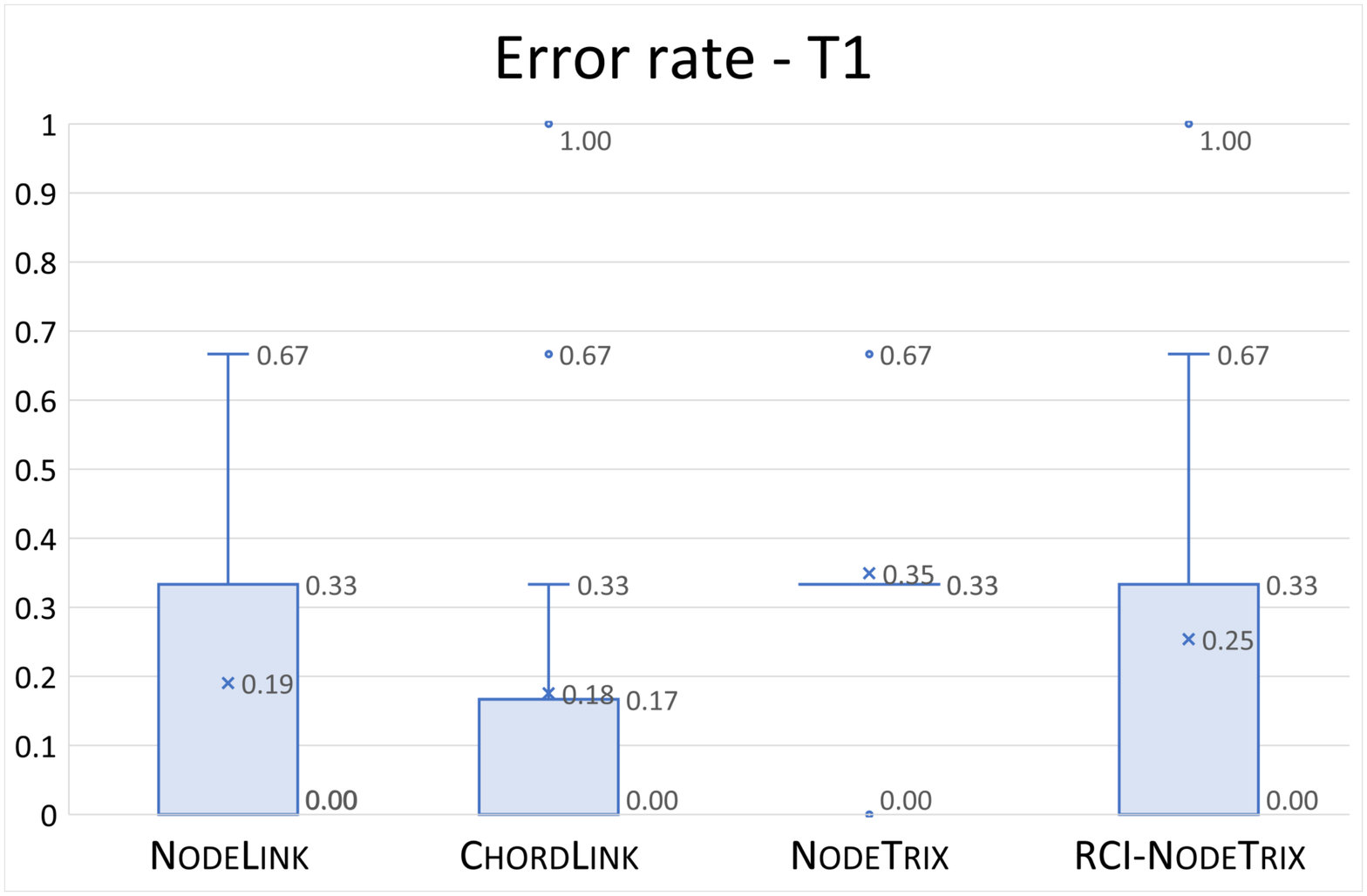}}
	\fbox{\includegraphics[width=0.48\textwidth,trim=20mm 25mm 25mm 20mm, clip]{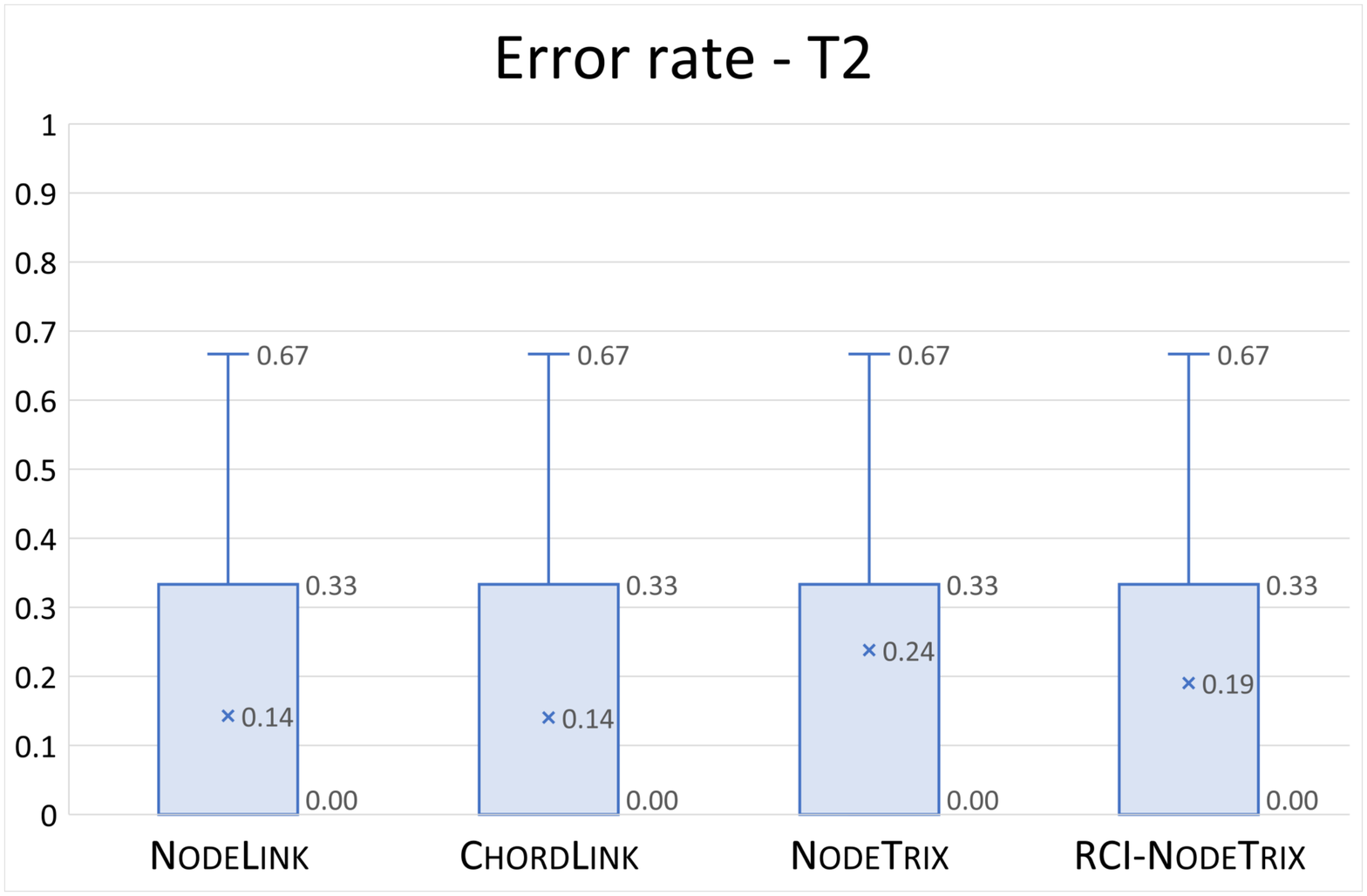}}
	\fbox{\includegraphics[width=0.48\textwidth,trim=20mm 25mm 25mm 20mm, clip]{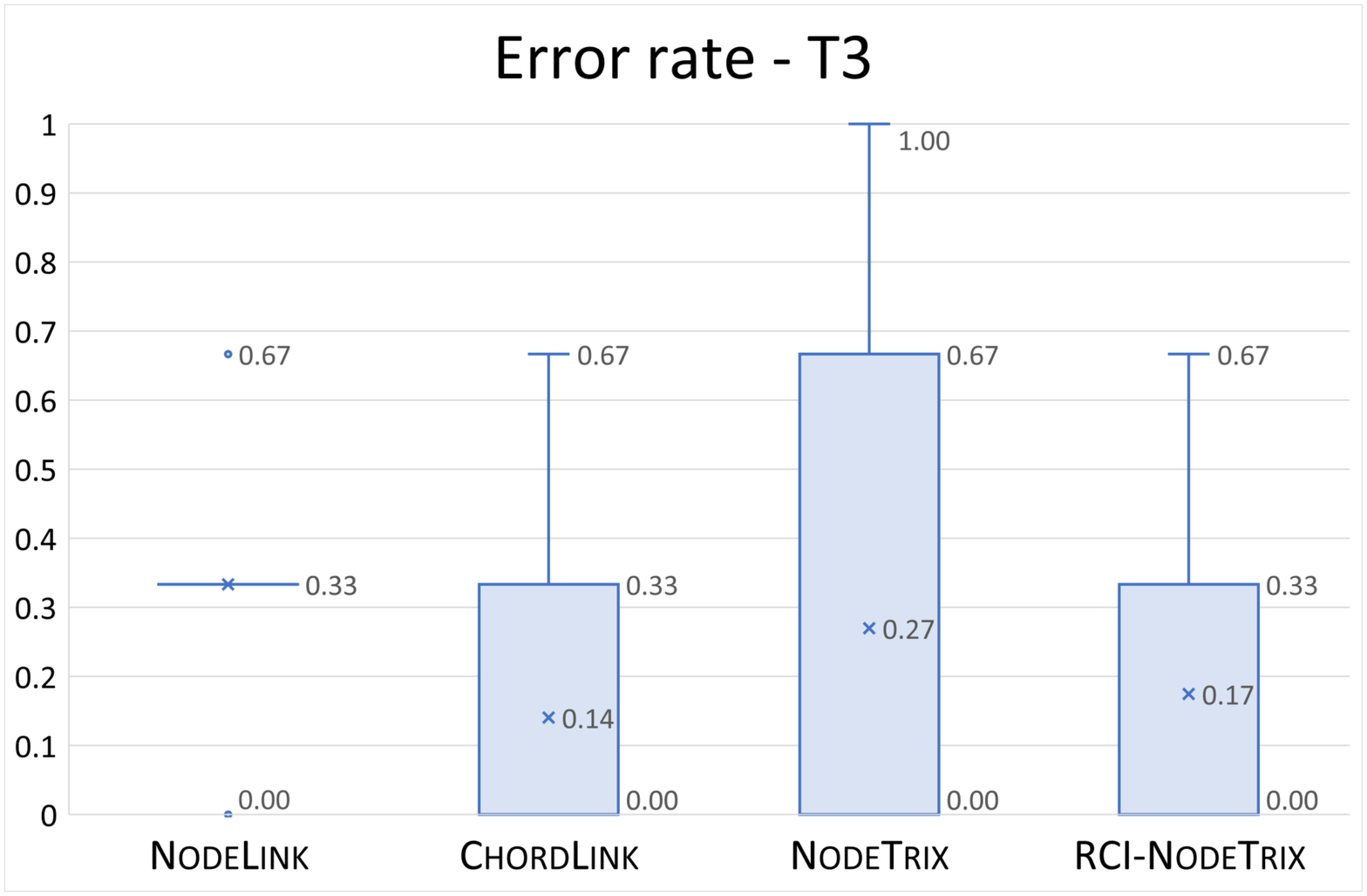}}
	\fbox{\includegraphics[width=0.48\textwidth,trim=20mm 25mm 25mm 20mm, clip]{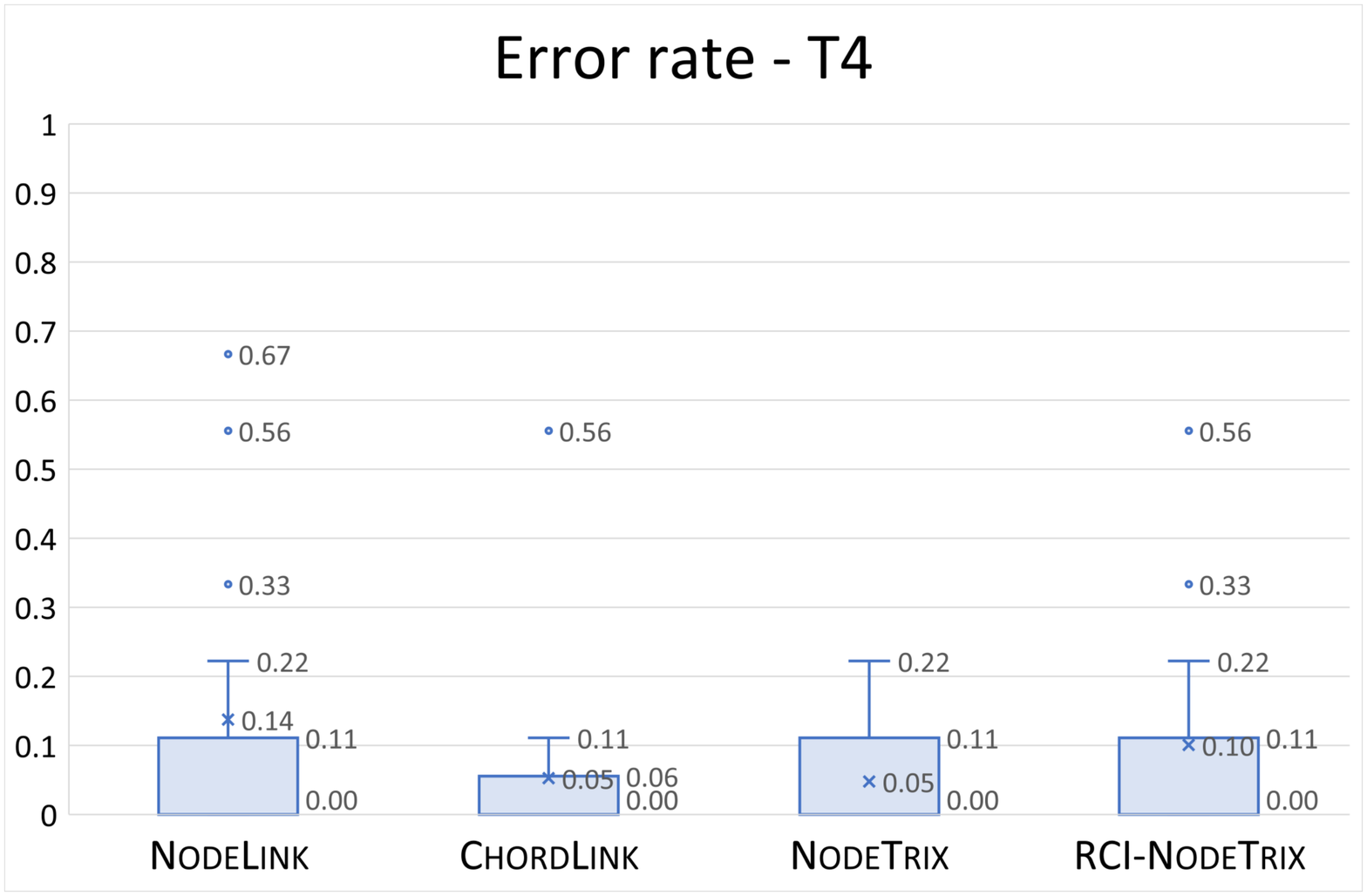}}
	\fbox{\includegraphics[width=0.48\textwidth,trim=20mm 25mm 25mm 20mm, clip]{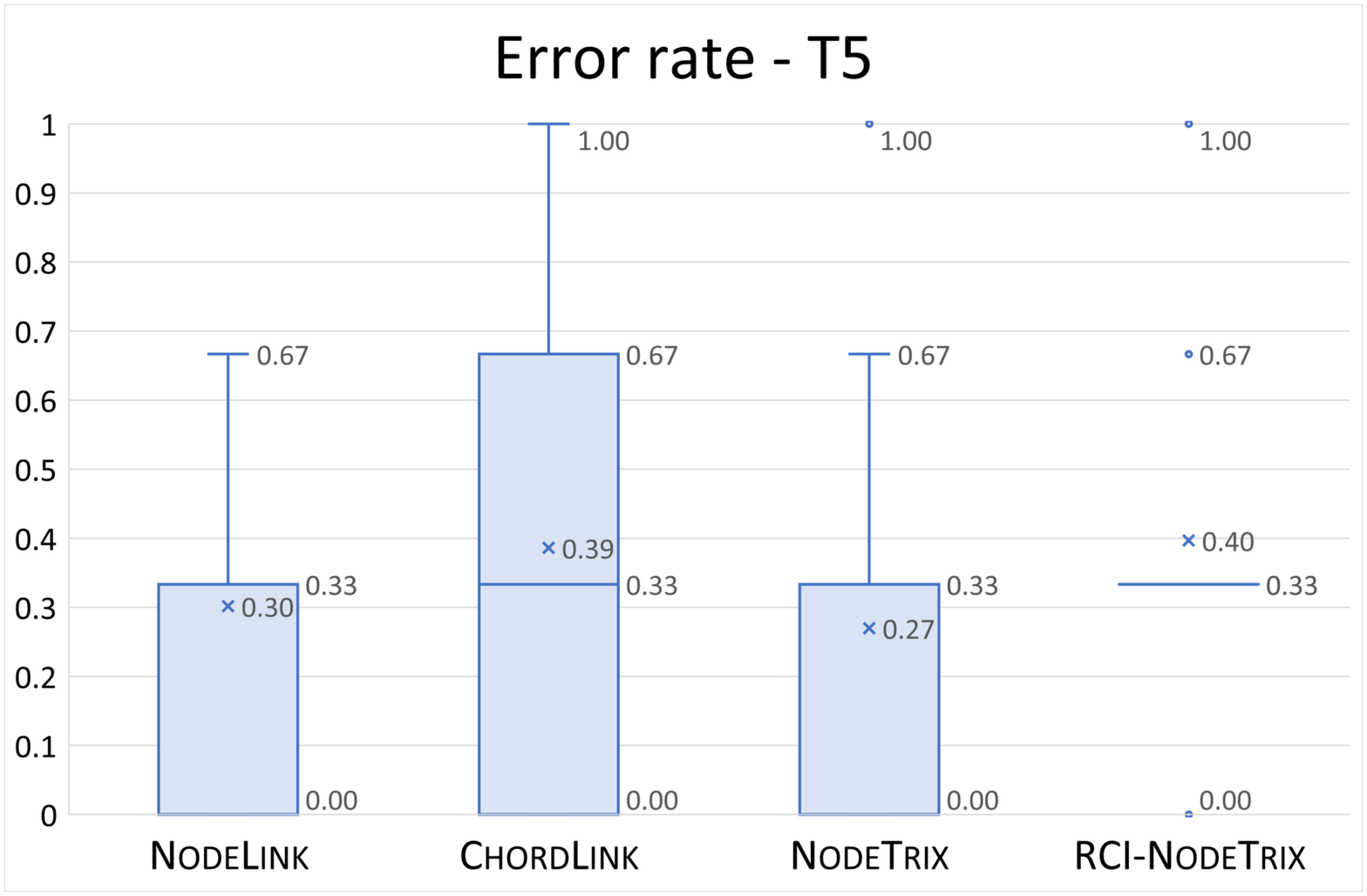}}
	\fbox{\includegraphics[width=0.48\textwidth,trim=20mm 25mm 25mm 20mm, clip]{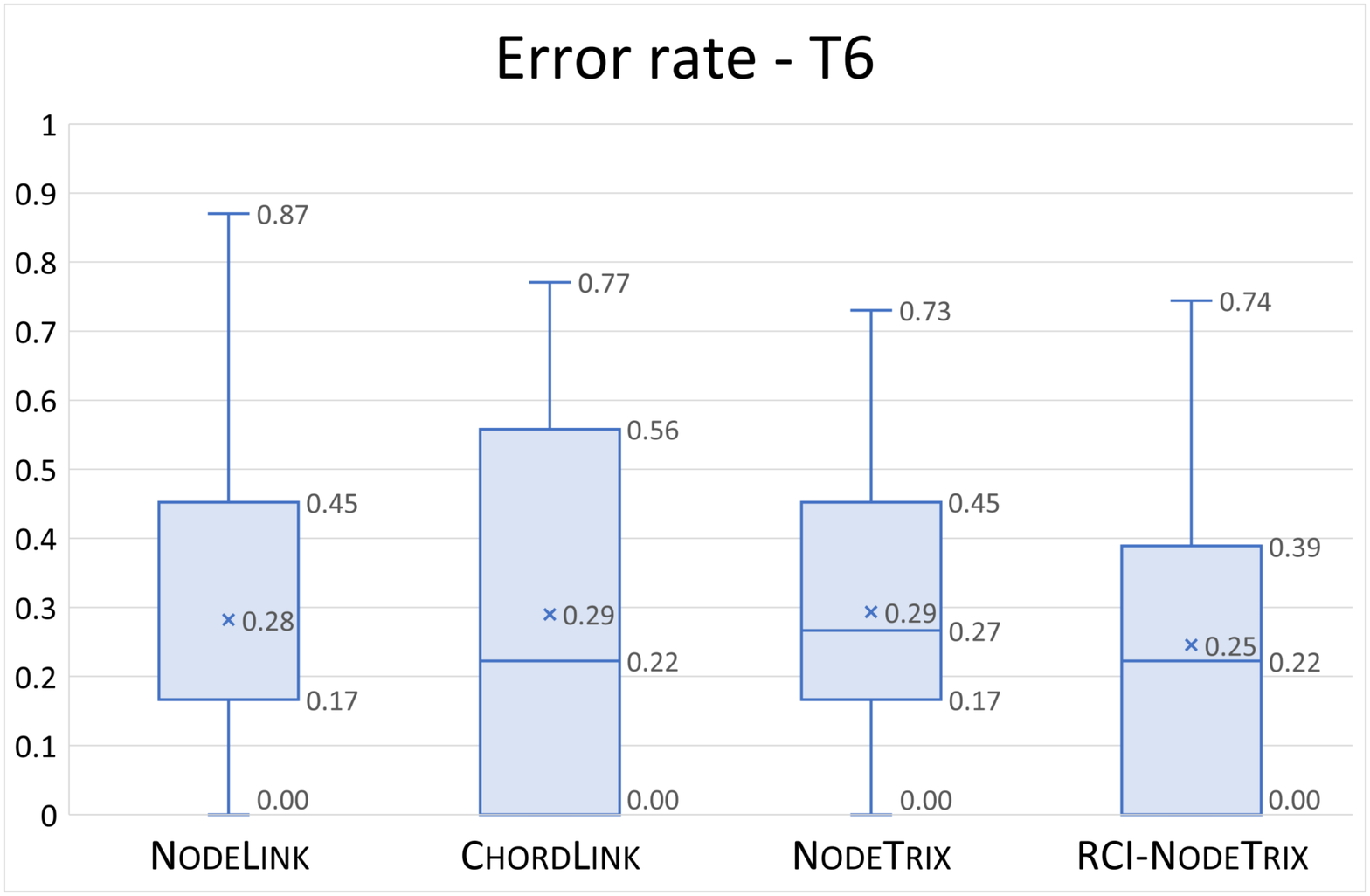}}
	\caption{Error rate aggregated by task.\label{fig:er-task}}
\end{figure}

\begin{figure}
	\fbox{\includegraphics[width=0.48\textwidth,trim=20mm 25mm 25mm 20mm, clip]{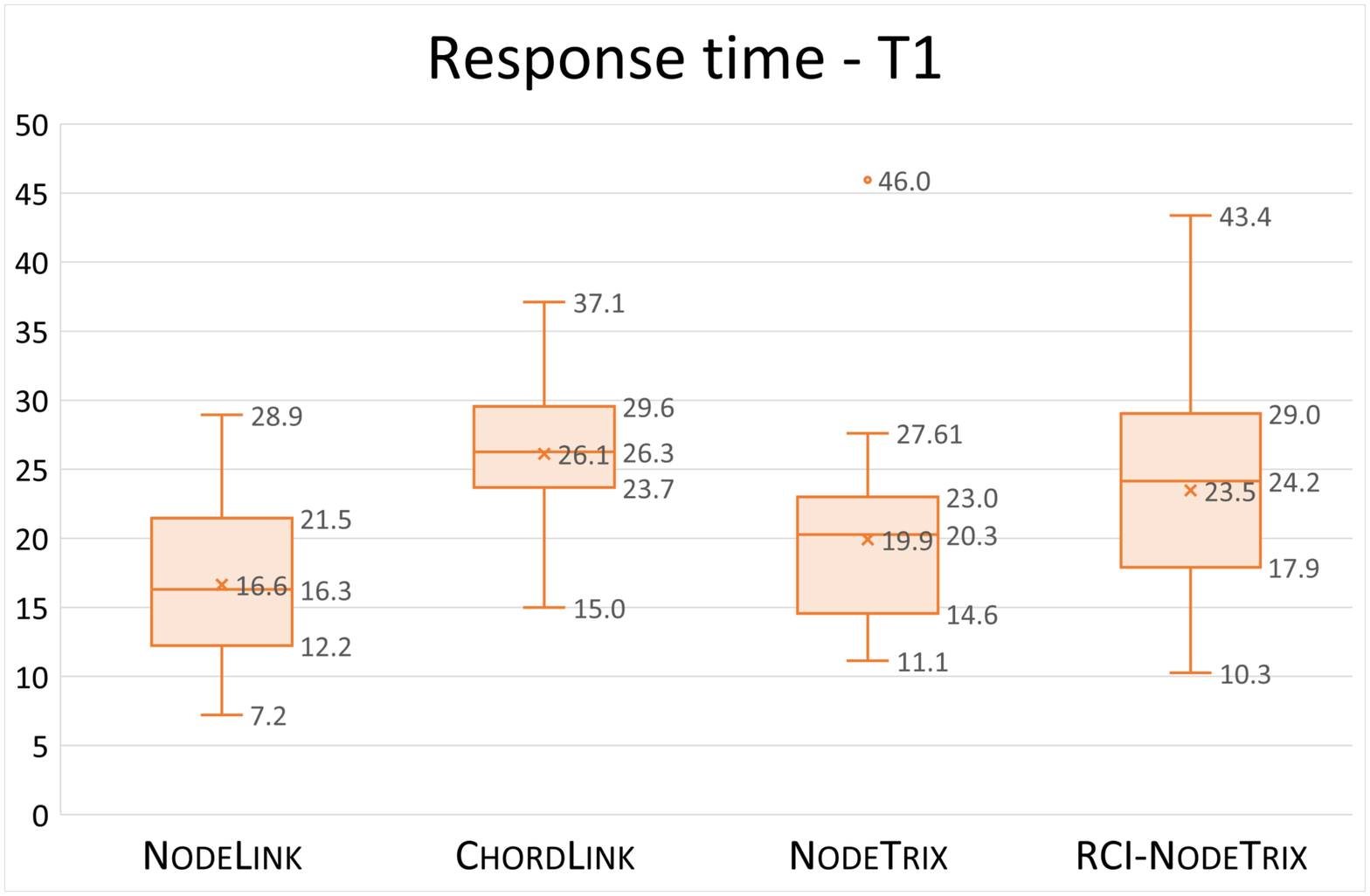}}
	\fbox{\includegraphics[width=0.48\textwidth,trim=20mm 25mm 25mm 20mm, clip]{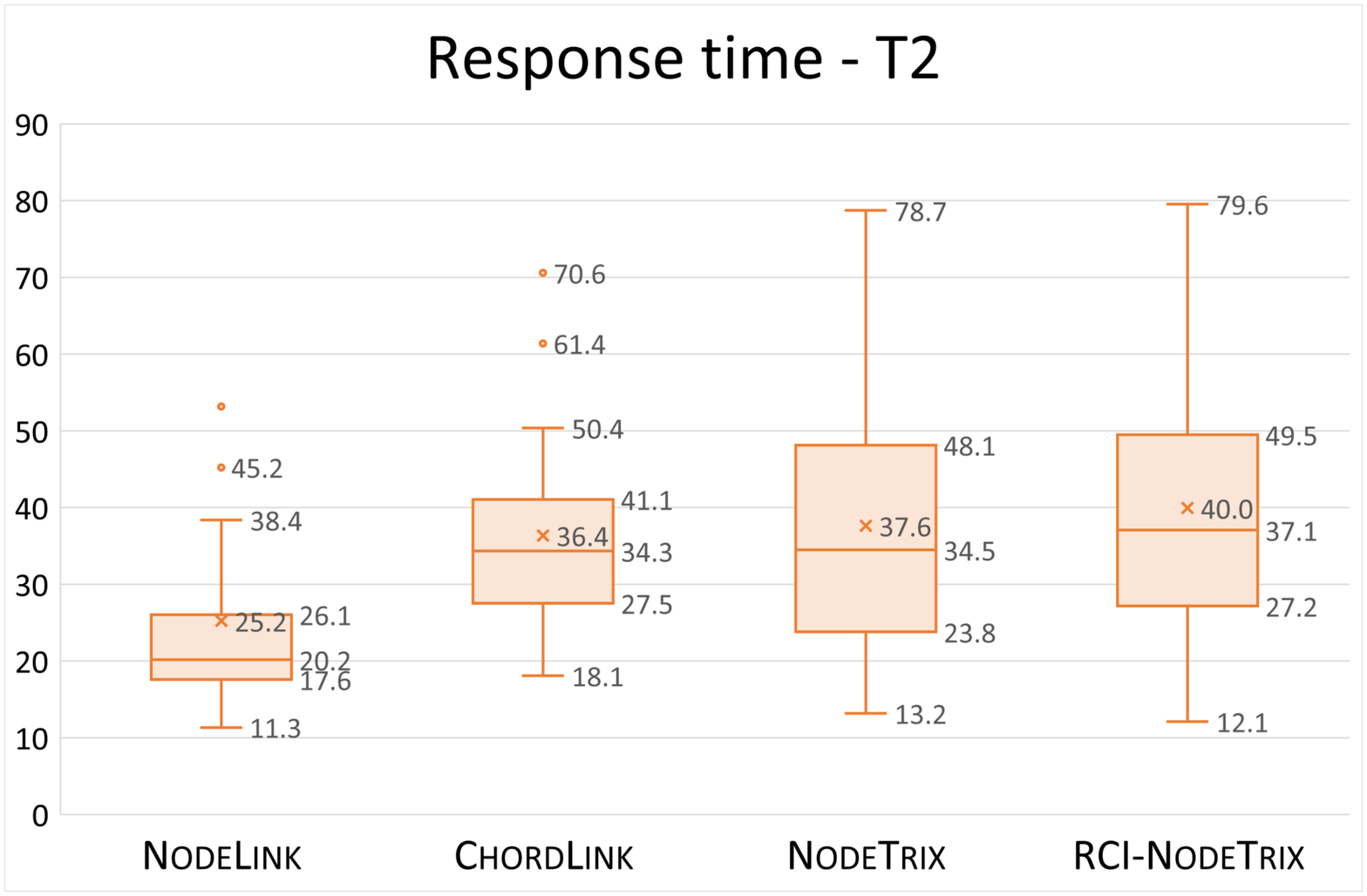}}
	\fbox{\includegraphics[width=0.48\textwidth,trim=20mm 25mm 25mm 20mm, clip]{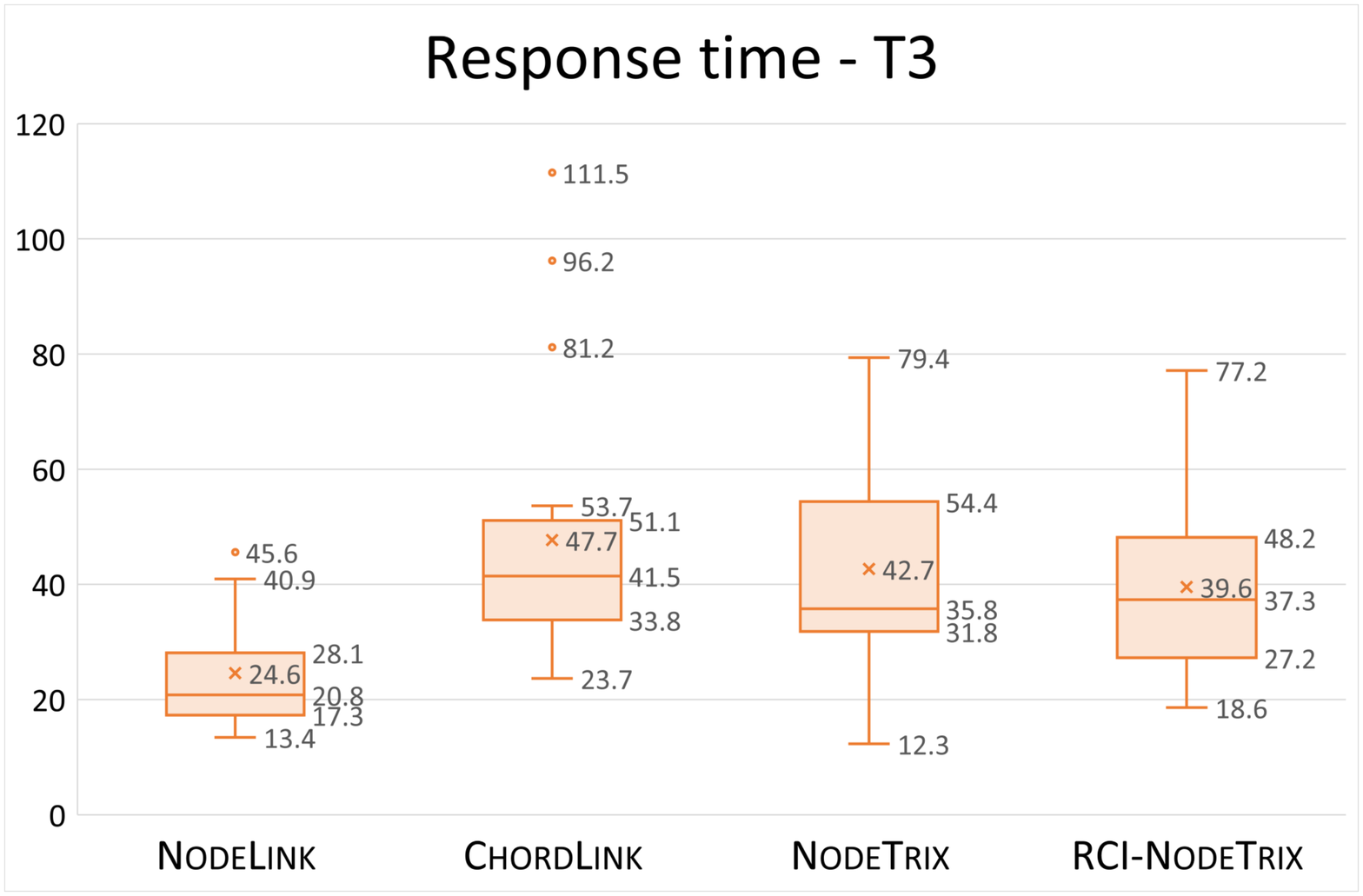}}
	\fbox{\includegraphics[width=0.48\textwidth,trim=20mm 25mm 25mm 20mm, clip]{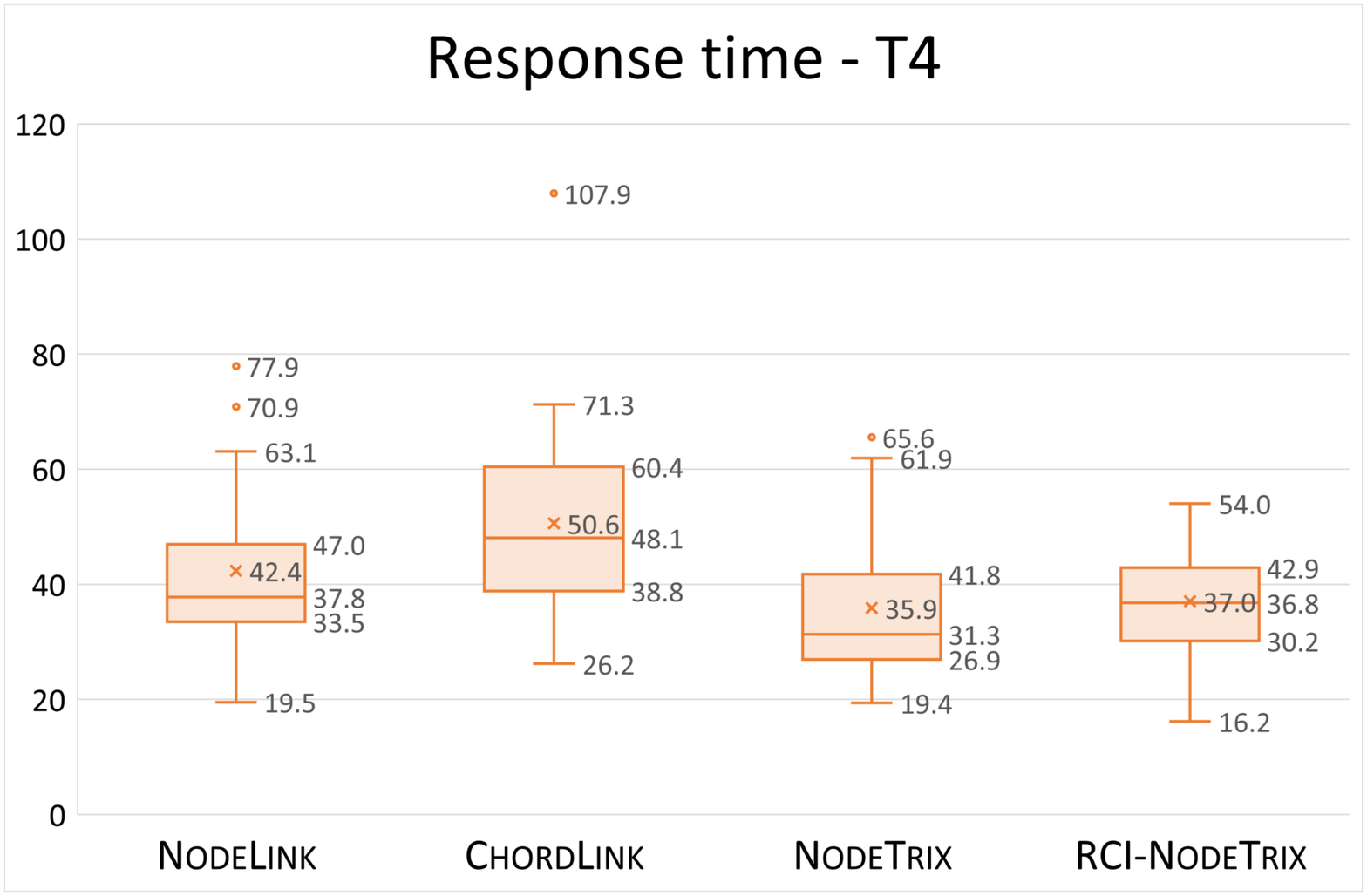}}
	\fbox{\includegraphics[width=0.48\textwidth,trim=20mm 25mm 25mm 20mm, clip]{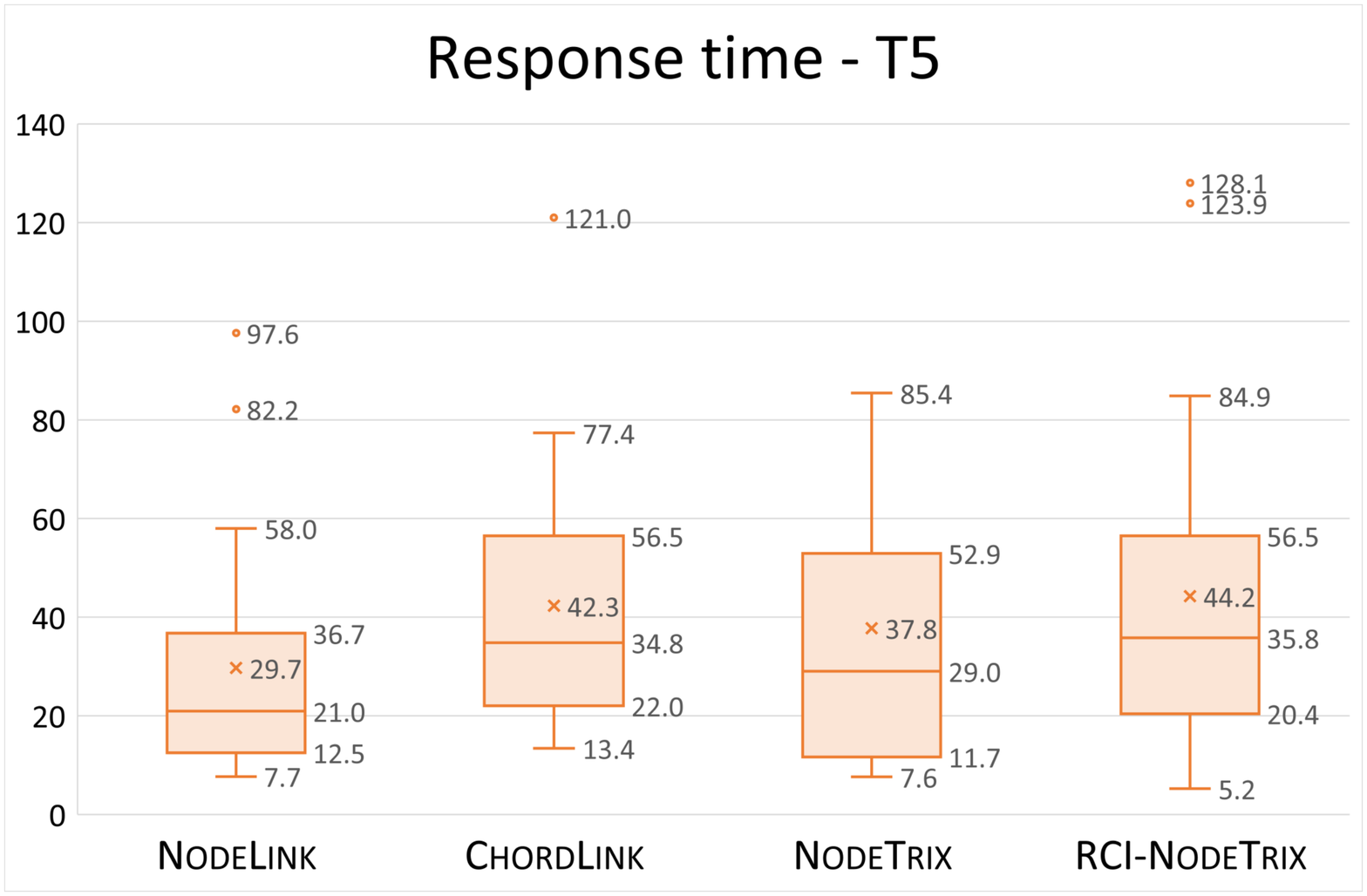}}
	\fbox{\includegraphics[width=0.48\textwidth,trim=20mm 25mm 25mm 20mm, clip]{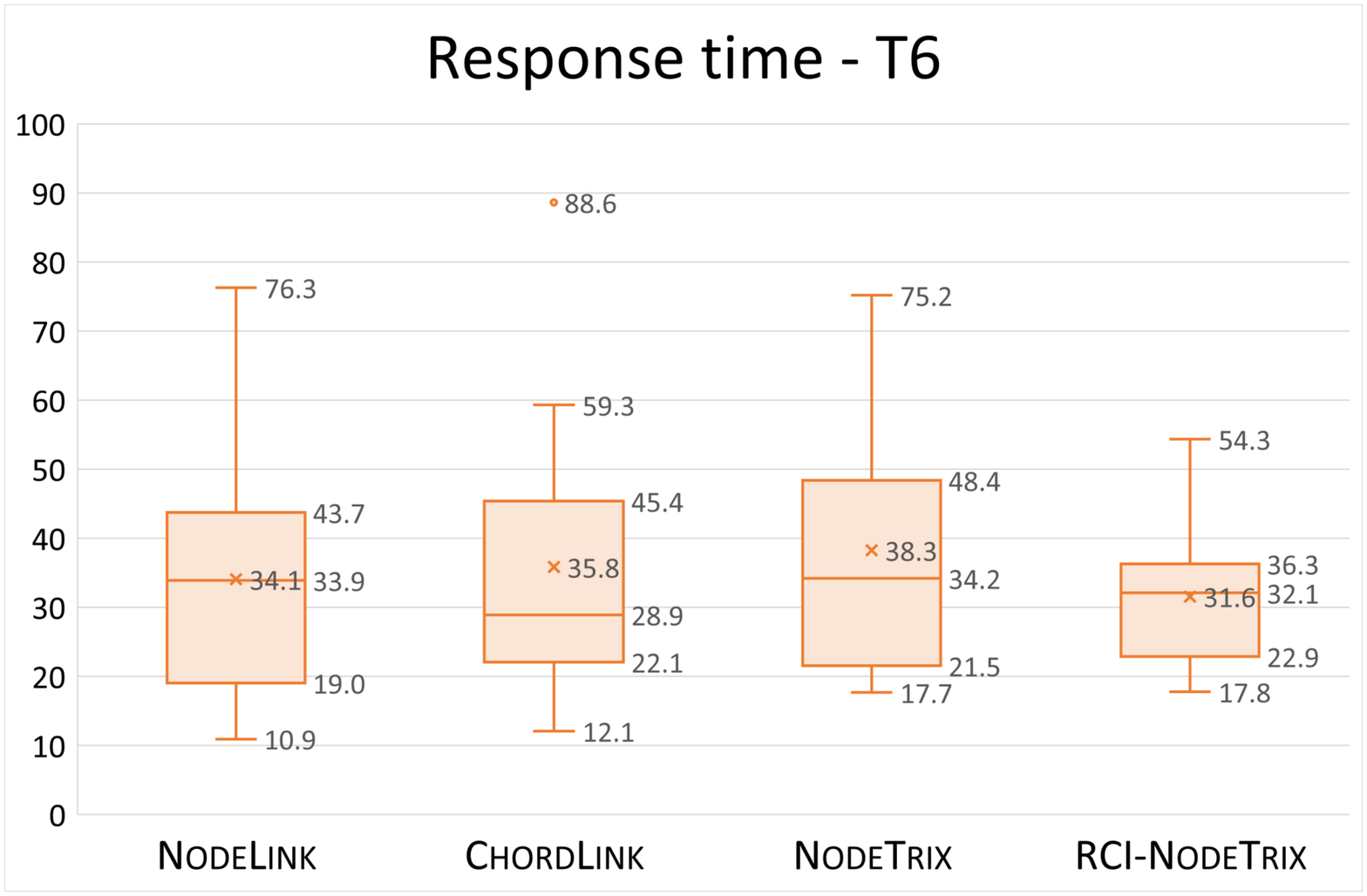}}
	\caption{Response time aggregated by task.\label{fig:time-task}}
\end{figure}

\Cref{ta:errorate-time} summarizes the results of our analysis both for the error rate (top) and for the response time (bottom). For each task, we list the models sorted by increasing values of the average error rate or response time. These values are shown in parentheses together with the model. \Cref{ta:errorate-time}  reports the statistic (column $H(3)$) and the $p$-value of the Kruskal-Wallis test. Finally, we report the adjusted (after Bonferroni corrections) significance for each pairwise comparison. Values that are statistically significant are highlighted in bold.
\Cref{fig:er-task,fig:time-task} depict the box-plots of the error rate and response time for all the tasks.

\begin{figure}[tb]
	\centering
	\fbox{\includegraphics[width=0.43\textwidth,trim=20mm 25mm 25mm 20mm, clip]{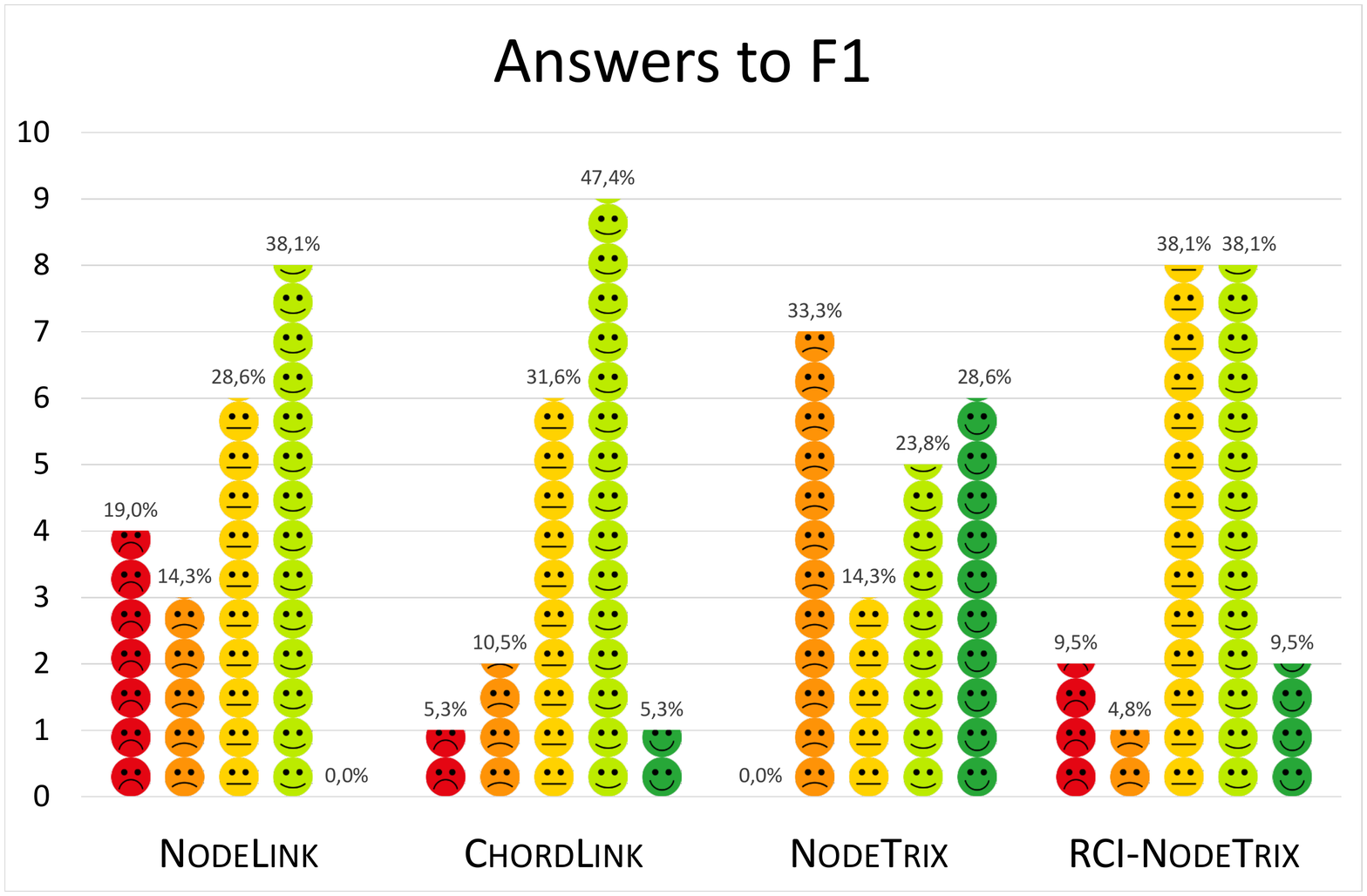}}
	\fbox{\includegraphics[width=0.43\textwidth,trim=20mm 25mm 25mm 20mm, clip]{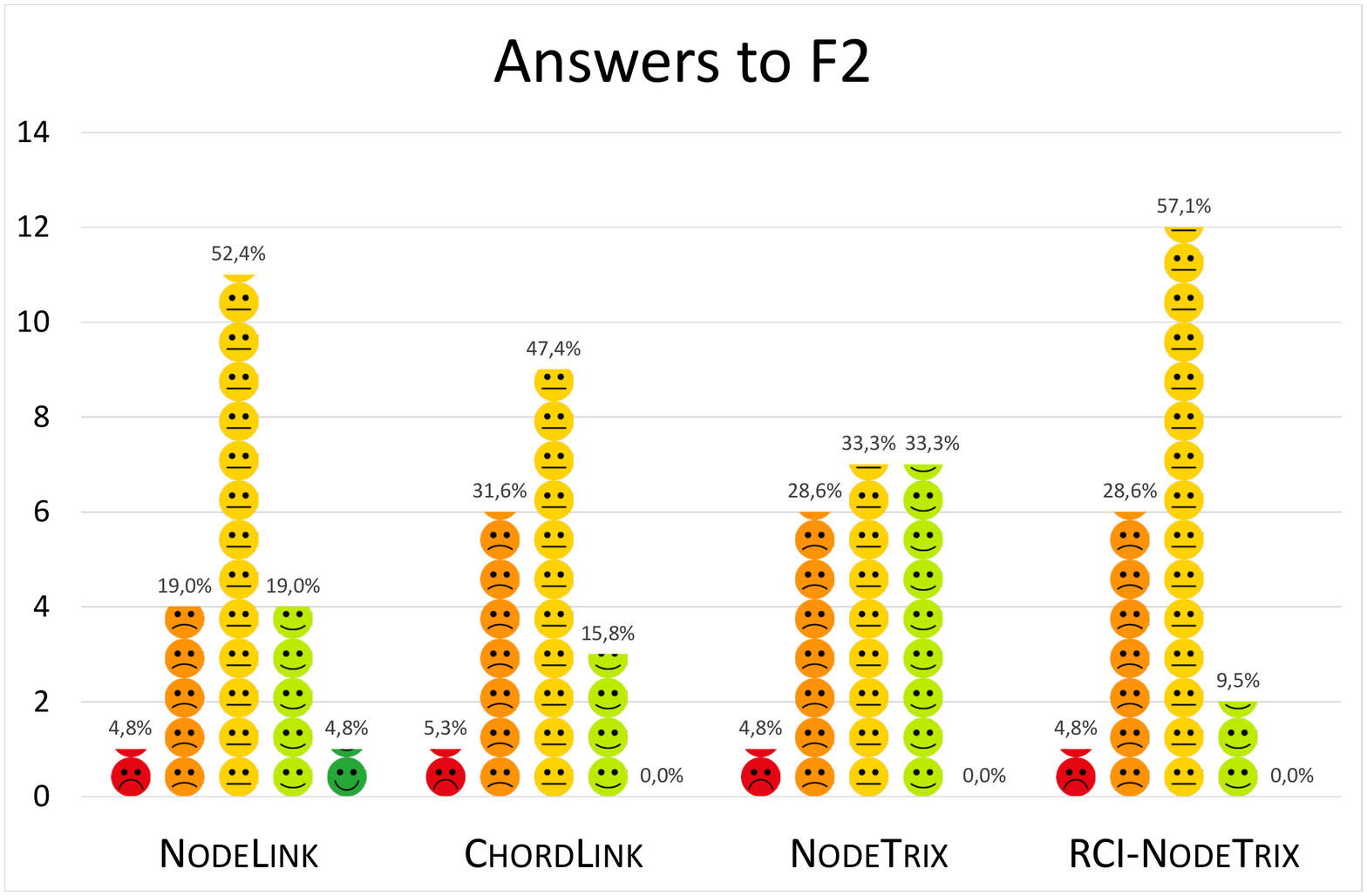}}\\
	\fbox{\includegraphics[width=0.43\textwidth,trim=20mm 25mm 25mm 20mm, clip]{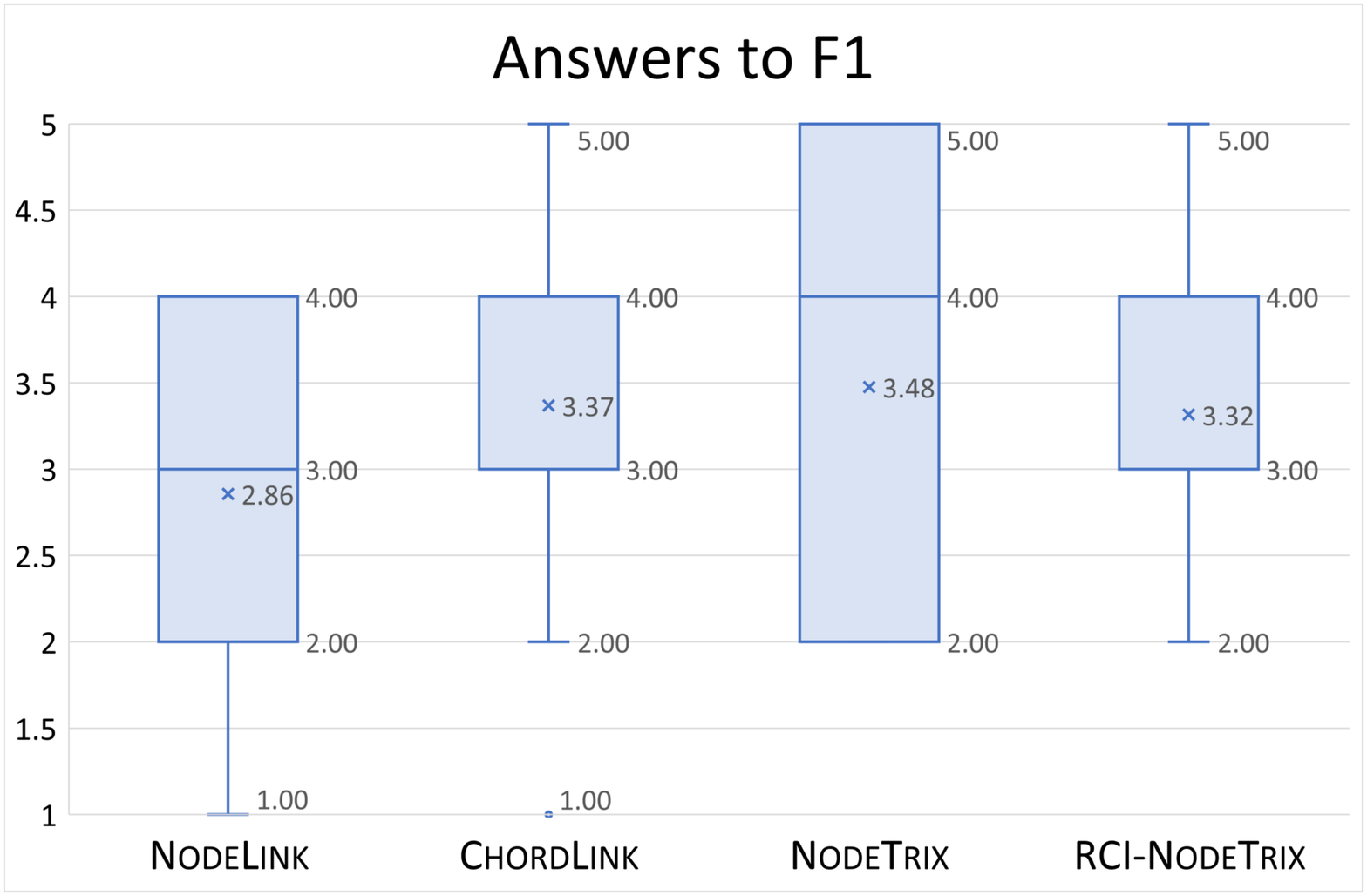}}
	\fbox{\includegraphics[width=0.43\textwidth,trim=20mm 25mm 25mm 20mm, clip]{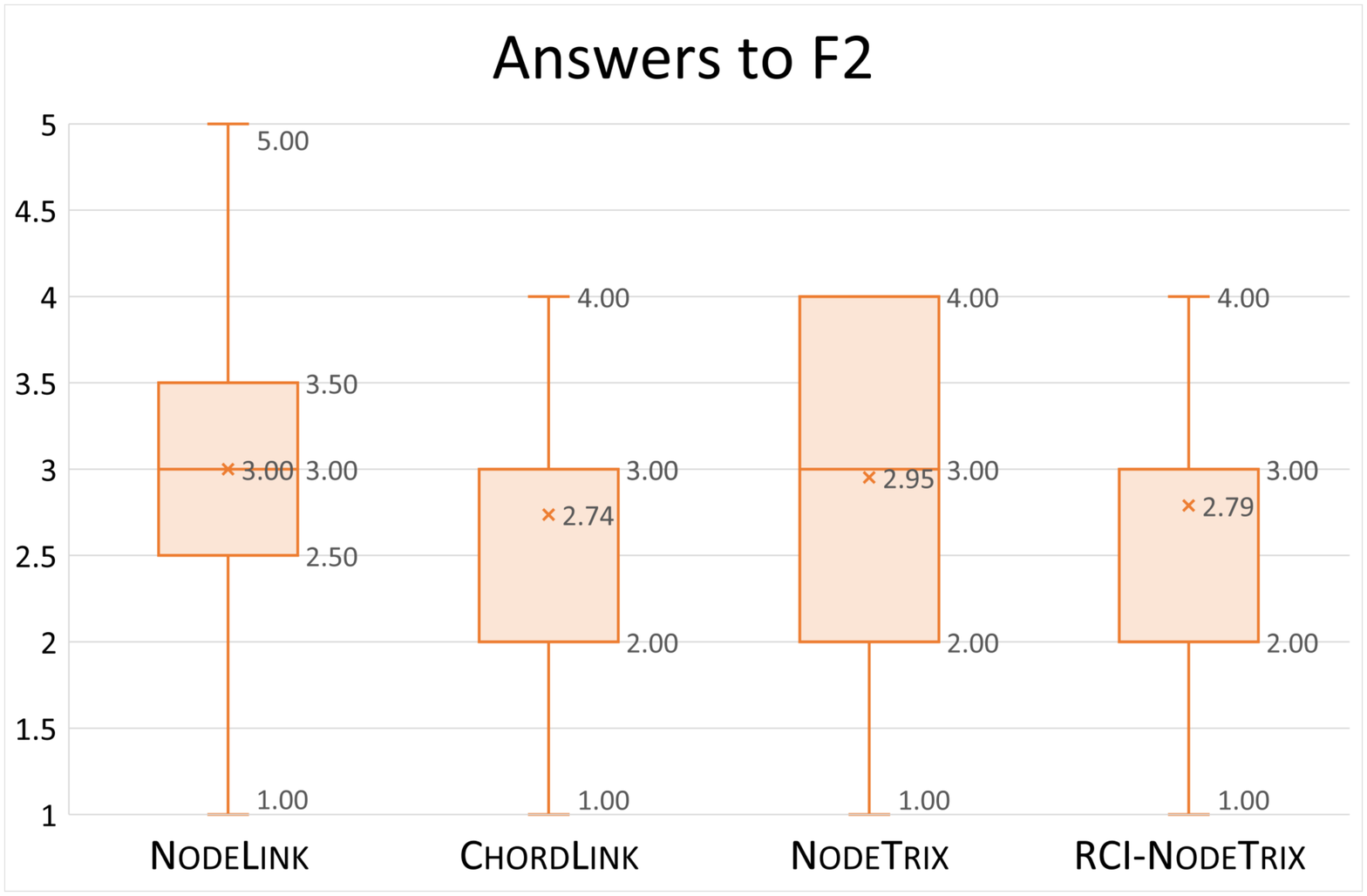}}
	\caption{Qualitative results.\label{fi:opinion}}
\end{figure}

\myparagraph{Qualitative Results.}
At the end of the test, we presented to the users the following questions:  \textsf{(F1)} \emph{``How much do you like the diagrams you have seen?''} and \textsf{(F2)} \emph{``How easy did you find answering the test questions?''}. The answers, given in a 5-point Likert scale, are summarized in \Cref{fi:opinion}, where  
we also report the answer distributions as box-plots; we assigned a score from 1 (lowest) to 5 (highest) to each answer. While there is no statistically significant difference among the models, about \textsf{(F1)} \nl received the highest percentage of strongly negative appreciations and \nt received the highest percentage of strongly positive appreciations, although with high variance. About \textsf{(F2)}, the easiness of answering was judged medium on average for all the models. 
%
%
%
In \Cref{app:freecomments} we discuss the free comments received by the participants.

%
%
%
%

\smallskip
\myparagraph{Discussion and Limitations.}
%
The following highlights summarize our results:

	\smallskip\noindent -- Hypothesis~\textbf{H1} is largely supported by the results in terms of response time. More precisely, \nl behaves better (with statistical significance) than: \cl and \rci for task~\textsf{T1}; \rci for task~\textsf{T2}; and for all hybrid models for~\textsf{T3}. In particular, the slower performance of \rci with respect to \nl for all the topology-based tasks seems to confirm the difficulty pointed out by some participants about dealing with non-symmetric matrices. In terms of error rate, \textbf{H1} is partially supported. Indeed, while there is no statistically significant difference for task~\textsf{T2}, we observe that for task~\textsf{T3} both \cl and \rci yield better accuracy than \nl, and for task~\textsf{T1} \cl behaves better than \nt. One may wonder why the same behavior is not observable between \cl and \rci on task~\textsf{T1}; our interpretation is that this might depend on the smaller number of crossings that \rci usually causes between edges that are incident to the matrices with respect to \nt.  
		
	\smallskip\noindent -- Hypothesis~\textbf{H2} is also partially supported by the results (see task~\textsf{T4}). In terms of response time, the two models based on matrices seem to lead to better performance than the other two models, with statistical significance when comparing \nt and \cl. In terms of error rate, we do not observe any statistically significant difference; in particular, the high accuracy achieved with all models seems to reveal that this task was generally easy to perform.
		
	\smallskip\noindent -- Hypothesis~\textbf{H3} is not supported by our results, as we do not observe any statistically significant difference among all the four models.          

\smallskip
We conclude by discussing the limits of our study. The choice of not allowing interaction implied to use networks of small/medium size that fit into the screen window; also, it required to have a set of predefined clusters that the user cannot change. On the other hand, a non-interactive environment facilitated the execution of an on-line test; we believe that enabling visual interaction for the considered models would require a different study design, preferably based on a controlled experiment. Further, interactions may introduce confounding factors and it is difficult to design interaction features that are fair to all models.
%
The number of tasks was limited to six, which is in line with many previous studies. Although some works use a larger number of tasks (see, e.g.,~\cite{ojk-nlamoqni-tvcg2019}), we believe that more tasks may cause long execution times and a high fatigue effect for the users, which may result in less reliable data. 
Finally, the visualization models that we compare may be sensitive to the specific algorithms used to produce the drawings. This justifies further investigation with different layout algorithms.

\section{Conclusions and Future Research}\label{se:conclusions}

We presented a user study that compares different hybrid visualization models and the popular node-link model. As a preliminary answer to \textsf{RQ1}, the results suggest that hybrid visualizations may help to overcome some limits of node-link diagrams in accurately executing topology-based tasks on globally sparse but locally dense networks, at the expenses of the execution time. About \textsf{RQ2}, we could not conclude that any of the considered hybrid models is superior; however, for some topology-based tasks, we observed better accuracy with \cl and faster execution with \nt. Our study has some limitations and cannot be generalized to settings significantly different from ours. This motivates further experiments with larger networks, interaction features, and additional tasks. 
Enlarging the set of participants is also an interesting future objective. In this direction, we tried to collect an additional data set by using the Amazon's Mechanical Turk crowdsourcing service. To keep the population sample homogeneous, we required users with background in computer science. However, we report a negative response on this side: the participation was low and the collected data was strongly unreliable (the~error rates were very high and the response times too short for~a~reasoned~answer). For these reasons, we decided to discard this additional data set.

\myparagraph{Acknowledgements.}
We thank Giuseppe~Liotta for useful discussions, Lorenzo Angori for his help in the implementation, and all the participants to the study.

\bibliography{biblio}
\bibliographystyle{splncs04}

\newpage

\appendix

\section{Further Details on the User Study Design}\label{app:design}

\subsection{Stimuli}\label{app:stimuli}

For each of the four visualization models described in~\Cref{sse:models}, we produced a diagram of the three networks described in~\Cref{sse:stimuli}. The diagrams for the \nl model are computed through the force-directed algorithm available in the D3 library~\cite{d3}. Starting from these drawings, we defined some geometric clusters with the technique based on the $K$-means algorithm~\cite{k-means} described in~\cite{admpt-hgvcl-tvcg20}. As explained in \Cref{sse:models}, the system presented in~\cite{admpt-hgvcl-tvcg20} is used to compute the diagrams in the \cl, \nt, and \rci models with the same sets of clusters. The algorithm for \nt is based on~\cite{nt-system,hfm-dhvsn-07} and uses the \emph{leaf order method} to compute the row/column order~\cite{DBLP:conf/ismb/Bar-JosephGJ01}. The algorithm for \rci is a variant of the one for \nt, where the orders for the rows and the columns are independently computed to reduce the number of crossings between inter-cluster edges; this is done by an adaptation of the sifting algorithm for layered drawings~\cite{BACHMAIER2010159,msm-usklscm}. 
In all four diagrams of the same network, we labeled all the nodes that belong to clusters and few high-degree nodes outside the clusters. We avoid label duplication in all the drawings; this reduces visual clutter and suffices to correctly interpret the data in all models; in particular, in \nt and \rci we filled each cell whose row and column refer to the same node with a color distinct from black and white (these cells correspond to the main matrix diagonal in \nt). Also, we use numerical id labels instead of real names to guarantee anonymity and to avoid that  users could be influenced by their knowledge about the network.

\subsection{Pilot Study}\label{app:pilot} 

The actual experiment was preceded by a pilot study with 19 participants, mostly colleagues and students in computer engineering, who are representatives of our class of target users. Based on the feedback received from the pilot study, we made some small changes to the survey. More precisely, for task \textsf{T4} we increased the number of labels to be found from one to three, because we had 100\% correct answers. For task \textsf{T6} we changed the type of question from a single choice question (the user selects the answer from a fixed set of values) to a free text answer. We made this change because the limited number of options helped the users to guess the right answer; in particular few participants reported that when they had in mind a wrong value that was not present among the options, they selected the closest value.

\clearpage

\subsection{Trials for the Network \textsf{weavers}}\label{app:trials}

\begin{figure}[h!]
	\centering
	\subfigure[\textsf{T1}]{%
		\centering
		\includegraphics[width=0.38\columnwidth]{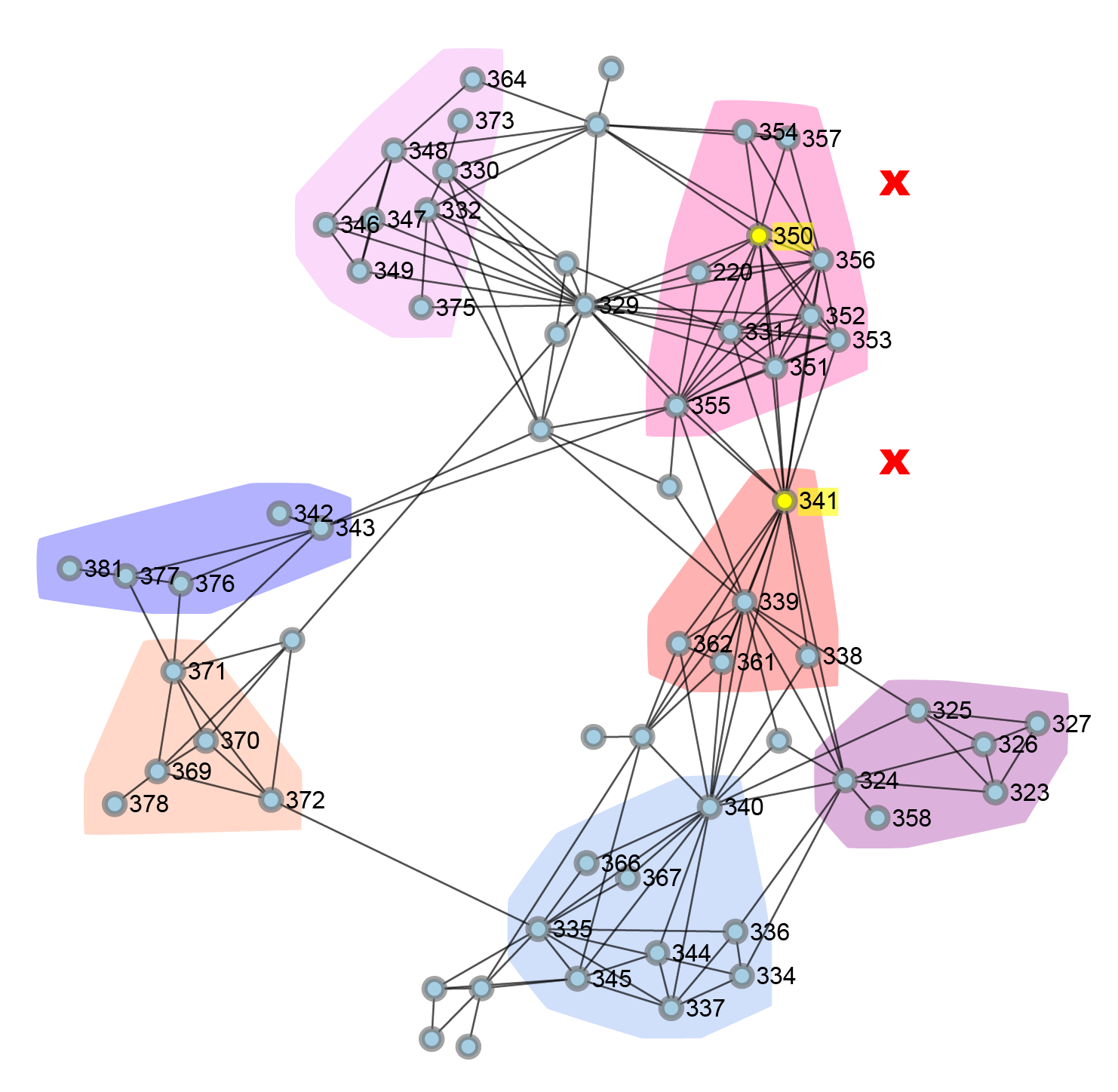}
		\label{fi:social-nl-t1}
	}\hfill
	\subfigure[\textsf{T2}]{%
		\centering
		\includegraphics[width=0.38\columnwidth]{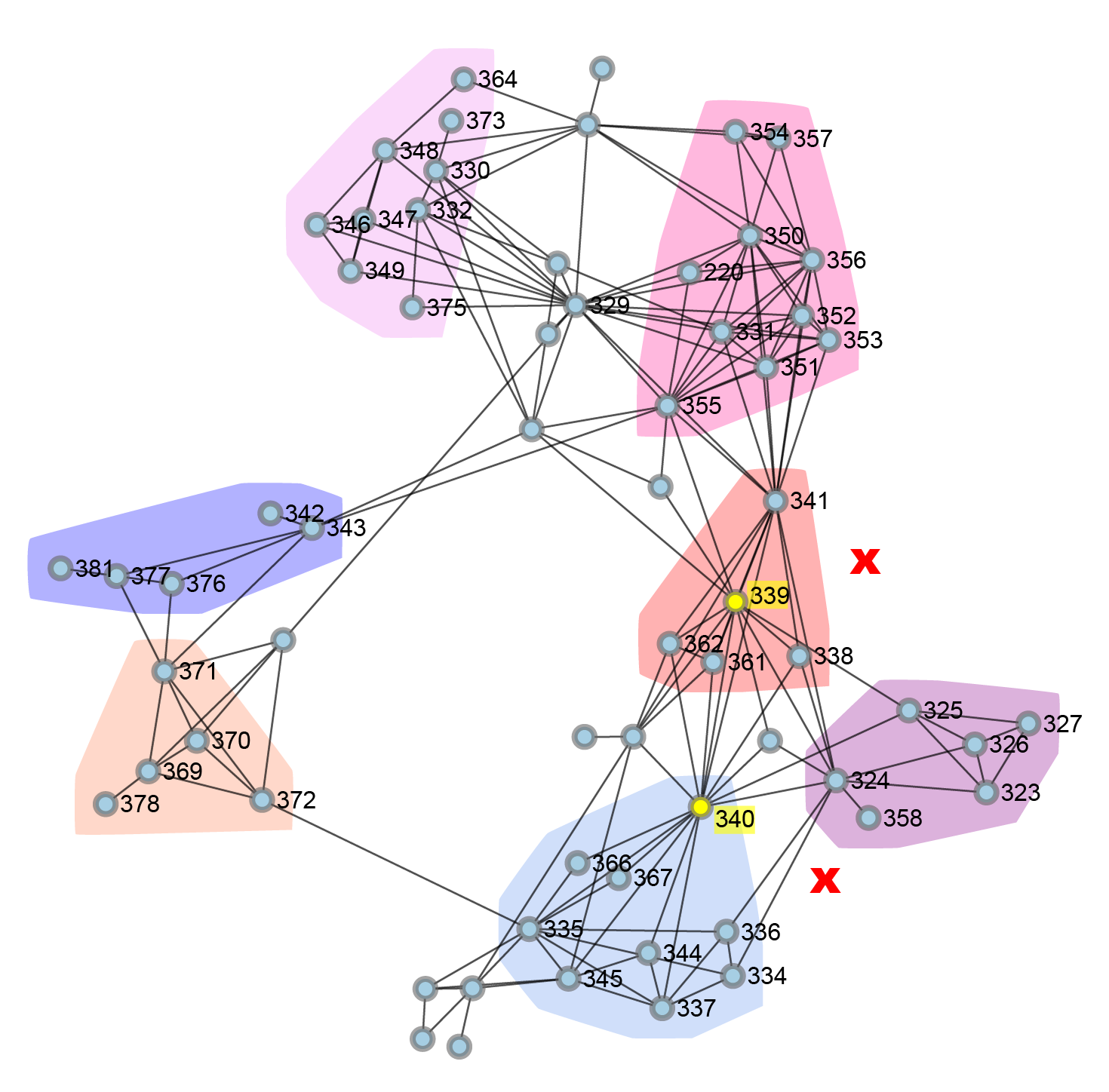}
		\label{fi:social-nl-t2}
	}\hfill
	\subfigure[\textsf{T3}]{%
		\centering
		\includegraphics[width=0.38\columnwidth]{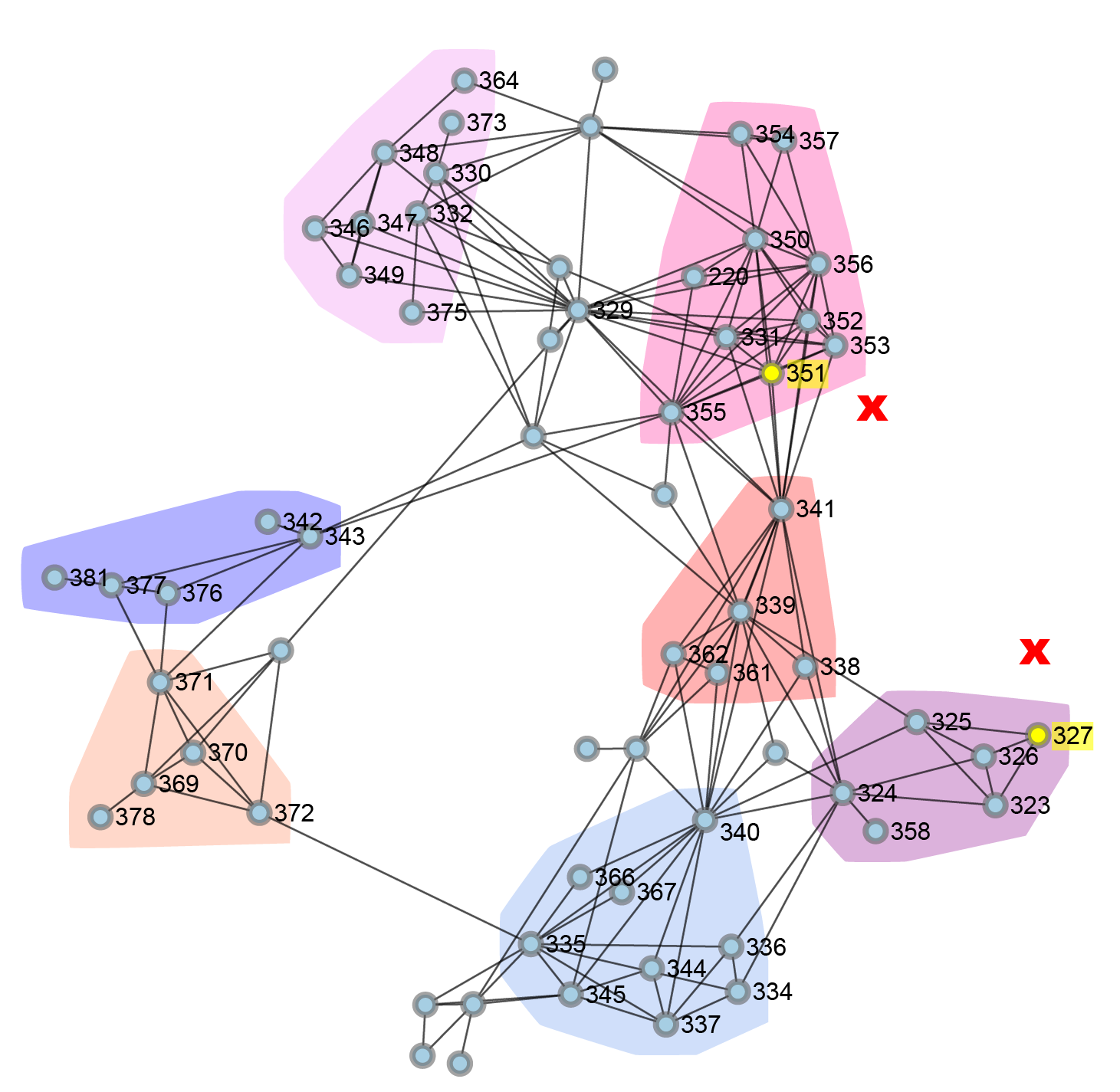}
		\label{fi:social-nl-t3}
	}\hfill
	\subfigure[\textsf{T4}]{%
		\centering
		\includegraphics[width=0.38\columnwidth]{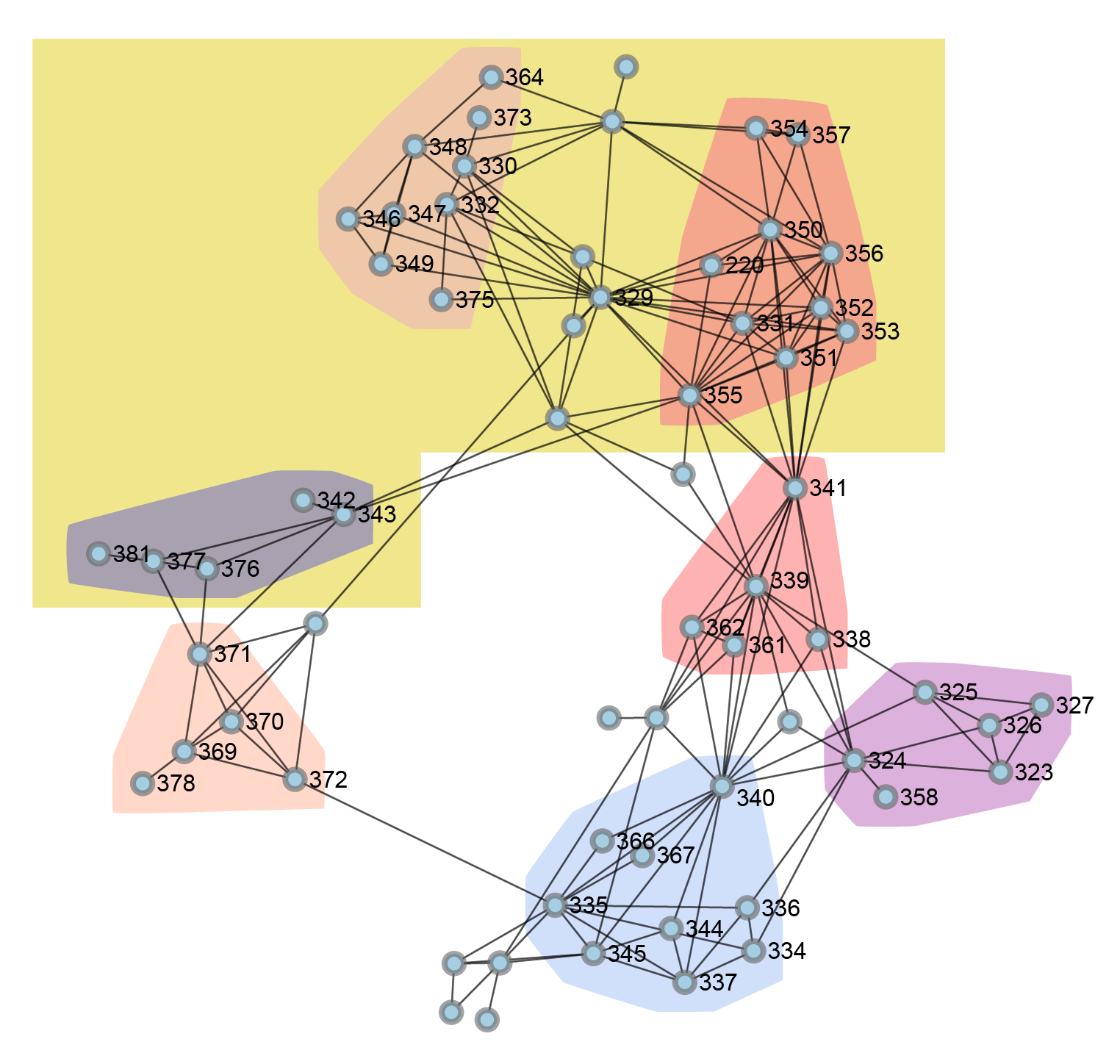}
		\label{fi:social-nl-t4}
	}\hfill
	\subfigure[\textsf{T5}]{%
		\centering
		\includegraphics[width=0.38\columnwidth]{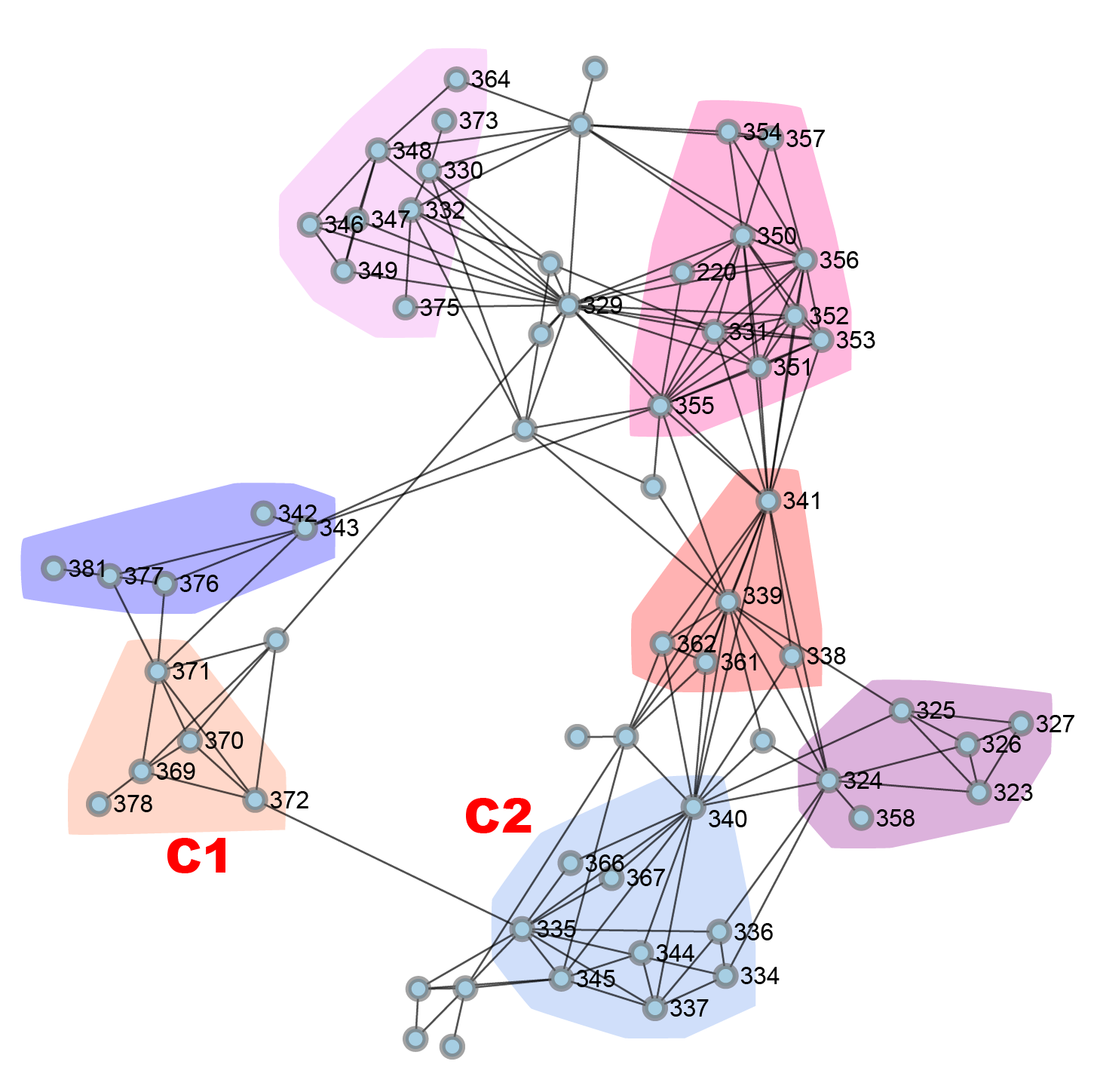}
		\label{fi:social-nl-t5}
	}\hfill
	\subfigure[\textsf{T6}]{%
		\centering
		\includegraphics[width=0.38\columnwidth]{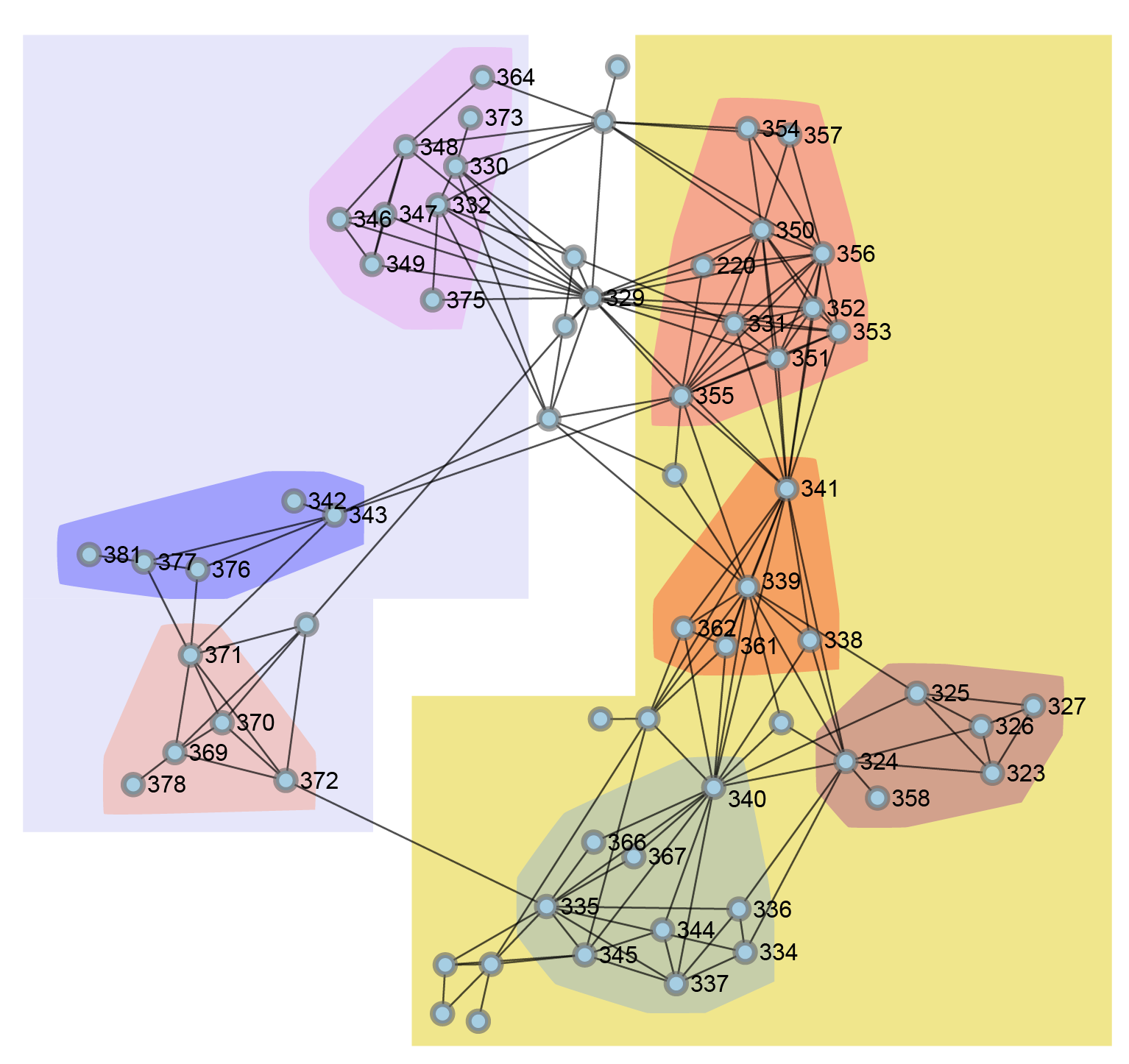}
		\label{fi:social-nl-t6}
	}\hfill
	\caption{Trials for the network \textsf{weavers} in the \nl model. For task \textsf{T3} $k=4$; for task \textsf{T4} the three labels are: 374, 334, 220.}\label{fi:trials-nl}
\end{figure}

\begin{figure}[h!]
	\centering
	\subfigure[\textsf{T1}]{%
		\centering
		\includegraphics[width=0.38\columnwidth]{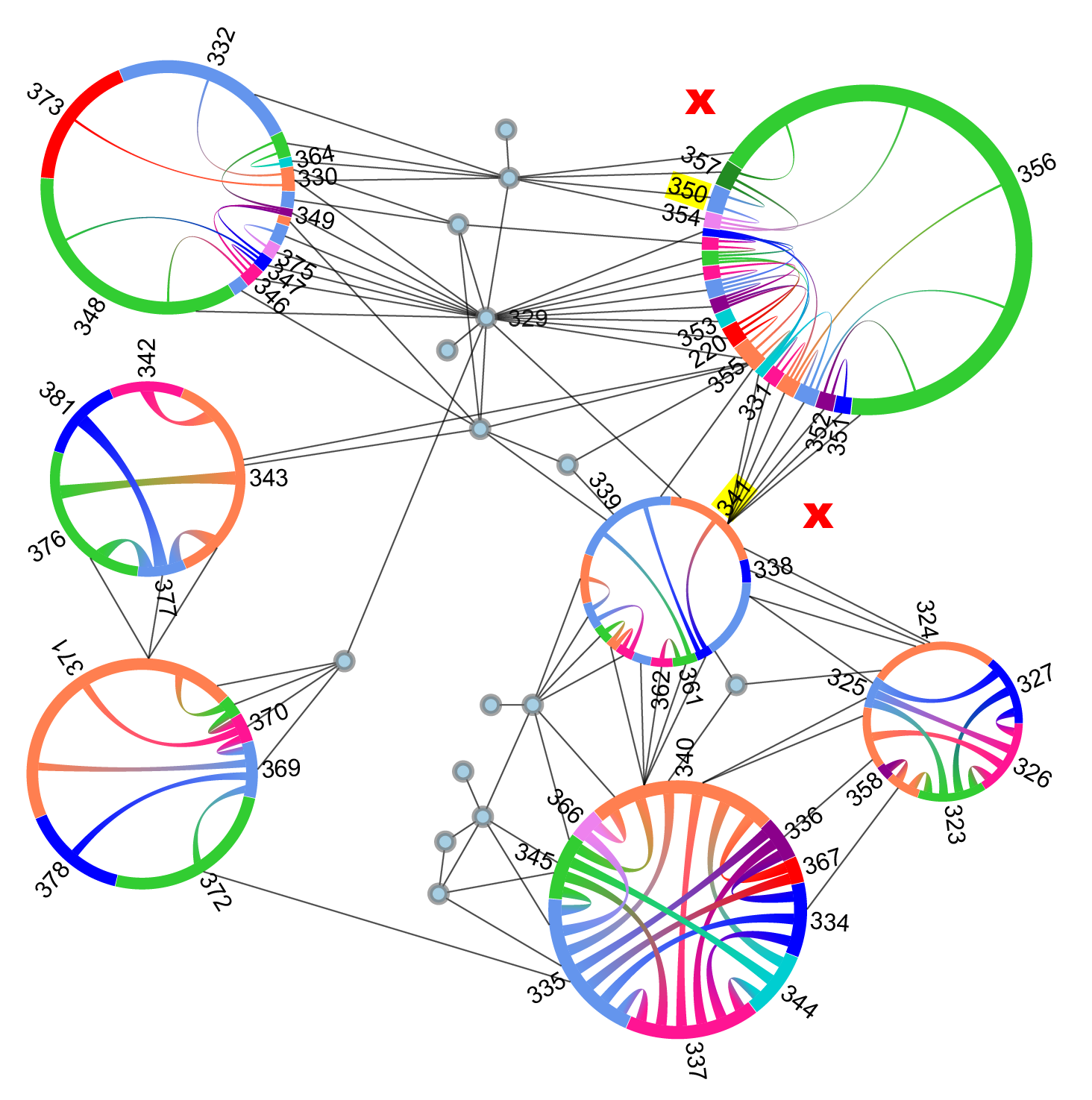}
		\label{fi:social-cl-t1}
	}\hfill
	\subfigure[\textsf{T2}]{%
		\centering
		\includegraphics[width=0.38\columnwidth]{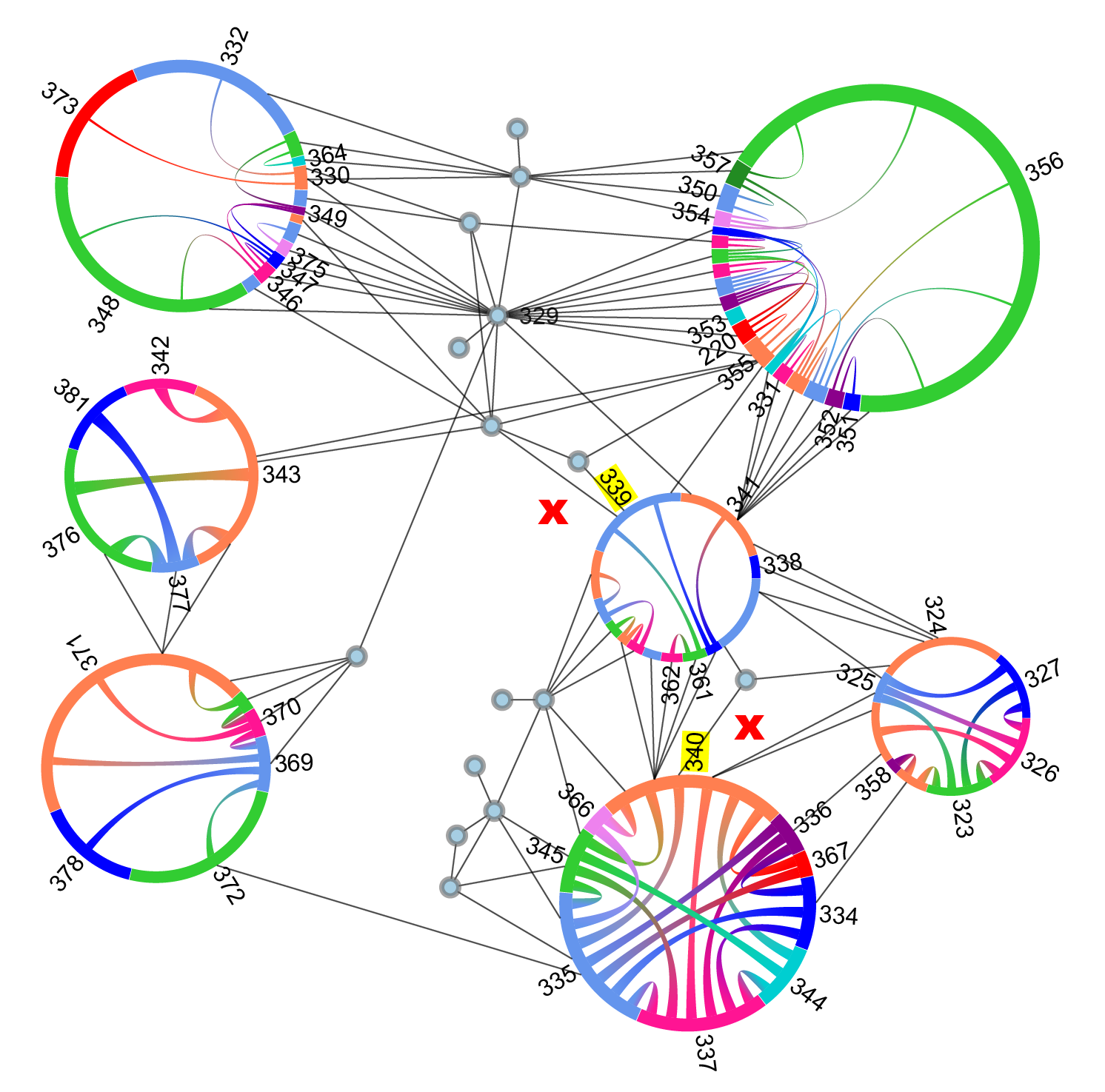}
		\label{fi:social-cl-t2}
	}\hfill
	\subfigure[\textsf{T3}]{%
		\centering
		\includegraphics[width=0.38\columnwidth]{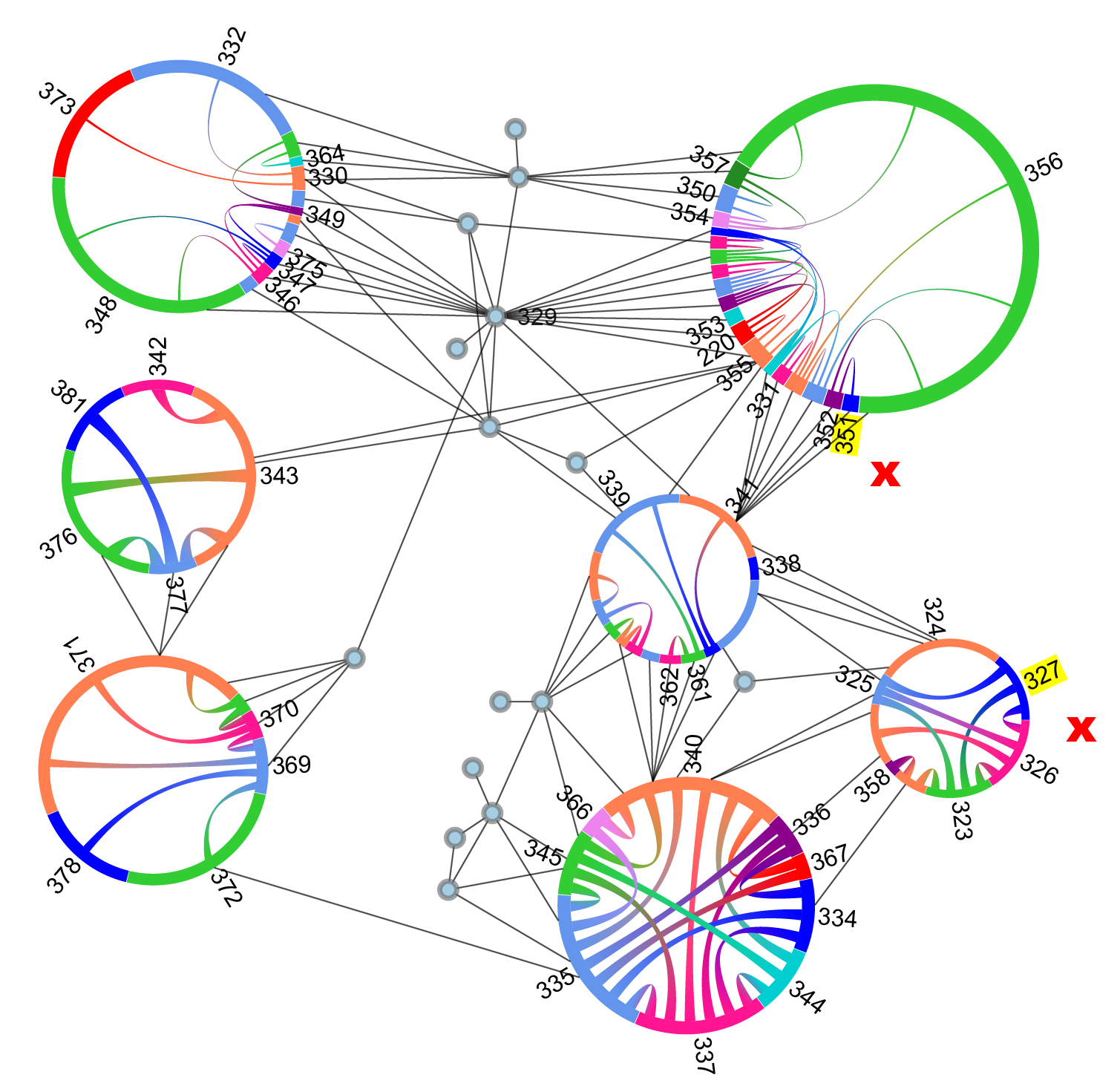}
		\label{fi:social-cl-t3}
	}\hfill
	\subfigure[\textsf{T4}]{%
		\centering
		\includegraphics[width=0.38\columnwidth]{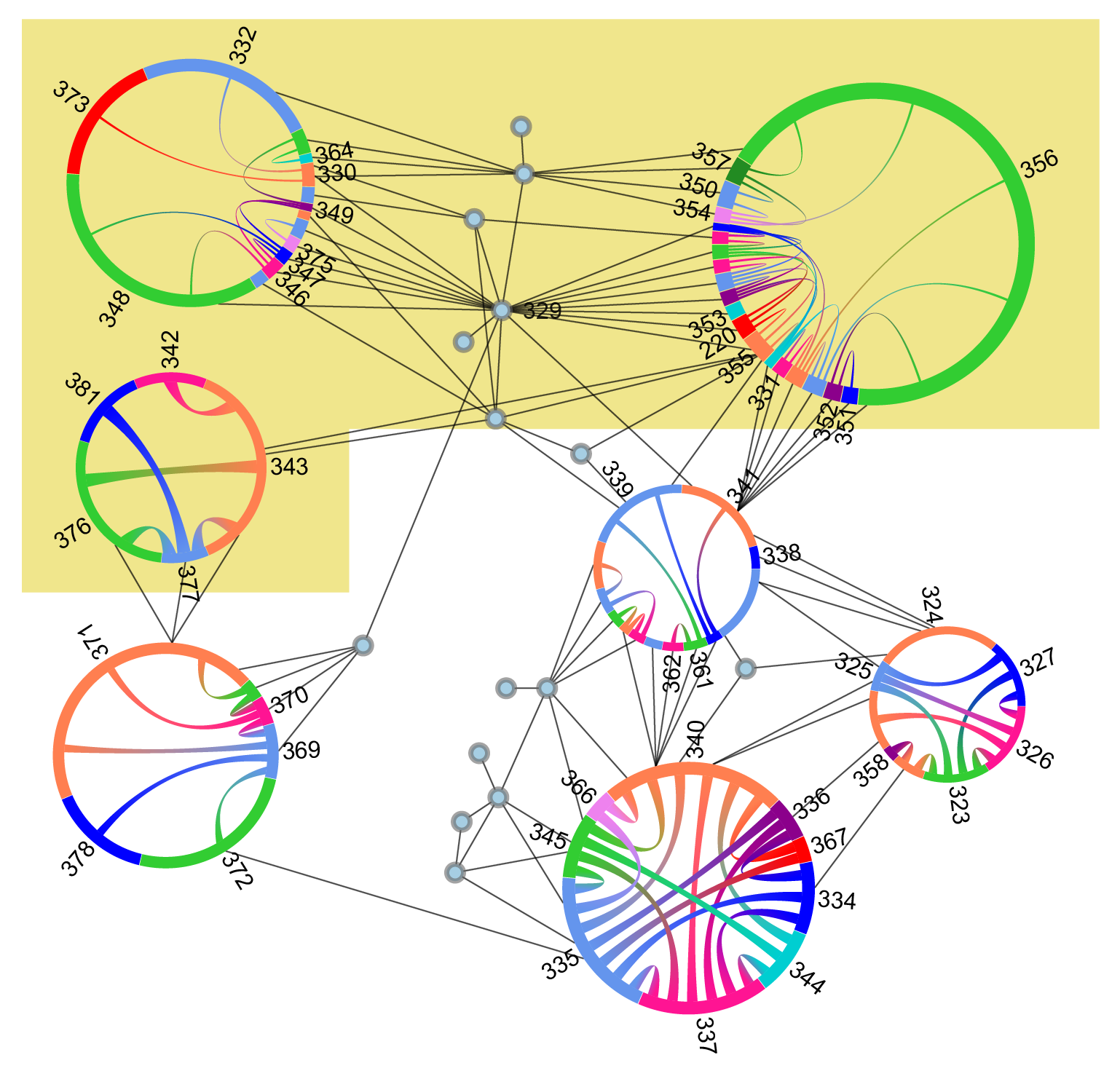}
		\label{fi:social-cl-t4}
	}\hfill
	\subfigure[\textsf{T5}]{%
		\centering
		\includegraphics[width=0.38\columnwidth]{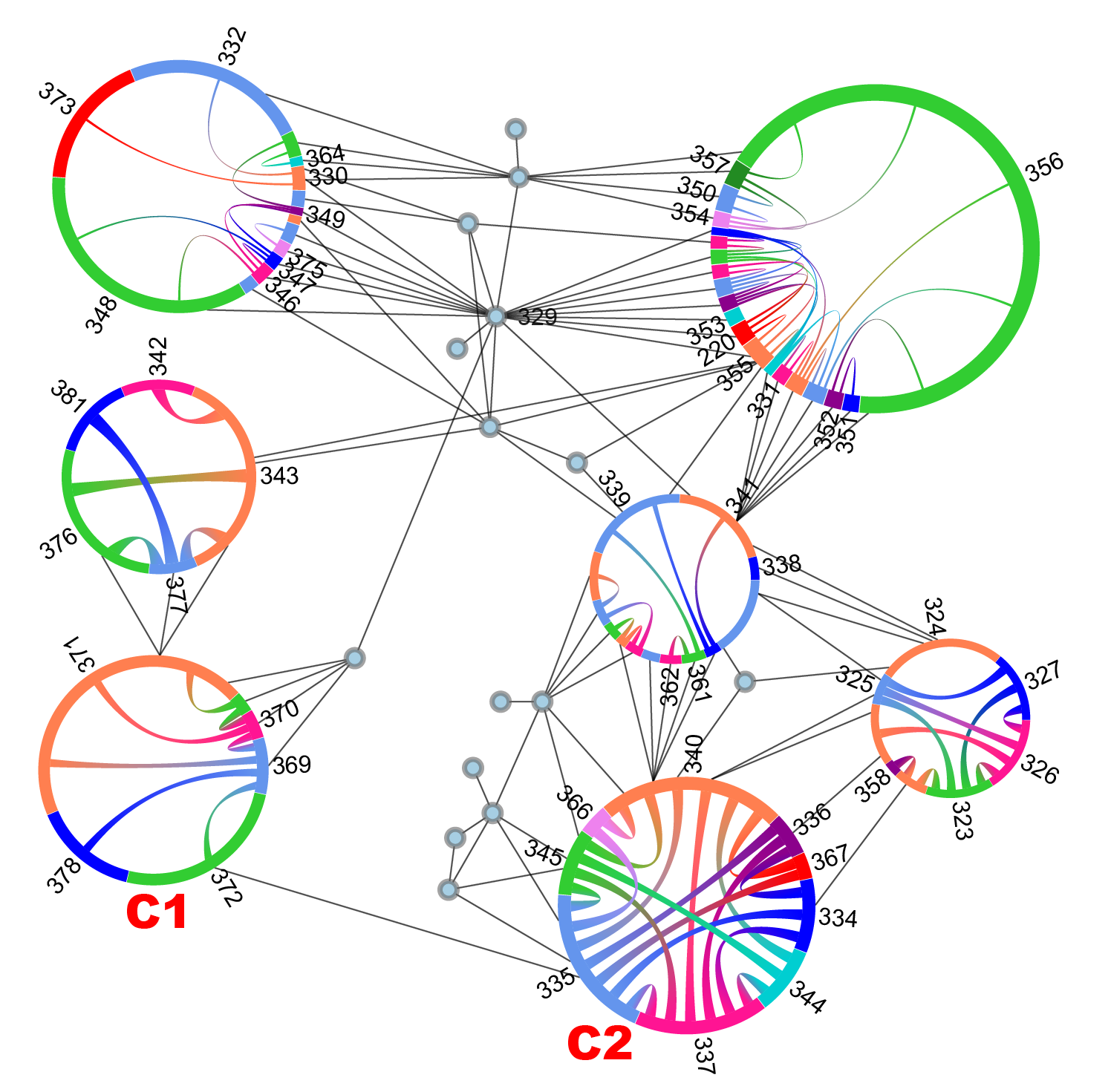}
		\label{fi:social-cl-t5}
	}\hfill
	\subfigure[\textsf{T6}]{%
		\centering
		\includegraphics[width=0.38\columnwidth]{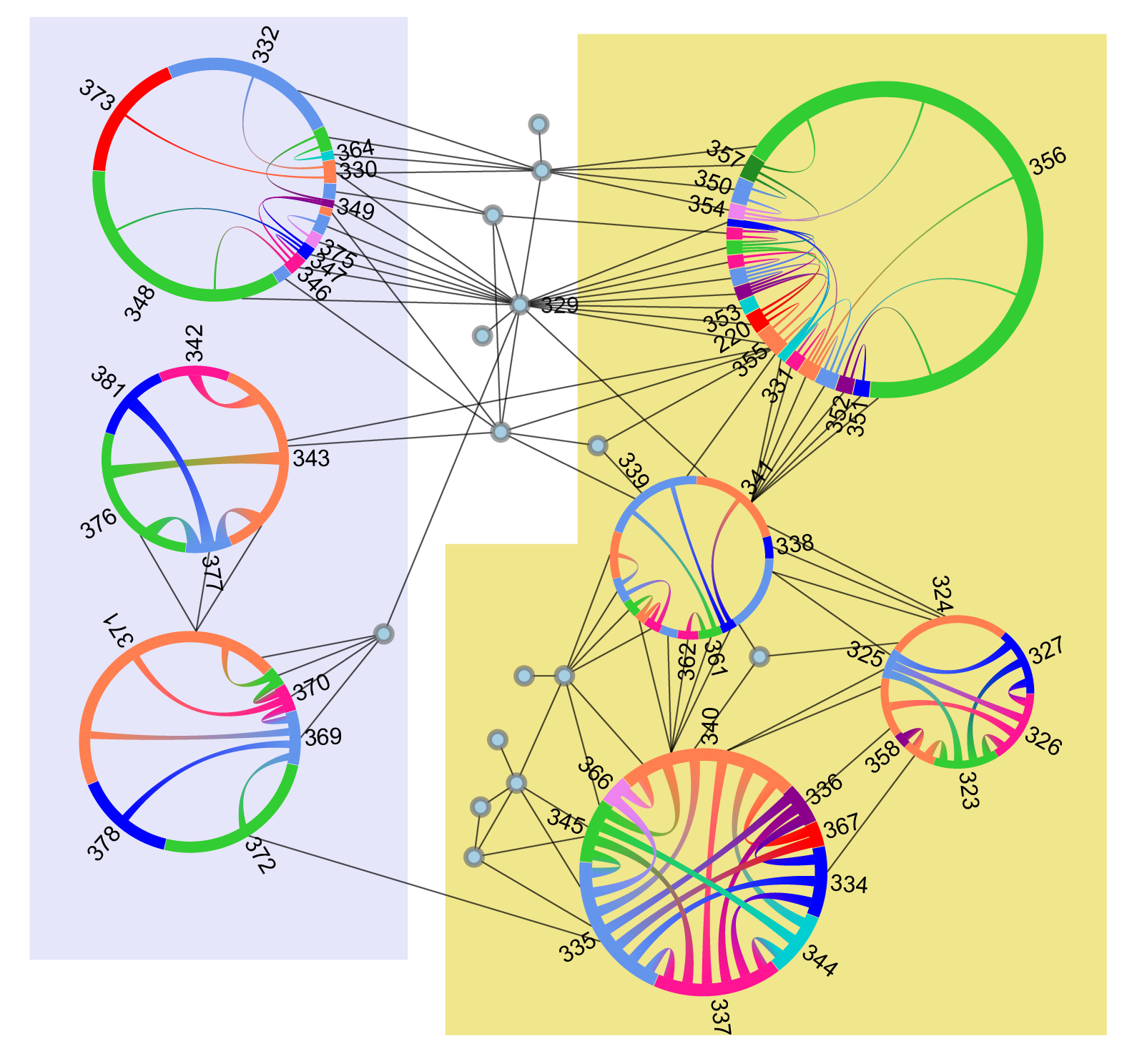}
		\label{fi:social-cl-t6}
	}\hfill
	\caption{Trials for the network \textsf{weavers} in the \cl model. For task \textsf{T3} $k=4$; for task \textsf{T4} the three labels are: 374, 334, 220.}\label{fi:trials-cl}
\end{figure}

\begin{figure}[h!]
	\centering
	\subfigure[\textsf{T1}]{%
		\centering
		\includegraphics[width=0.38\columnwidth]{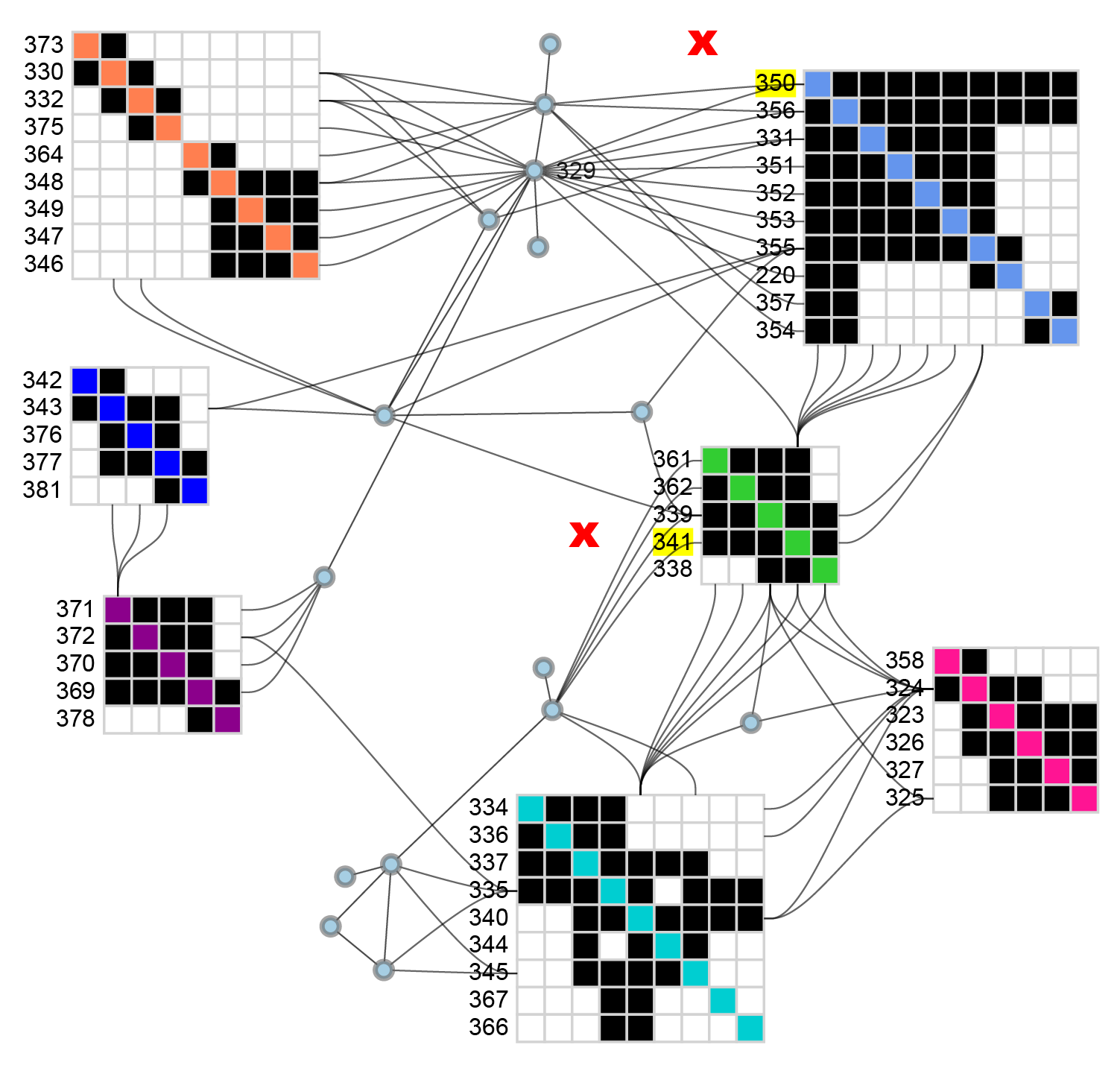}
		\label{fi:social-nt-t1}
	}\hfill
	\subfigure[\textsf{T2}]{%
		\centering
		\includegraphics[width=0.38\columnwidth]{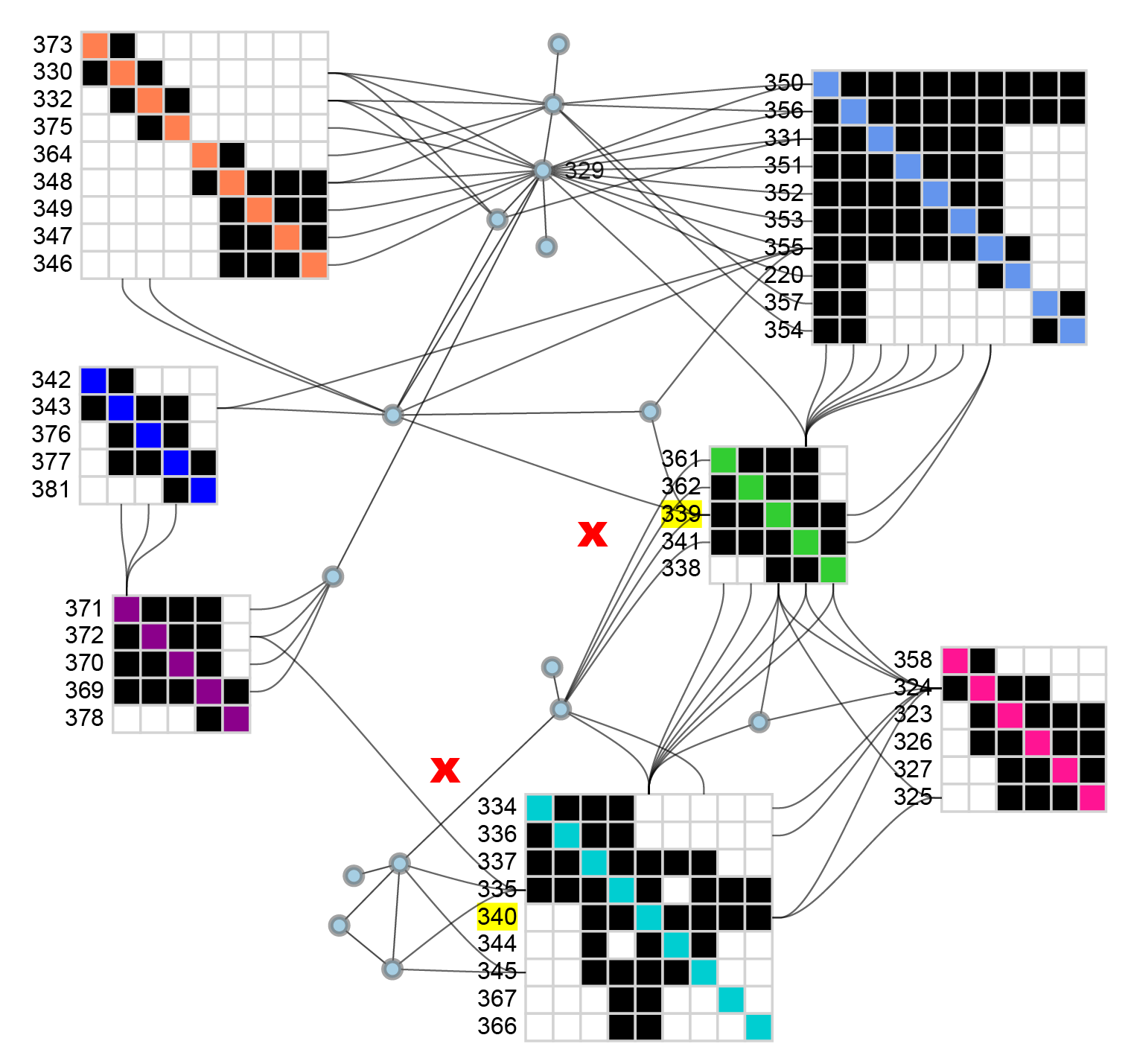}
		\label{fi:social-nt-t2}
	}\hfill
	\subfigure[\textsf{T3}]{%
		\centering
		\includegraphics[width=0.38\columnwidth]{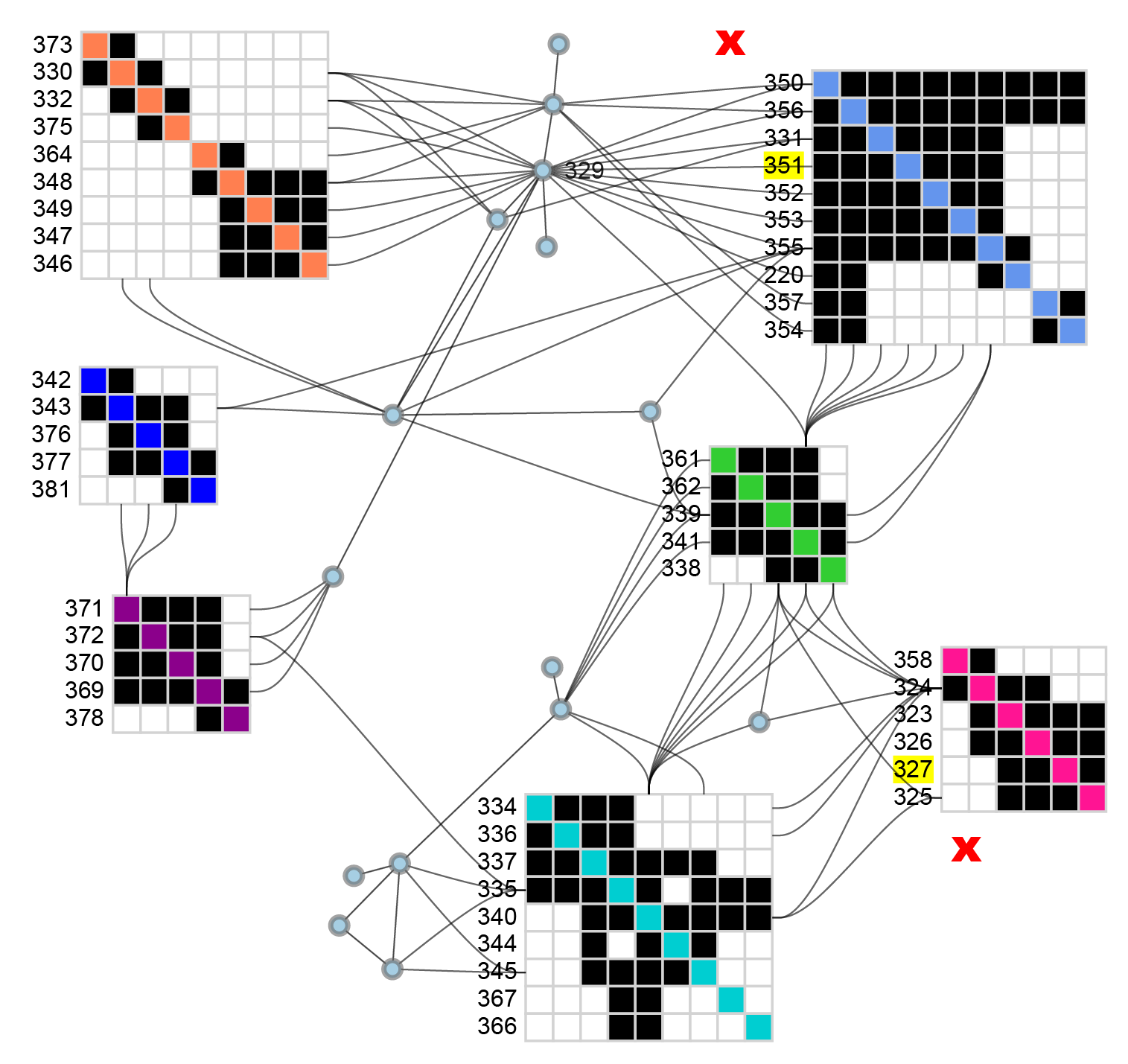}
		\label{fi:social-nt-t3}
	}\hfill
	\subfigure[\textsf{T4}]{%
		\centering
		\includegraphics[width=0.38\columnwidth]{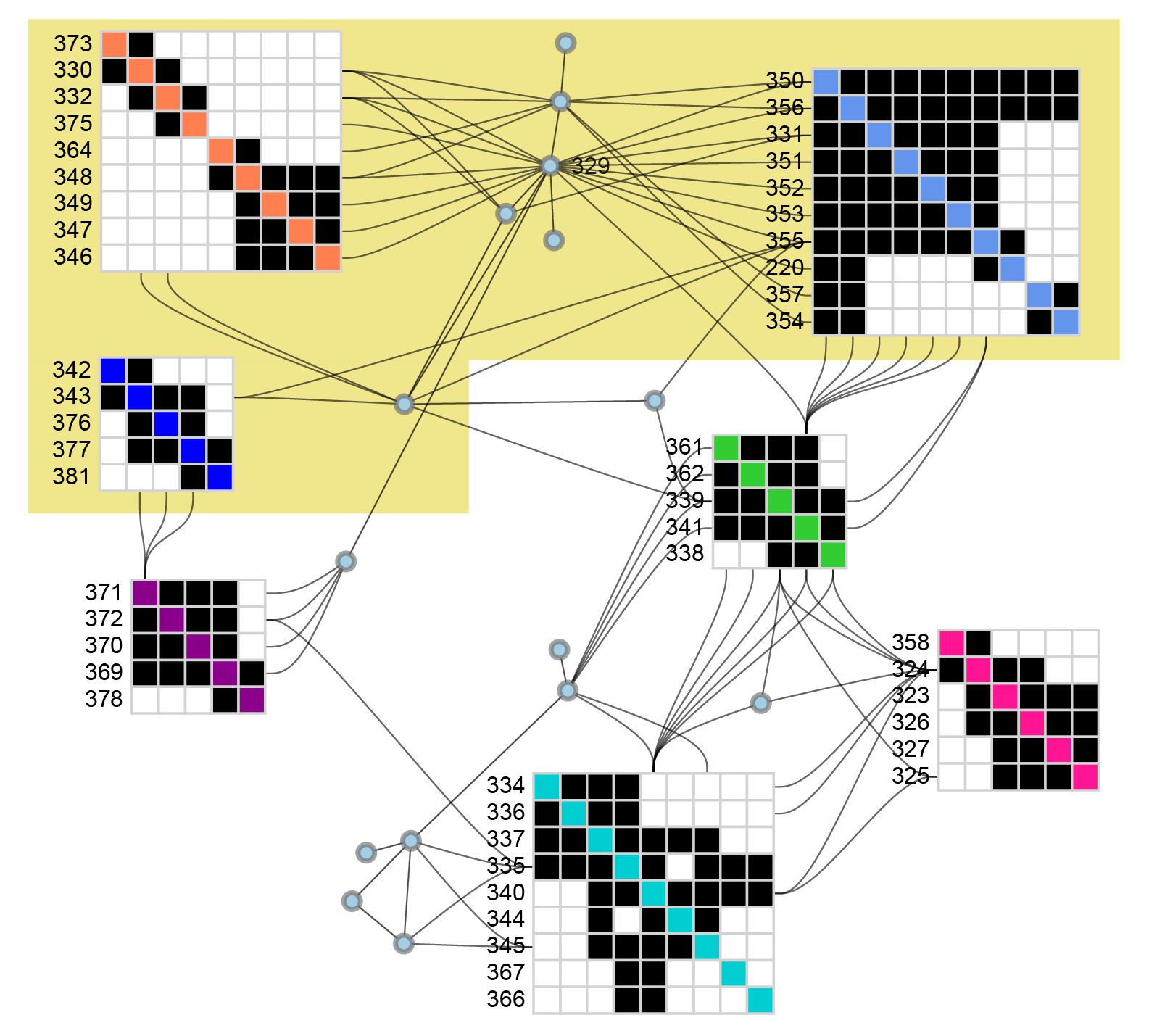}
		\label{fi:social-nt-t4}
	}\hfill
	\subfigure[\textsf{T5}]{%
		\centering
		\includegraphics[width=0.38\columnwidth]{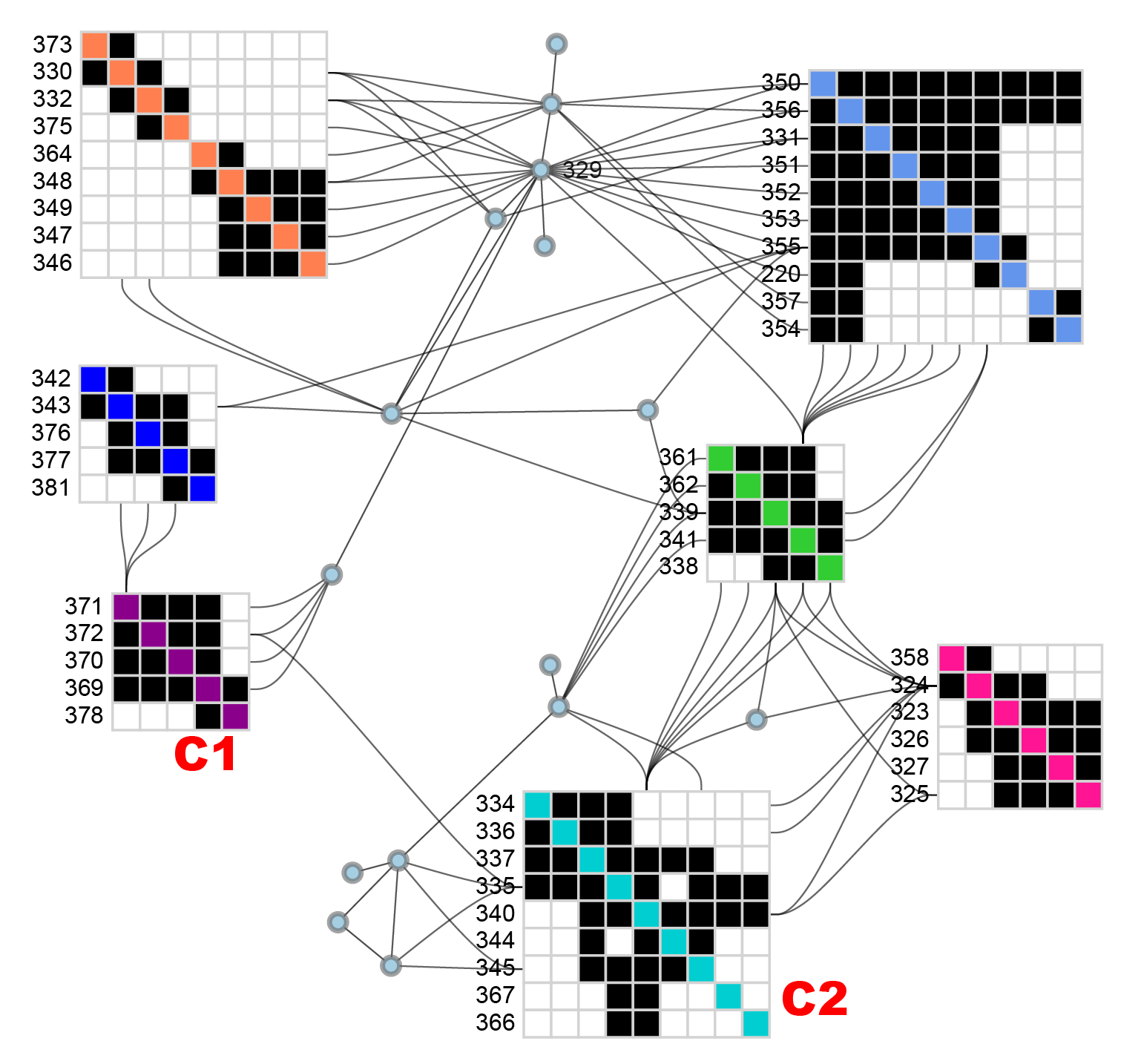}
		\label{fi:social-nt-t5}
	}\hfill
	\subfigure[\textsf{T6}]{%
		\centering
		\includegraphics[width=0.38\columnwidth]{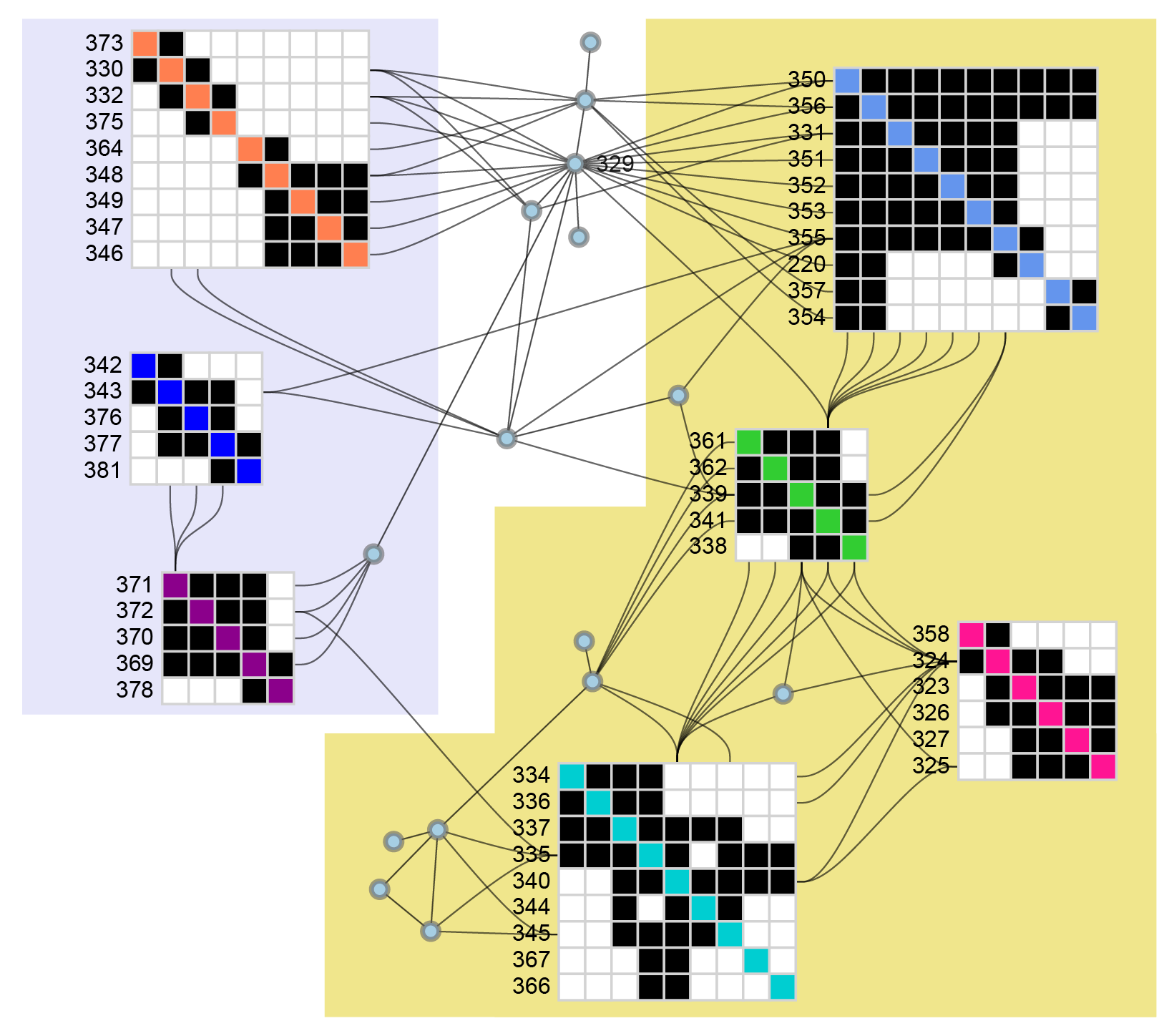}
		\label{fi:social-nt-t6}
	}\hfill
	\caption{Trials for the network \textsf{weavers} in the \nt model. For task \textsf{T3} $k=4$; for task \textsf{T4} the three labels are: 374, 334, 220.}\label{fi:trials-nt}
\end{figure}

\begin{figure}[h!]
	\centering
	\subfigure[\textsf{T1}]{%
		\centering
		\includegraphics[width=0.38\columnwidth]{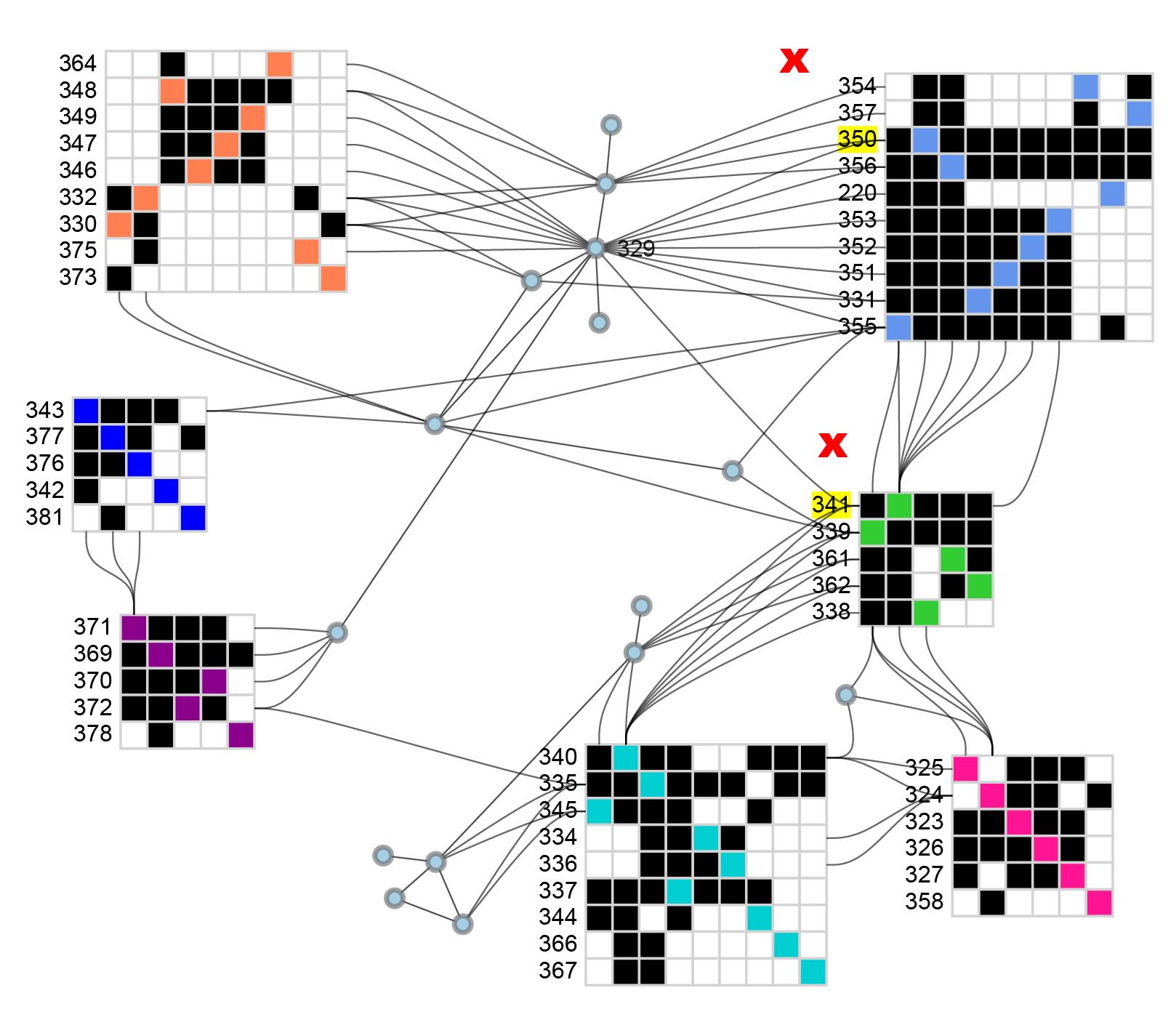}
		\label{fi:social-rci-t1}
	}\hfill
	\subfigure[\textsf{T2}]{%
		\centering
		\includegraphics[width=0.38\columnwidth]{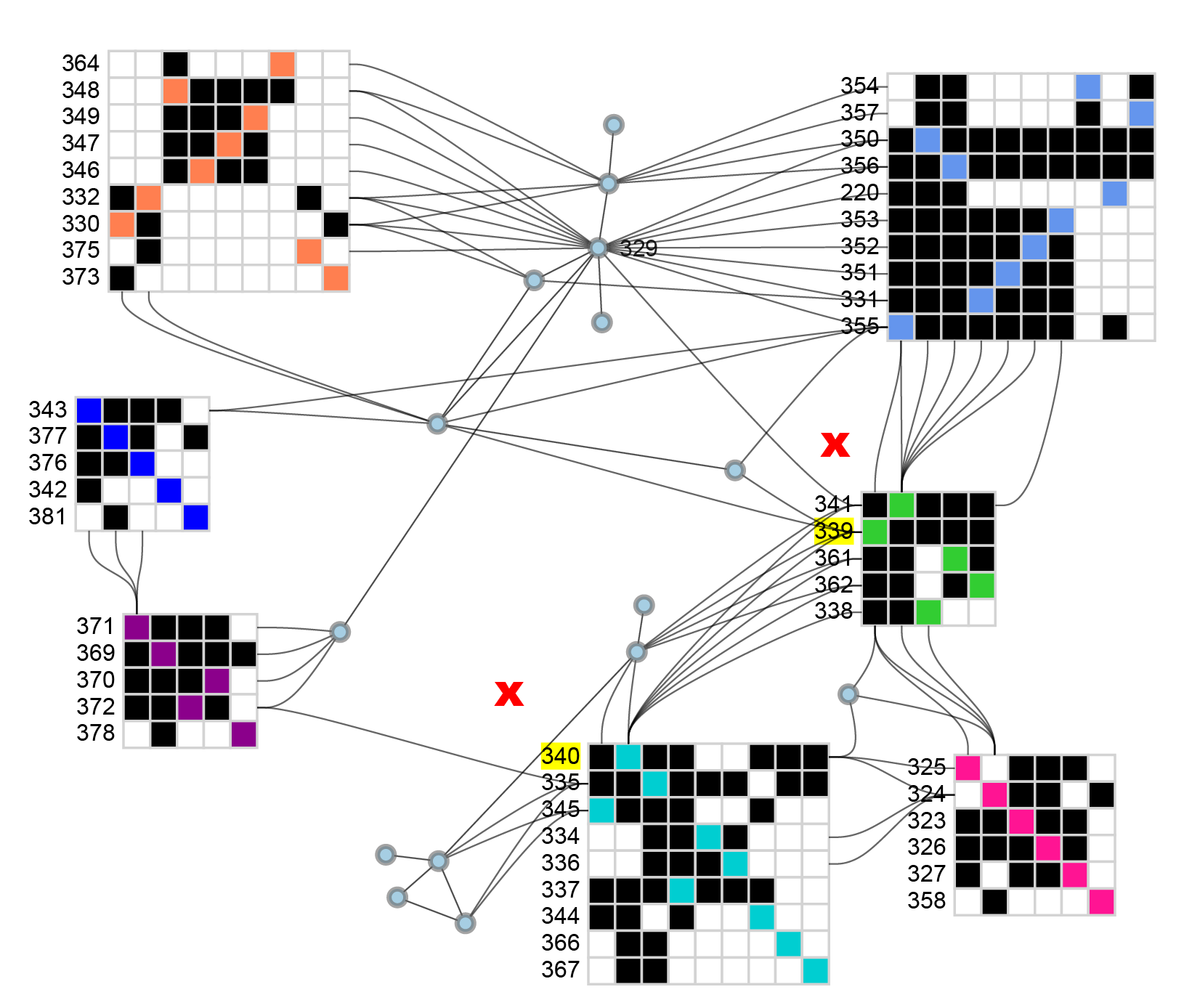}
		\label{fi:social-rci-t2}
	}\hfill
	\subfigure[\textsf{T3}]{%
		\centering
		\includegraphics[width=0.38\columnwidth]{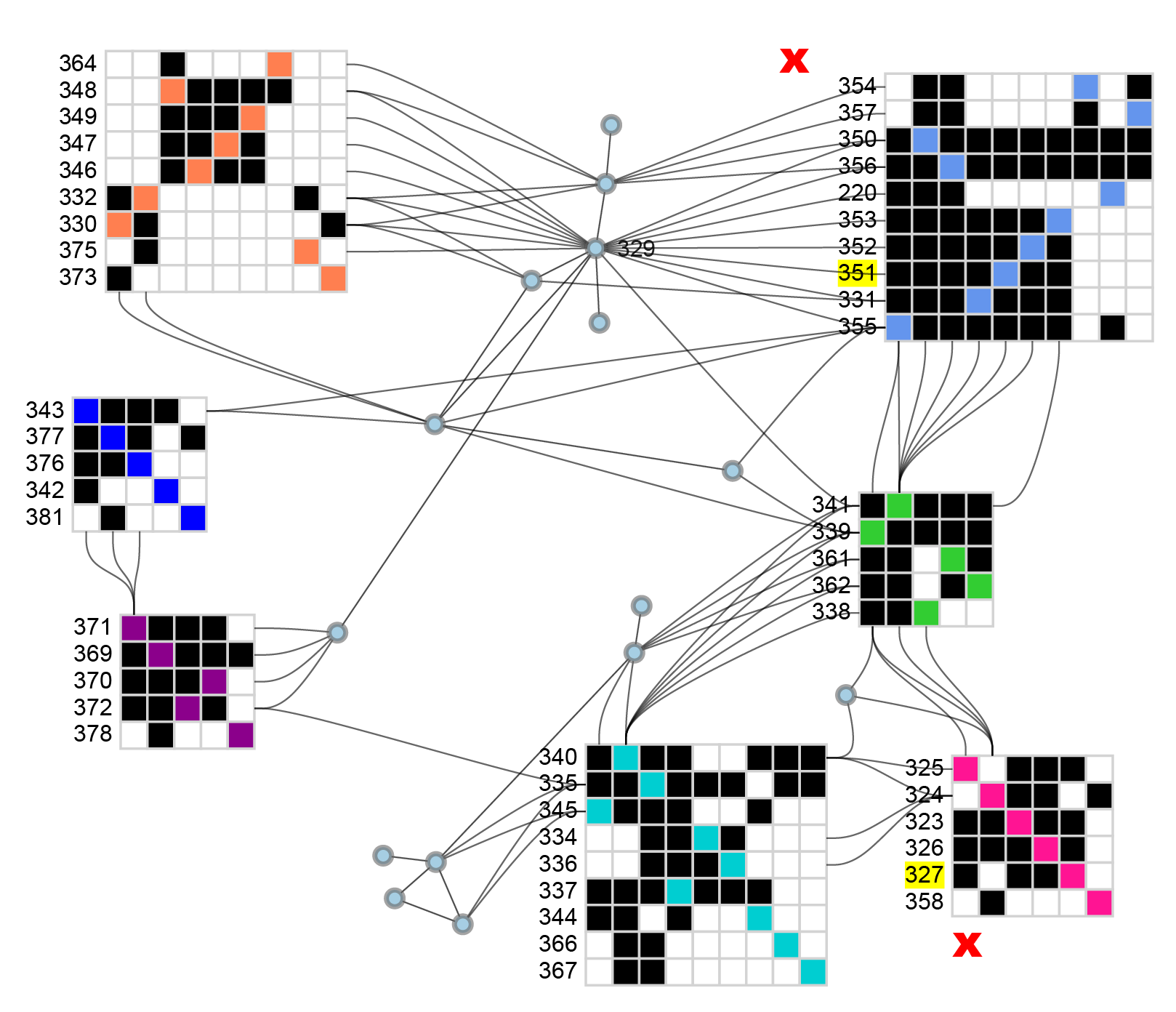}
		\label{fi:social-rci-t3}
	}\hfill
	\subfigure[\textsf{T4}]{%
		\centering
		\includegraphics[width=0.38\columnwidth]{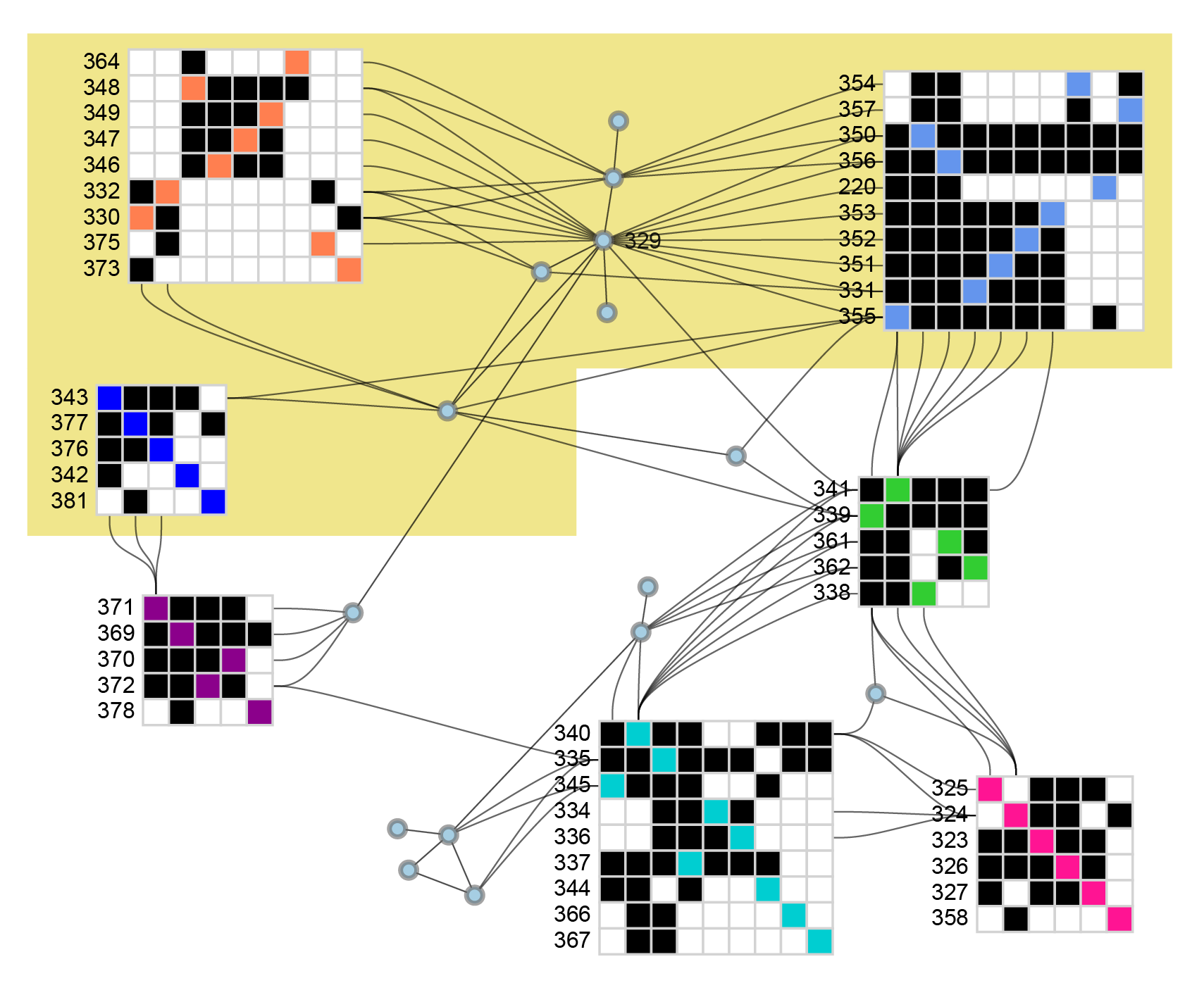}
		\label{fi:social-rci-t4}
	}\hfill
	\subfigure[\textsf{T5}]{%
		\centering
		\includegraphics[width=0.38\columnwidth]{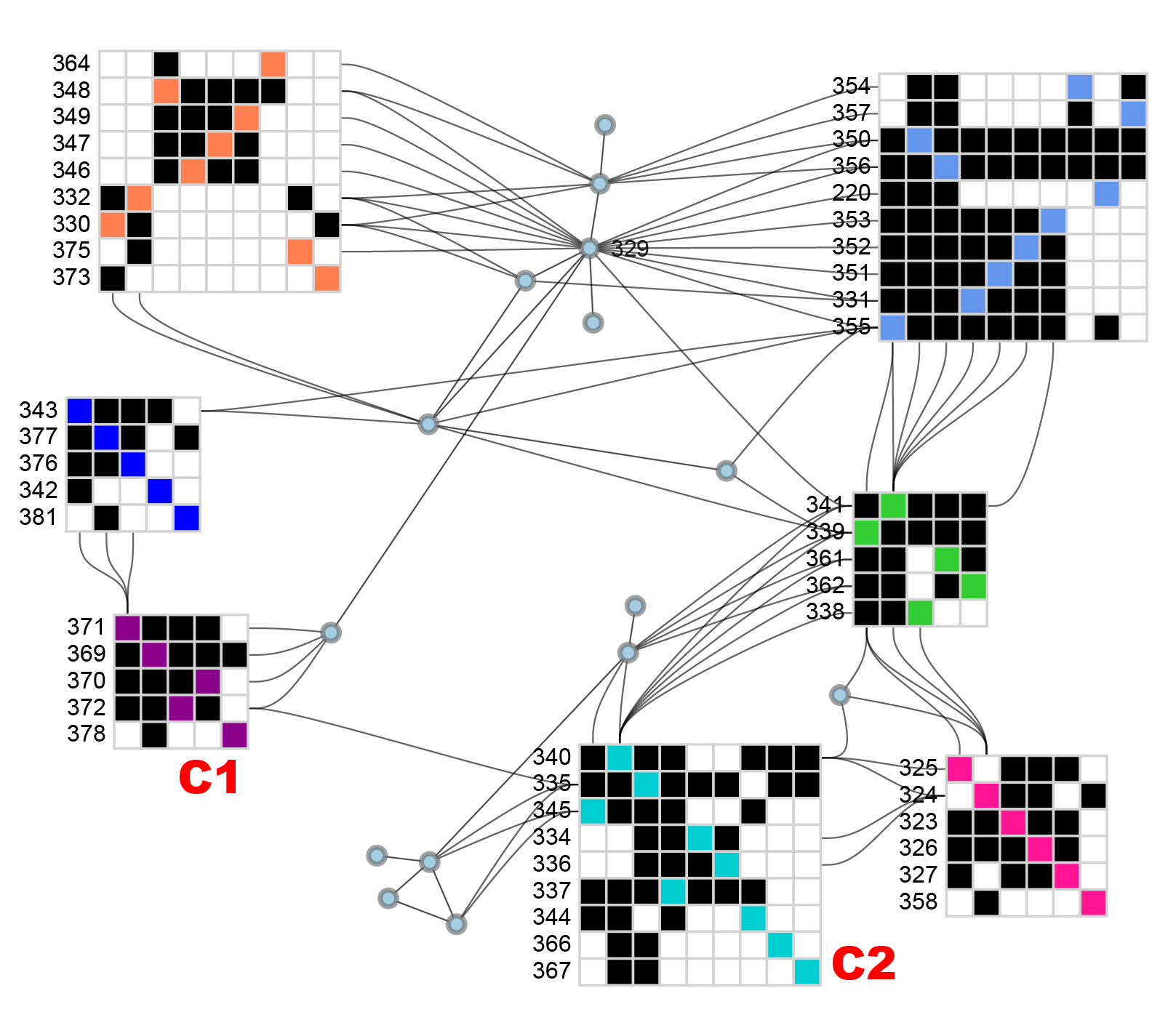}
		\label{fi:social-rci-t5}
	}\hfill
	\subfigure[\textsf{T6}]{%
		\centering
		\includegraphics[width=0.38\columnwidth]{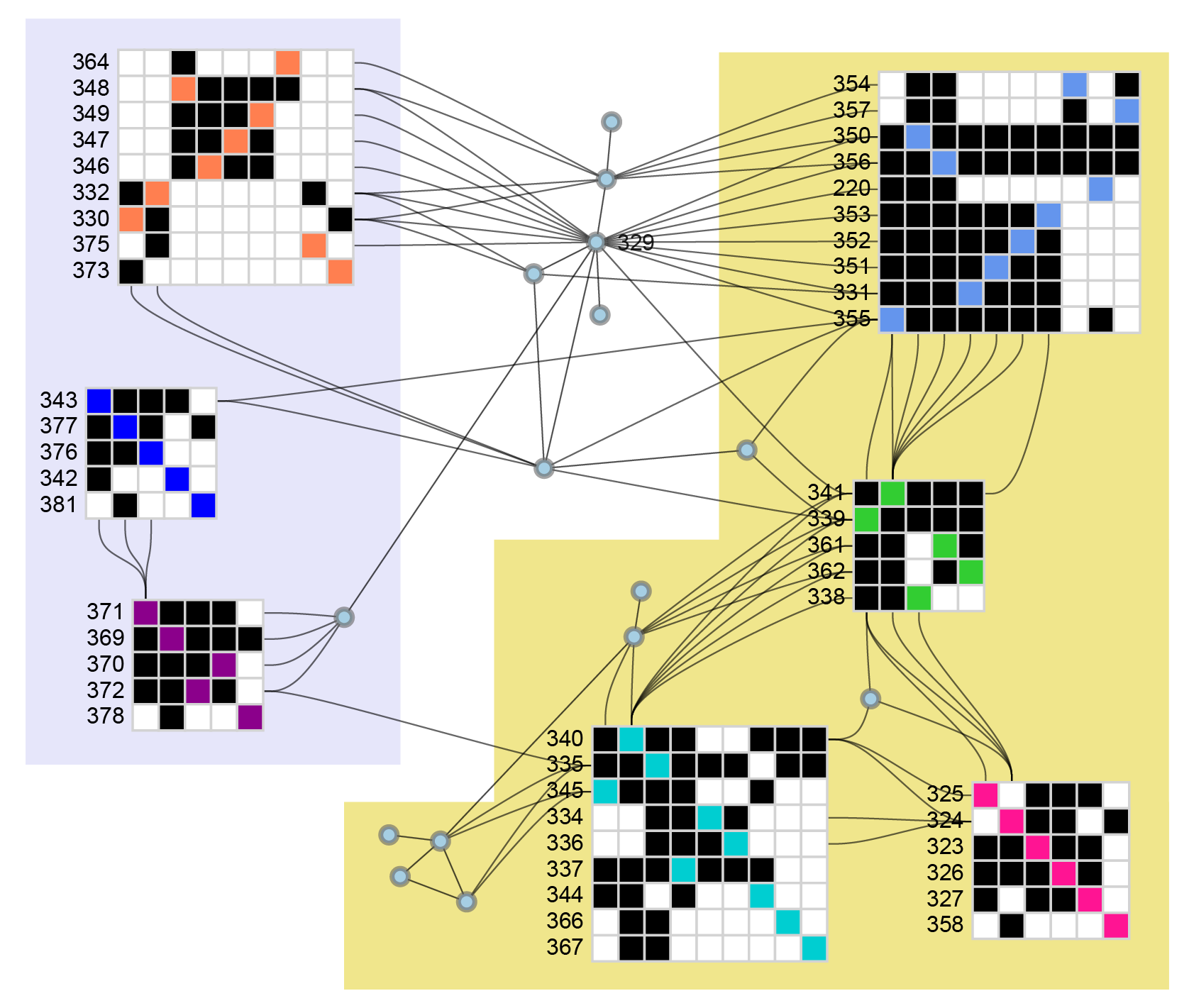}
		\label{fi:social-rci-t6}
	}\hfill
	\caption{Trials for the network \textsf{weavers} in the \rci model. For task \textsf{T3} $k=4$; for task \textsf{T4} the three labels are: 374, 334, 220.}\label{fi:trials-rci}
\end{figure}

\clearpage

\section{Additional Material about the Study Results}\label{app:morestudy}

\subsection{Data about the Participants to the Study}\label{app:charts}

\begin{figure}
	\centering
	\fbox{\includegraphics[width=0.46\textwidth,trim=20mm 25mm 25mm 20mm, clip]{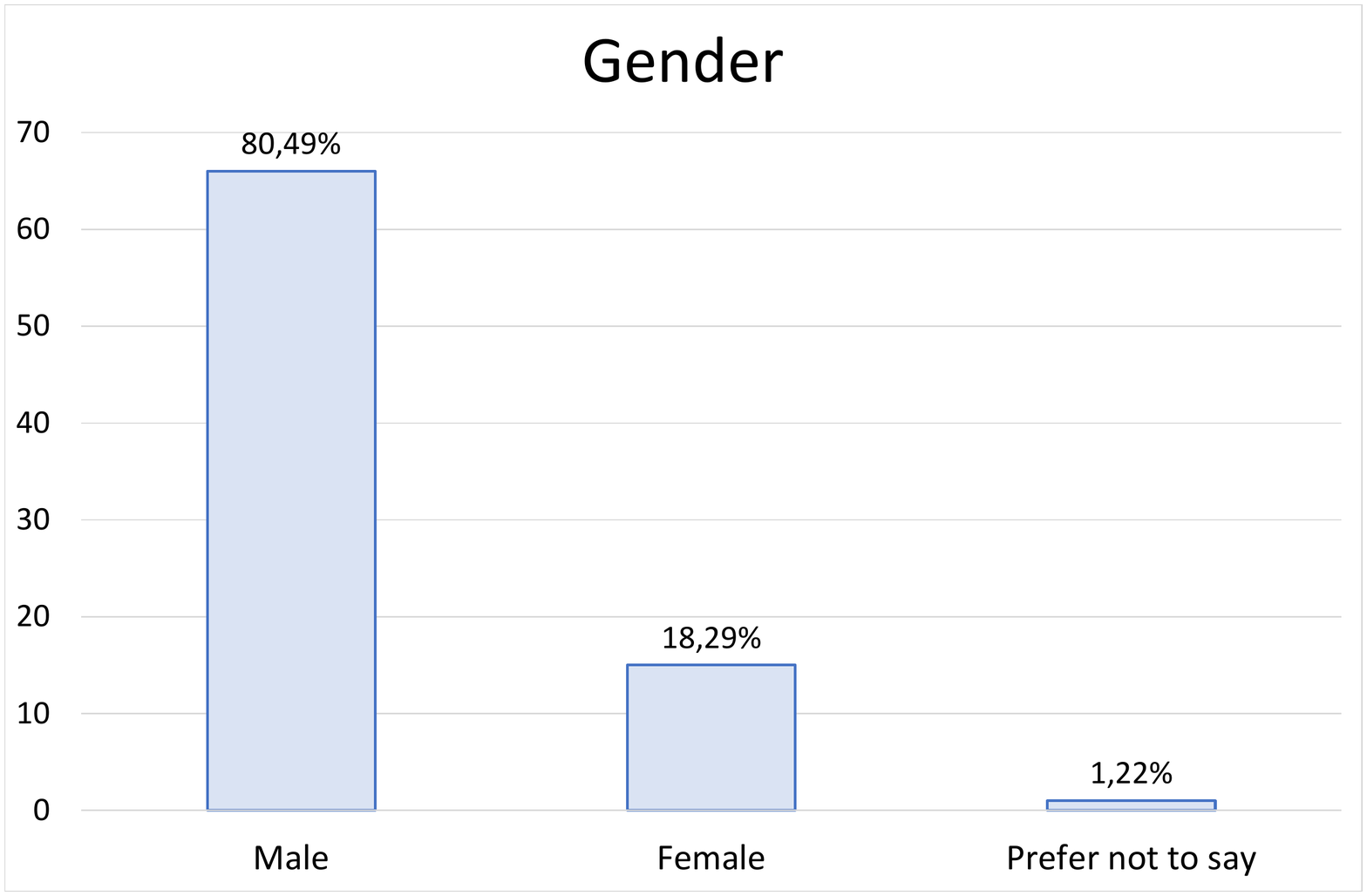}}
	\fbox{\includegraphics[width=0.46\textwidth,trim=20mm 25mm 25mm 20mm, clip]{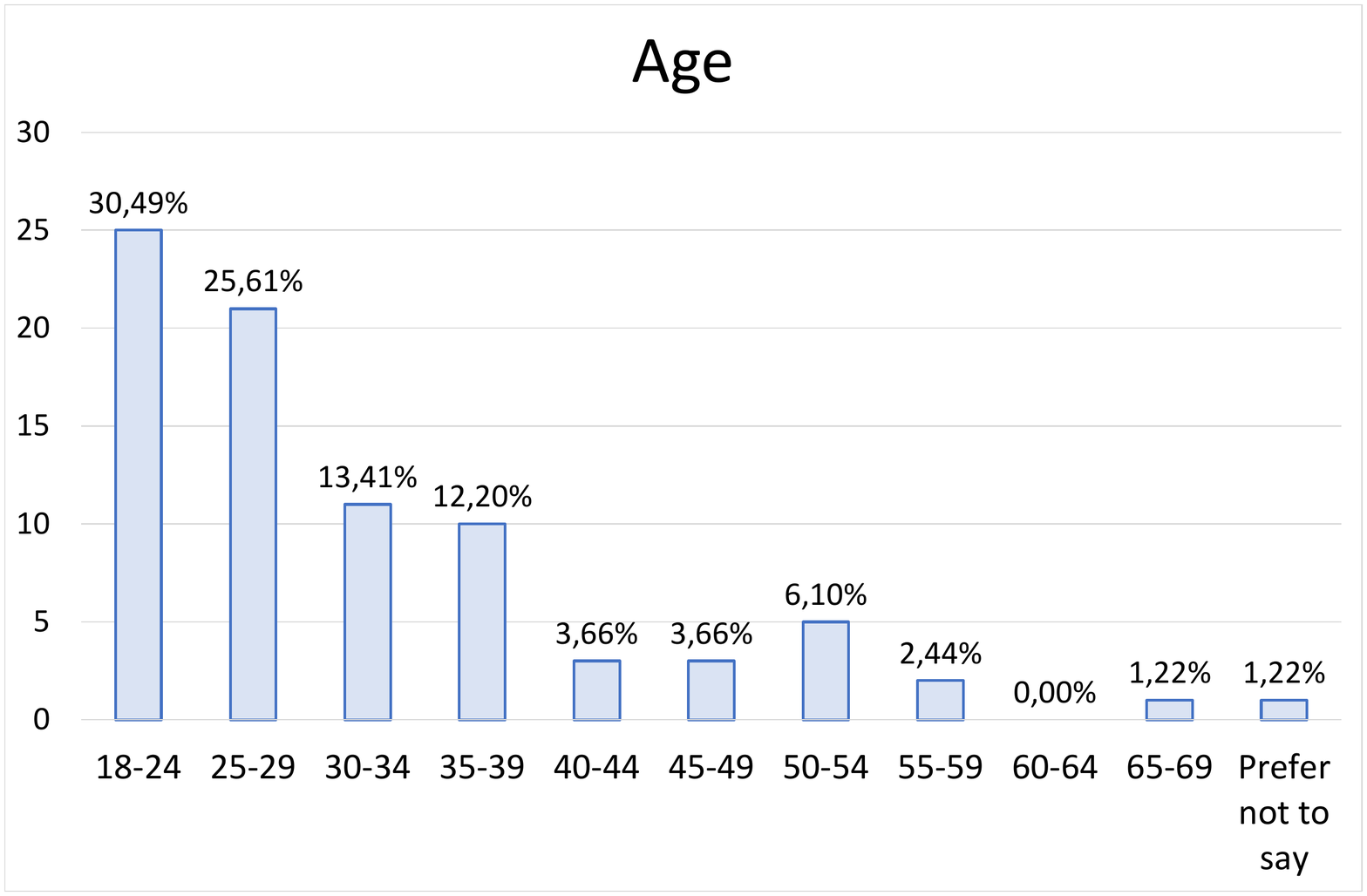}}\\
	\fbox{\includegraphics[width=0.46\textwidth,trim=20mm 25mm 25mm 20mm, clip]{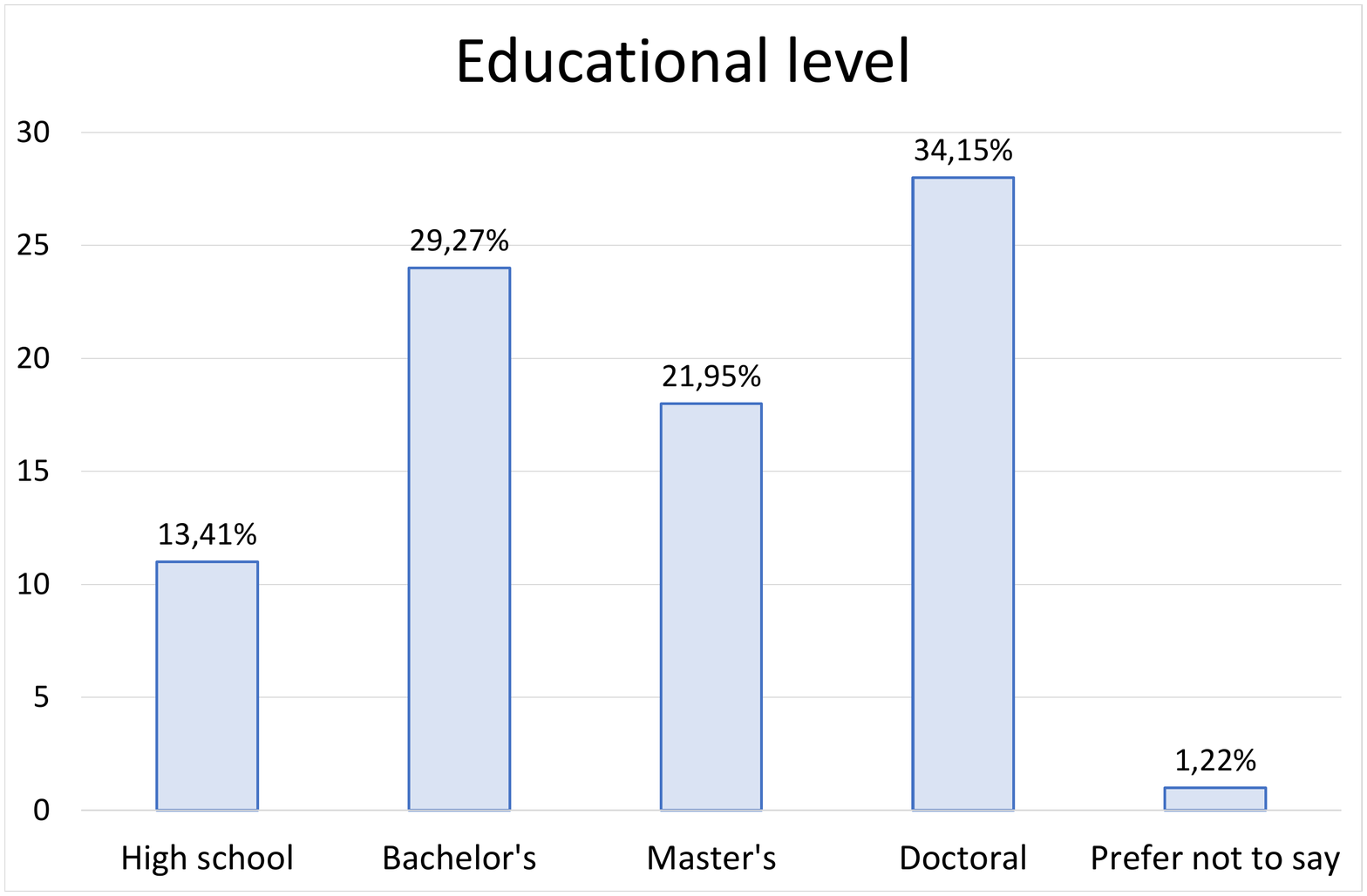}}
	\fbox{\includegraphics[width=0.46\textwidth,trim=20mm 25mm 25mm 20mm, clip]{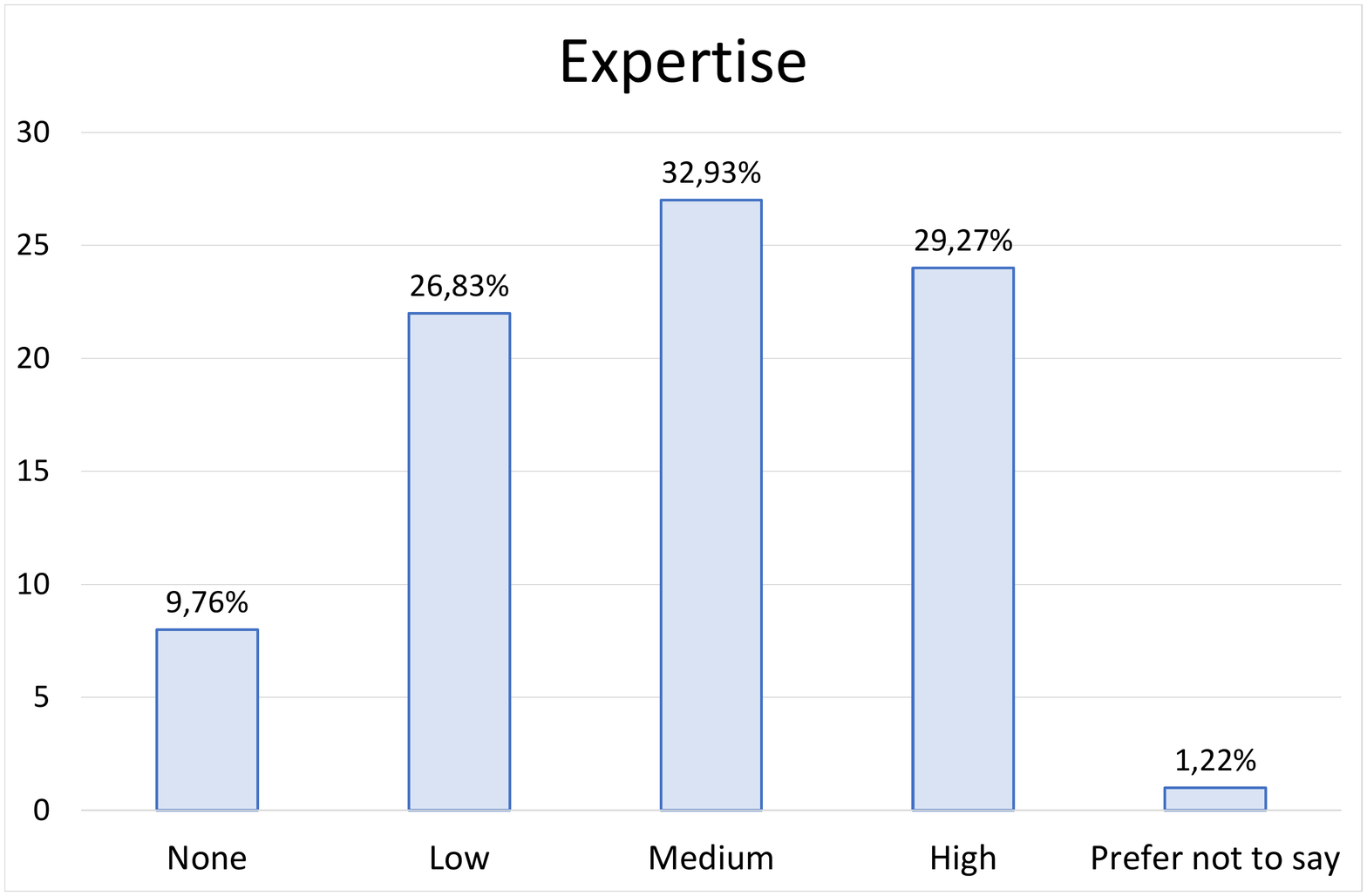}}\\
	\fbox{\includegraphics[width=0.46\textwidth,trim=20mm 25mm 25mm 20mm, clip]{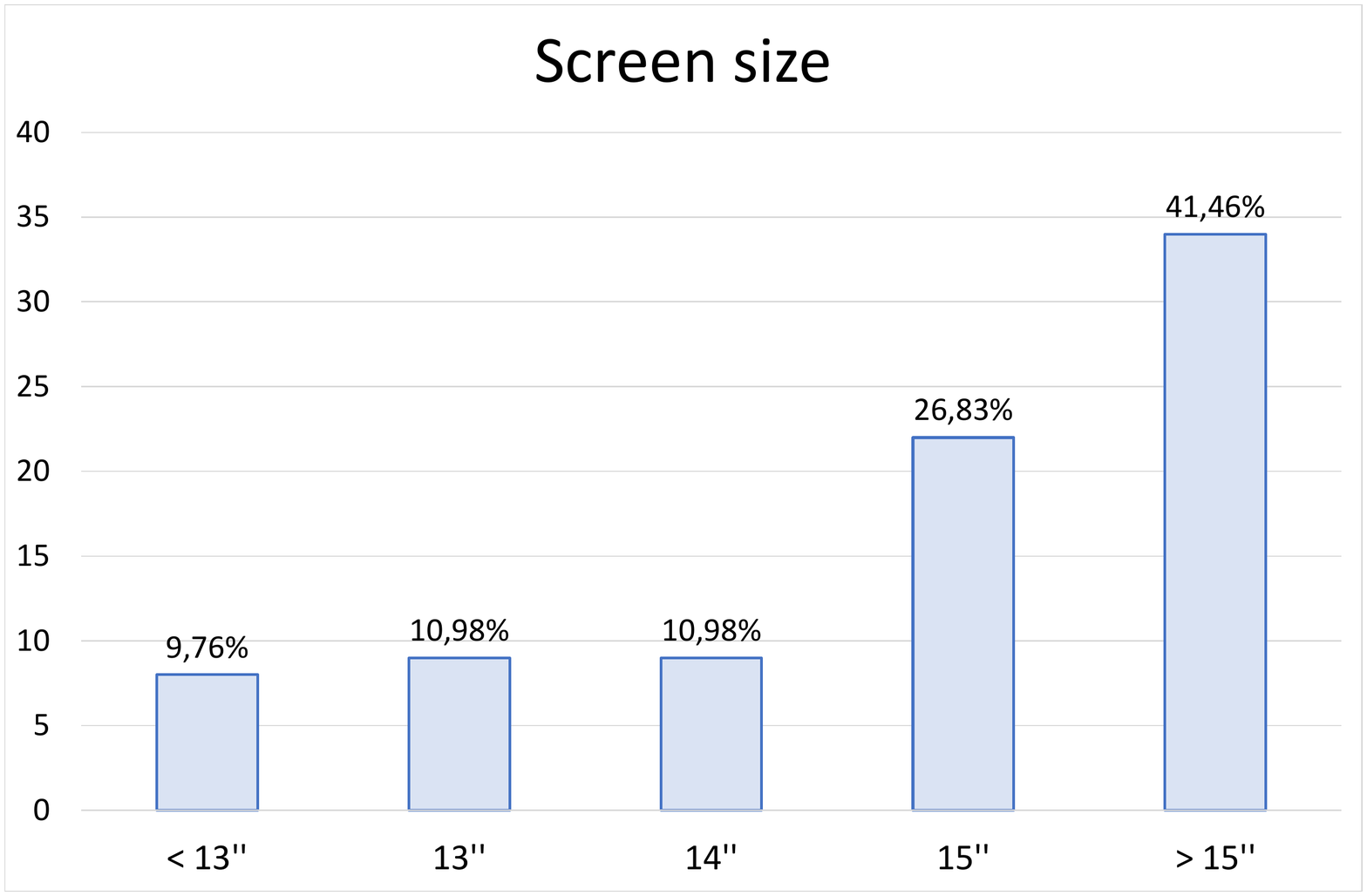}}
	\caption{Data about the participants to the study.\label{fi:demo}}
\end{figure}

\clearpage

%
%
%
%

\subsection{Discarded Questionnaires}\label{app:discarded} 

We collected questionnaires from 89 participants. We discarded seven tests for various reasons. One of the participants indicated in the free comments area that some images were not shown properly. Four participants indicated to have some color vision deficiency. Since they happened to be all assigned to the same model, we decided to discard their tests to avoid an unbalanced effect of this factor on the results of the experiment. Finally, since the experiment was fully online and thus not controlled, we discarded two tests whose total response time (i.e., the total time spent to answer the 18 trials) was an outlier. According to common practice, we consider the total response time of a test an outlier if it falls more than $1.5$ times IQR below the first quartile or above the third quartile. 

%

\subsection{Free Comments}\label{app:freecomments}

In what follows we report, for each model, a summary of the main free comments posted by the participants at the end of the study.

\smallskip\noindent 
\nl: All comments point out that the visual clutter caused by dense portions of the network makes the execution of some tasks difficult. This is coherent with the motivation behind the introduction of hybrid graph~visualizations.

\smallskip\noindent 
\cl: The comments point out that node duplication affects the drawing readability and makes it difficult to perform tasks related to cluster density. This is coherent with our rationale about Hypothesis~\textbf{H3}, i.e., node duplication may give the impression that a cluster is sparser than it actually is. Other comments highlight that cluster regions represented by circles with small diameter make intra-cluster edges difficult to distinguish in some cases.  

\smallskip\noindent 
\nt: Here, the main comments report a difficulty in reading inter-cluster edges and, in particular, their incidence to the matrices. 

\smallskip\noindent 
\rci: The main comment here is that it is somewhat counter-intuitive that the matrices do not use the same row and column order, and this has a negative impact on following paths in the network.

\smallskip
We further report that two participants found our definition of density (denoted as $d_1$ in \Cref{sse:tasks}) less intuitive than the alternative one (denoted as $d_2$ in \Cref{sse:tasks}).

\end{document}